\documentclass[onecolumn]{aastex63}

\usepackage[english]{babel}

\newcommand{\equ}[1]{\begin{eqnarray}
        #1
\end{eqnarray}}

\newcommand{\tianshu}[1]{{#1}}

\begin{document}

\title{Effects of Different Closure Choices in Core-Collapse Supernova Simulations}

\correspondingauthor{Tianshu Wang}
\email{tianshuw@princeton.edu}

\author[0000-0002-0042-9873]{Tianshu Wang}
\affil{Department of Astrophysical Sciences, Princeton, NJ, 08544}
\author[0000-0002-3099-5024]{Adam Burrows}
\affil{Department of Astrophysical Sciences, Princeton, NJ, 08544}

%\author{}
\date{September 26, 2022}

\begin{abstract}
The two-moment method is widely used to approximate the full neutrino transport equation in core-collapse supernova (CCSN) simulations, and different closures lead to subtle differences in the simulation results. In this paper, we compare the effects of closure choices on various physical quantities in {1D and 2D} time-dependent CCSN simulations with our multi-group radiation hydrodynamics code F{\sc{ornax}}. We find that choices of the 3rd-order closure relations influence the time-dependent simulations only slightly. Choices of the 2nd-order closure relation have larger consequences than choices of the 3rd-order closure do, but these are still small compared to the remaining variations due to ambiguities in some physical inputs such as the nuclear equation of state. We also find that deviations in Eddington factors are not monotonically related to deviations in physical quantities, which means that simply comparing the Eddington factors does not inform one concerning which closure is better.
\end{abstract}

\section{Introduction}
Neutrino transport has long been known to play an important role in the explosion of core-collapse supernovae (CCSN). Although some tests and simulations have been done by solving the angle-dependent transport equation directly \citep{smit2000,richers2017,nagakura2018,harada2019,harada2020,iwakami2020,iwakami2022}, the high computational cost of this method makes it very expensive to be employed for high spatial resolution, multi-dimensional (especially 3D), time-dependent CCSN simulations today.
One widely used approximation to the full transport equation is the two-moment method, also known as the M1 method \citep{murchikova2017} or the algebraic Eddington factor method \citep{just2015}. The 0th and 1st moments of the transport equation are solved together with a closure relation, in which the 2nd moment is approximated in terms of lower order moments. For non-static moving media, a closure relation for the 3rd angular moment is also needed when a Lorentz transformation is applied between co-moving and lab frames and/or transport calculations are done in the co-moving frame \citep{just2015,skinner2019}.

Various closures have been proposed in the literature. To test the performance of these closures, two types of comparisons have in the past been made. One was to compare the two-moment method with Monte Carlo (MC) simulations \citep{janka1992,murchikova2017,richers2017}; another was to compare with the discrete-ordinate (SN) method \citep{smit2000,richers2017,nagakura2018,harada2019,harada2020,iwakami2020,iwakami2022}. Differences due different closure choices have been noticed. For example, many of these works pointed out that there are regions at early times after bounce with Eddington factor below $1/3$ in the supernova core which can't be reproduced by most of the closures\footnote{except the Wilson and the Fermi-Dirac maximum entropy closures \citep{iwakami2022}}.

However, due to the high computation cost of the MC and the SN methods, these tests are either done at several snapshots of a CCSN simulation (e.g. \citet{murchikova2017}) or are carried out to only a few tens of milliseconds after bounce (e.g. \citet{iwakami2022}). As a result, the comparisons are limited to quantities like Eddington factors or tensors, and the hydrodynamic consequences remain unknown. Therefore, one has little knowledge concerning how the physical quantities of an explosion are influenced by a closure choice or the two-moment approximation to the transport equation \citep{mezzacappa2022}. In addition, comparisons in the literature of the 3rd moment closures are rare. Moreover, some closures don't have self-consistent 3rd-order moment relations and workers often interpolate between the optically-thin and -thick limits for the 3rd moment in the same way as the 2nd-order moment \citep{shibata2011}. Whether this is a proper way to deal with the 3rd moment remains to be tested. 

In this paper, we address some of these unanswered questions by comparing 1D {and 2D} time-dependent simulations using different closures carried out to at least 700 milliseconds after bounce. We find that most closures behave similarly, but that the Kershaw and the Wilson closures deviate noticably from the others by $5\%-10\%$ in shock radius {before the explosion}. Differences that arise in other physical quantities are all closely related to this shock position difference. In addition, we notice that deviations in Eddington factor profiles are not monotonically related to deviations in other quantities. This means that one cannot compare only the Eddington factors calculated using various closure relations to the results given by the MC and SN methods, since small deviations in the Eddington factor can still result in large differences in hydrodynamic quantities in a time-dependent CCSN simulation. However, we suggest that differences caused by various closure choices are generally small compared to differences due to remaining uncertainties in various physical inputs such as the choice of the equation of state {which may even change the explodability} \citep{couch2013,yasin2020}. Furthermore, we compare two different treatments of the 3rd-order closure and find that the differences caused by them can be neglected.

This paper is arranged as follows: Section \ref{sec:closure_summary} summarizes the closures used in this paper, Section \ref{sec:method} describes the numerical method and the simulations, and Section \ref{sec:result} compares various physical quantities of simulations done with different closures and progenitors. In Section \ref{sec:conclusion}, we summarize our salient results and conclude with various observations. The appendix derives many of the more challenging closures studied in the paper.

\section{List of Closures}
\label{sec:closure_summary}
Based on symmetries, the pressure and heat tensor can be written in the following forms \citep{just2015}:
\equ{
P_{ij}&=&ED_{ij} \nonumber\\
D_{ij}%&=\frac{3p-1}{2}D_{ij}^{\rm thin}+\frac{3(1-p)}{2}D_{ij}^{\rm thick}\\
&=&\frac{3p-1}{2f^2}f_if_j+\frac{3(1-p)}{2}\frac{1}{3}\delta_{ij} \nonumber\\
Q_{ijk}&=&cEH_{ijk} \nonumber\\
H_{ijk}%&=\frac{5q-3f}{2}H_{ijk}^{\rm thin}+\frac{f-q}{2}H_{ijk}^{\rm thick}\\
&=&\frac{5q-3f}{2}\frac{f_if_jf_k}{f^3}+\frac{f-q}{2}\frac{f_i\delta_{jk}+f_j\delta_{ki}+f_k\delta_{ij}}{f}\, ,
}
where $E$ is the energy density, $f_i=\frac{F_i}{E}$ is the normalized flux component, and $f=\frac{|\vec{F}|}{E}$ is the flux factor. Therefore, the pressure and heat tensor are determined by the parameter $p$ (also called the  Eddington factor) and by $q$, respectively. In 1D, the pressure and heat tensor have simple expressions: $P=pE$ and $Q=qcE$.

Given the angular distribution of a given energy group $\mathcal{F}(\theta,\phi)$, the occupancy rate $e$, the flux factor $f$, the Eddington factor $p$, and the $q$ parameter are given by:
\equ{
e&&=\int d\Omega \mathcal{F}(\theta,\phi) \nonumber\\
f&&=\frac{1}{e}\int d\Omega \cos(\theta)\mathcal{F}(\theta,\phi)\nonumber\\
p&&=\frac{1}{e}\int d\Omega \cos(\theta)^2\mathcal{F}(\theta,\phi) \nonumber\\
q&&=\frac{1}{e}\int d\Omega \cos(\theta)^3\mathcal{F}(\theta,\phi)\, .
}
In a multi-group context, different closure values are relevant for each energy group, because in general the angular distributions for different energy groups at the same spatial position are different.

For closures that don't have 3rd-order moment relations, it has been common to interpolate between the optically-thin and -thick limits using the Eddington factor in the same way as is done for the 2nd-order moment \citep{shibata2011}:
\equ{
H_{ijk}%&=\frac{5q-3f}{2}H_{ijk}^{\rm thin}+\frac{f-q}{2}H_{ijk}^{\rm thick}\\
&=&\frac{(3p-1)f}{2}\frac{f_if_jf_k}{f^3}+\frac{3(1-p)f}{10}\frac{f_i\delta_{jk}+f_j\delta_{ki}+f_k\delta_{ij}}{f}
}
which is equivalent to
\equ{
q&=&\frac{(3p+2)f}{5}\, .
}
{For convenience, in the rest of the paper we refer to this particular interpolation combining the optically-thin and -thick limits as the Shibata interpolation.}
The $p$ and $q$ parameters given by different closures are plotted in Figure \ref{fig:pq}.

In this section we list seven closures. Two of them (Levermore and maximum entropy) are physically motivated, while the others are either interpolations between the optically-thin and -thick limits or are fitted to Monte Carlo or SN simulations.

\subsection{Closures with 3rd-Order Moment Relation}
\subsubsection{Levermore Closure}
\label{sec:levermore}
The Levermore closure assumes that there is some inertial frame in which the radiation field is isotropic. It was first proposed in \citet{levermore1984}, and the 3rd-moment closure can be found in \citet{vaytet2011}:
\equ{
p&=&\frac{3+4f^2}{5+2\sqrt{4-3f^2}} \nonumber\\
q&=&\frac{1}{\left(-2+a\right)^5}\Bigg(4 f^3 \left(286-89 a\right)+576 f \left(-2+a\right)+3 f^5 \left(-80+9
a\right) \nonumber\\&&
\left.-48 \left(f^6+f^2 \left(42-15 a\right)+3 f^4 \left(-5+a\right)
+16 \left(-2+a\right)\right)
\text{\rm arctanh}\left[\frac{-2+a}{f}\right]\right)\nonumber\\
a&=&\sqrt{4-3 f^2}\, .
}
Although an isotropic distribution in the inertial frame is also the maximum entropy distribution in that frame, it is different from the one obtained from the maximum entropy closure. The reason is that the Lorentz transformation changes the neutrino energy along each direction differently.  The Levermore closure assumes the radiation field to be monochromatic in the new inertial frame in which the distribution is isotropic, while the maximum entropy closure assumes the radiation field to be monochromatic in the original one. Since the monochromatic energy groups are defined in the original frame, the physical motivation for the Levermore closure is not as strong as that for the maximum entropy closure. {However, the advantage of the Levermore closure is that it is covariant under a Lorentz transformation; thus, the relations between energy, flux, and pressure are consistent whether it is applied in the lab frame or in the comoving frame.}

\subsubsection{Minerbo and Maximum Entropy Closures}
These closures are derived from the idea that the most likely radiation angular distribution given energy density and flux is the one that maximizes the entropy. {By maximizing the entropy for Maxwell-Boltzmann (Minerbo) and Fermi-Dirac distributions (MEFD)}, \citet{minerbo1978} and \citet{cernohorsky1994} obtained the following closures:
\equ{
{\rm Minerbo:\ } &&p=\frac{1}{3}+\frac{2f^2}{15}(3-f+3f^2) \nonumber\\
&&q=\frac{f}{75}(45+10f-12f^2-12f^3+38f^4-12f^5+18f^6) \nonumber\\
{\rm MEFD:\ } &&p\approx\frac{1}{3}+\frac{2}{3}(1-e)(1-2e)\chi(\frac{f}{1-e}) \nonumber\\
&&\chi(x)=L^{-1}(x)\approx x^2(3-x+3x^2)/5\, ,
}
where $L(x)=\coth x-\frac{1}{x}$ is the Langevin function.
The Bose-Einstein case is not listed here since neutrinos are fermions. {The maximum packing closure in \citet{smit2000} is also not included because it's the forward-peak limit of the MEFD closure.} Note that the MEFD analytic formula provided in \citet{cernohorsky1994} actually depends on the approximation that $\frac{p-\frac{1}{3}}{p_{\rm max}-\frac{1}{3}}$ is a constant (where $p_{\rm max}=1-2e+\frac{4}{3}e^2$ and $e$ is the occupancy), which is accurate to $0.1\%$. To reach a higher accuracy or to calculate the 3rd-moment relation, the MEFD closure has to be calculated numerically.

The MEFD closure has in the past been preferred when the Eddington factors/tensors were compared to Monte Carlo or SN simulation results \citep{murchikova2017,iwakami2022}. Its motivation is physically straightforward, {but it is not Lorentz covarient.} %but the distribution it gives is different if it is applied in different frames. We note that \citet{iwakami2022} find some differences when the MEFD closure is applied in the lab and comoving frames.}

\subsubsection{Nagakura Closure}
In \citet{nagakura2021}, the angular distribution in a CCSN supernova simulation with the SN method is fitted using the following function
\equ{
\ln \mathcal{F}(\mu)= 
\left\{
	\begin{array}{c}
  a\mu^2+b\mu+c \,\,\,\,(\mu>\mu_0)\\
  d\mu^2+g\mu+h \,\,\,\,(\mu<\mu_0)\, ,\\
\end{array}
\right.
}
where $\mu=\cos(\theta)$. The seven parameters $a$, $b$, $c$, $d$, $g$, $h$ and $\mu_0$ are functions of the flux factors and are interpolated from the data table provided in \citet{nagakura2021}. The Eddington factor and $q$ parameter can then be calculated using:
\equ{
p=\frac{\int d\mu \mu^2\mathcal{F}(\mu)}{\int d\mu \mathcal{F}(\mu)} \nonumber\\
q=\frac{\int d\mu \mu^3\mathcal{F}(\mu)}{\int d\mu \mathcal{F}(\mu)}\, .
}
In Figure \ref{fig:pq}, we can see that the Nagakura closure almost overlaps the Levermore closure. Therefore, the SN simulations in \citet{nagakura2021} support the Levermore closure more than the maximum entropy closure. In addition to the code differences, \citet{nagakura2021} aims to reconstruct the full angular distribution of the neutrino field, while \citet{iwakami2022} aims to compare only the 2nd-order moments. This might be another explanation of their different preferences.

\subsection{Closures without 3rd-Order Moment Relations}
As described above, the $q$ factors of closures without intrinsic 3rd-order moment relations can for the purposes of simulation be given by $q=\frac{(3p+2)f}{5}$. This is what we do in the analysis below.
\subsubsection{Kershaw Closure}
The Kershaw closure \citep{kershaw1976} is a simple interpolation between the optically-thin and -thick limits. The closure reads
\equ{
p=\frac{1+2f^2}{3}\, .
}
\subsubsection{Wilson Closure}
Wilson proposed a flux-limiter for neutrino diffusion in \citet{wilson1975}. Since \citet{levermore1984} derives a one-to-one relation between closures and flux limiters, the Wilson flux limiter is equivalent to the following closure \citep{murchikova2017}:
\equ{
p=\frac{1}{3}-\frac{1}{3}f+f^2\, .
}

\subsubsection{Janka Closures}
In \citet{janka1991,janka1992}, Janka presented analytic fits to Monte Carlo neutrino transport calculations in PNS envelopes. his closure is parametrized as
\equ{
p=\frac{1}{3}(1+af^m+(2-a)f^n)\, .
}
Following \citet{murchikova2017}, we consider two sets of parameters: Janka1 with \{$a=0.5, m=1.3064, n=4.1342$\} and Janka2 with \{$a=1, m=1.345, n=5.1717$\}. The former is fitted from electron neutrino distributions, while the latter is obtained from the ``$\mu$" neutrino radiation field.

\section{Method}
\label{sec:method}
To investigate the influences of different closures on several physical quantities like shock radii and heating rates, we consider four massive-star stellar evolution progenitors from \citet{sukhbold2018} with ZAMS masses 13, 16, 20 and 25 $M_\odot$. Each progenitor was simulated with all closures listed in Section \ref{sec:closure_summary} using our multi-group radiation hydrodynamics code F{\sc{ornax}} \citep{skinner2019}. We used the SFHo equation of state (EOS) of \citet{steiner2013}, consistent with most known laboratory nuclear physics constraints \citep{tews2017}. All of these models were run with 1024 radial zones, and with 12 logarithmically-distributed energy groups for each of our three neutrino species (electron type, anti-electron type, and the rest bundled as ``$\mu$"-type). 

F{\sc{ornax}} solves the zeroth and first moments of the frequency-dependent comoving-frame radiation transport equation. Keeping all terms to $\mathcal{O}(v/c)$ and dropping terms proportional to the fluid acceleration, the
monochromatic radiation moment equations {without GR effects} can be written as \citep{skinner2019,vartanyan2019}:
\equ{
E_{\nu,t} + (F_{\nu}^{i}+v^iE_{\nu})_{;i} + v^i_{\,\,;j}[P_{\nu i}^j-\partial_\nu (\nu P_{\nu i}^j)]&=& R_{\nu E} \label{equ:3d-E}\\
F_{\nu j,t} + (c^2P_{\nu j}^{i}+v^iF_{\nu j})_{;i} + v^i_{\,\,;j}F_{\nu i}-v^i_{\,\,;k}\partial_\nu (\nu Q_{\nu ji}^k)&=& R_{\nu F_j}\, , \label{equ:3d-F}
}
where $i,j,k$ are spatial indices and {$\nu$ is the group neutrino energy}. $R_{\nu E}=j_\nu-c\kappa_\nu E_\nu$ and $R_{\nu F_j}=-c(\kappa_\nu+\sigma_\nu)F_{\nu j}$ are source terms that account for interactions between radiation and matter, where $j_\nu$ is the emissivity, $\kappa_\nu$ is the absorption coefficient, and $\sigma_\nu$ is the scattering coefficient. These interaction terms are independent on the closure choices. 

The second- and third-order moments in Equations \ref{equ:3d-E} and \ref{equ:3d-F}, $P_{\nu ij}$ and $Q_{\nu ijk}$, are given by the closure relations described in Section \ref{sec:closure_summary}. In the spherical symmetric case, the equations reduce to the following form:
\equ{
E_{\nu,t} + (F_{\nu}+vE_{\nu})_{;r} + v_{;r}[pE_{\nu}-\partial_\nu (\nu pE_{\nu})]&=& R_{\nu E} \label{equ:1d-E}\\
F_{\nu,t} + (c^2pE_{\nu}+vF_{\nu})_{;r} + v_{;r}[F_{\nu}-\partial_\nu (\nu qcE_{\nu})]&=& R_{\nu F}\, . \label{equ:1d-F}
}
We see in Eqs. \ref{equ:1d-E} and \ref{equ:1d-F} how the $p$ and $q$ factors are involved in neutrino transport. {Because the $\partial_\nu$ term containing $q$ disappears after integration over neutrino energy and $q$ is found only in this one term, we speculate that $q$ mostly affects the spectra of the neutrinos. However, $p$ occurs in both the $\partial_\nu$ and spatial derivative terms and it can directly modify both the spectra and the effective luminosities in the lab frame.}

We first do 1D simulations for all progenitors and closures. Although 1D simulations don't explode due to the absence of neutrino-driven turbulence, their non-chaotic nature enables us to measure and understand closure-induced differences in physical quantities. {We then choose one explosive progenitor and do 2D simulations with different closures. The 2D simulations are done with 128 zones along $\theta$ direction.}
%A few 2D simulations were done later to study how multi-dimensional effects may modify the closure effects. 
All these simulations were run to at least 0.7 second after bounce.

\section{Results}
\label{sec:result}

\subsection{1D Simulations}
\begin{figure}
    \centering
    \includegraphics[width=0.48\textwidth]{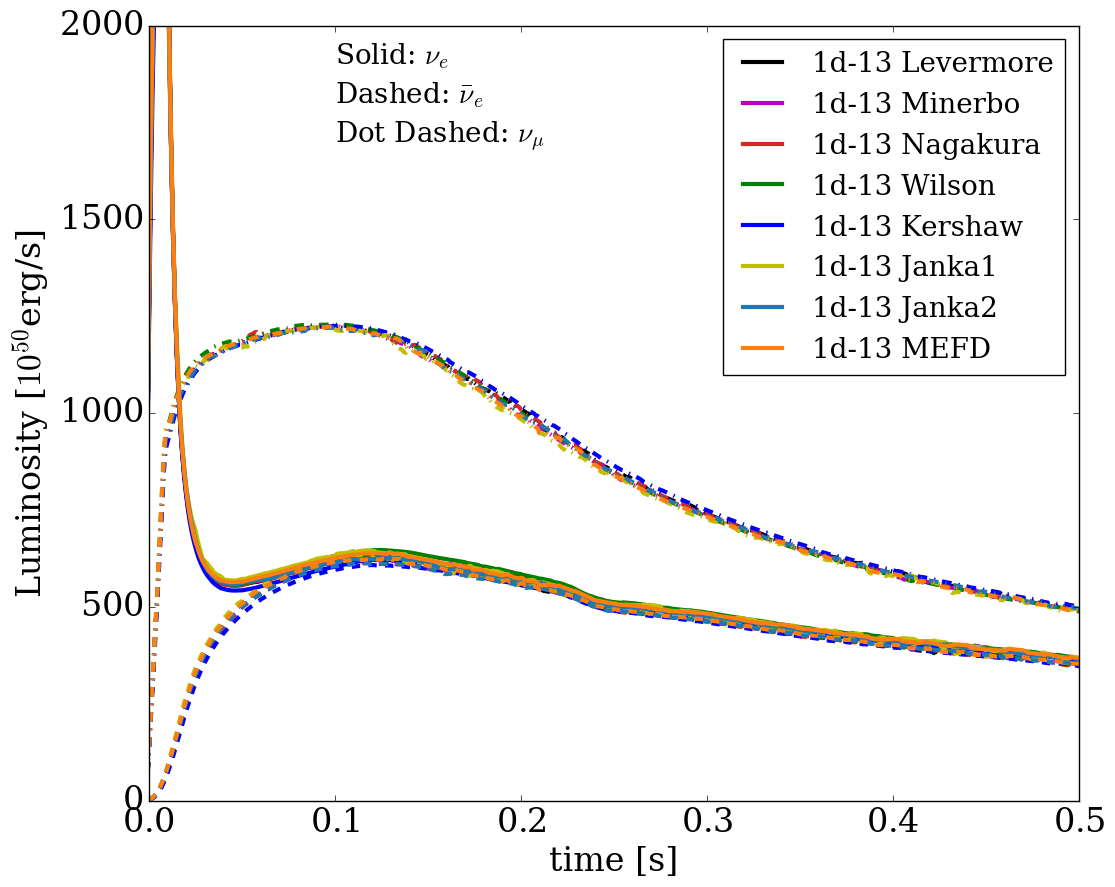}
    \includegraphics[width=0.48\textwidth]{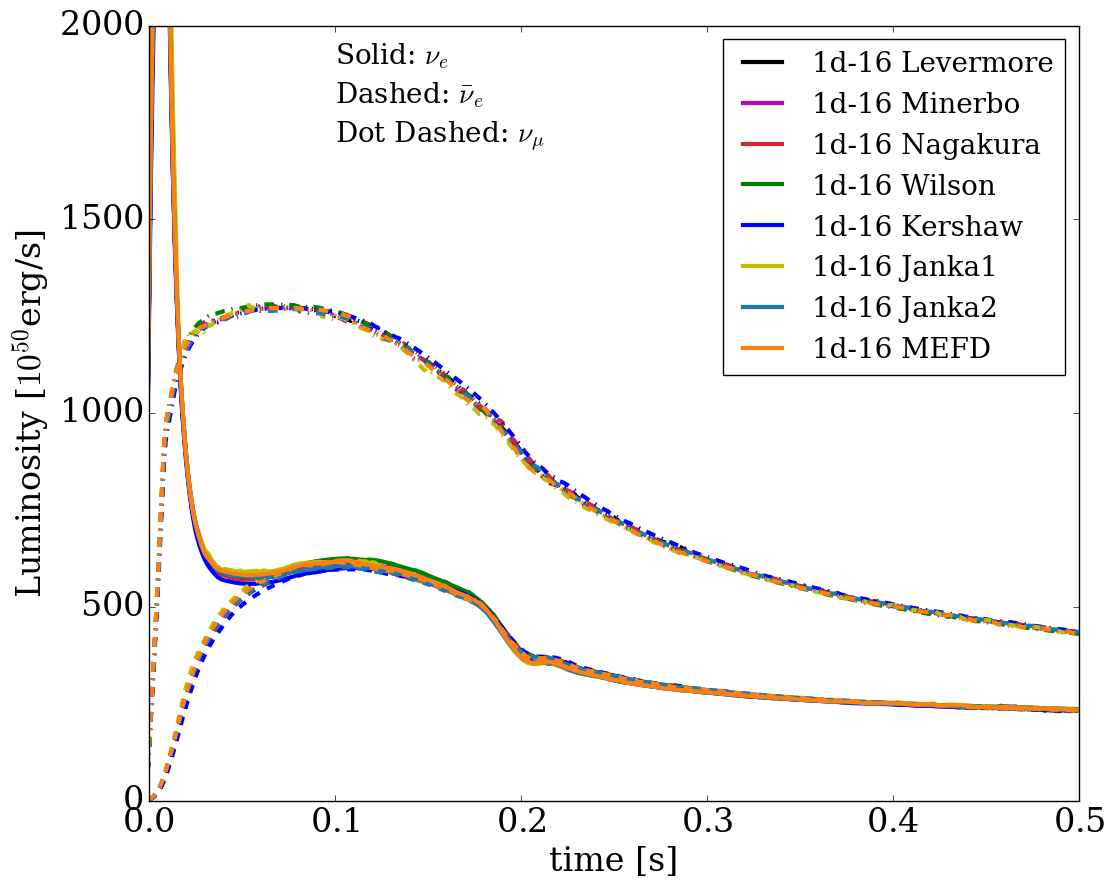}
    \includegraphics[width=0.48\textwidth]{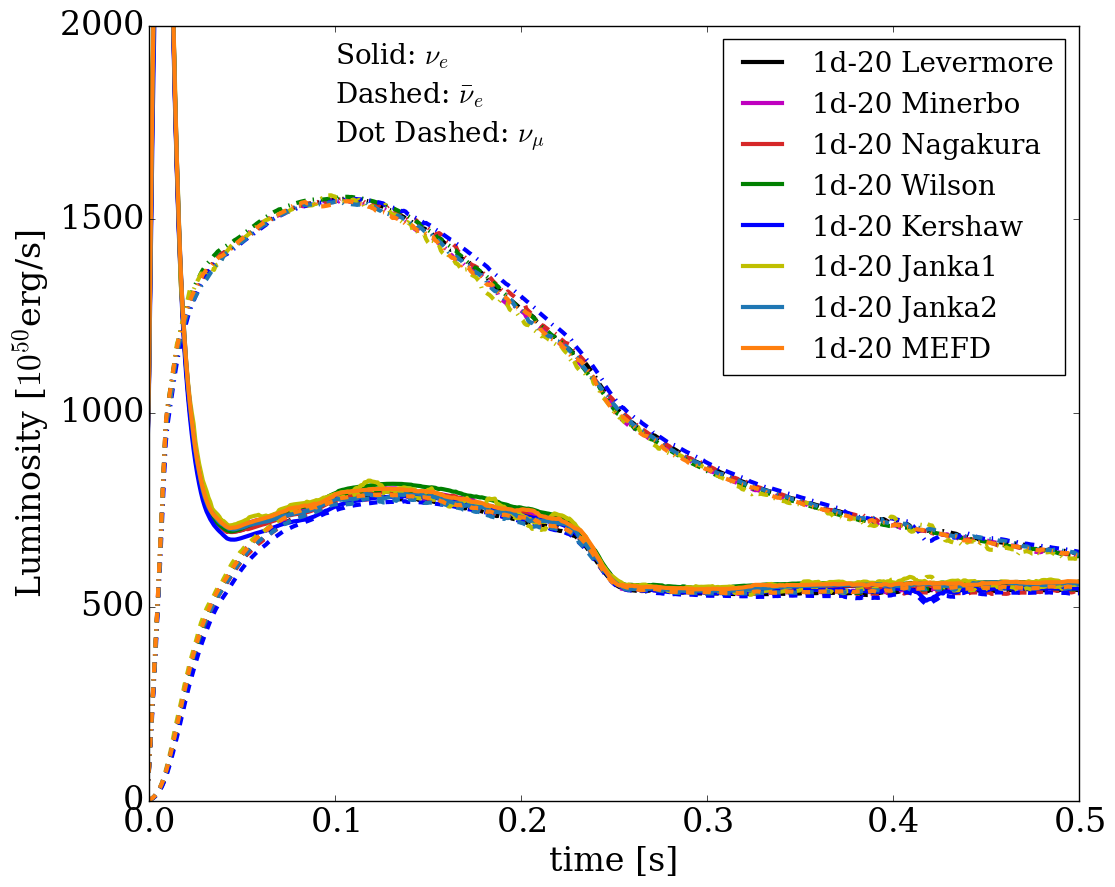}
    \includegraphics[width=0.48\textwidth]{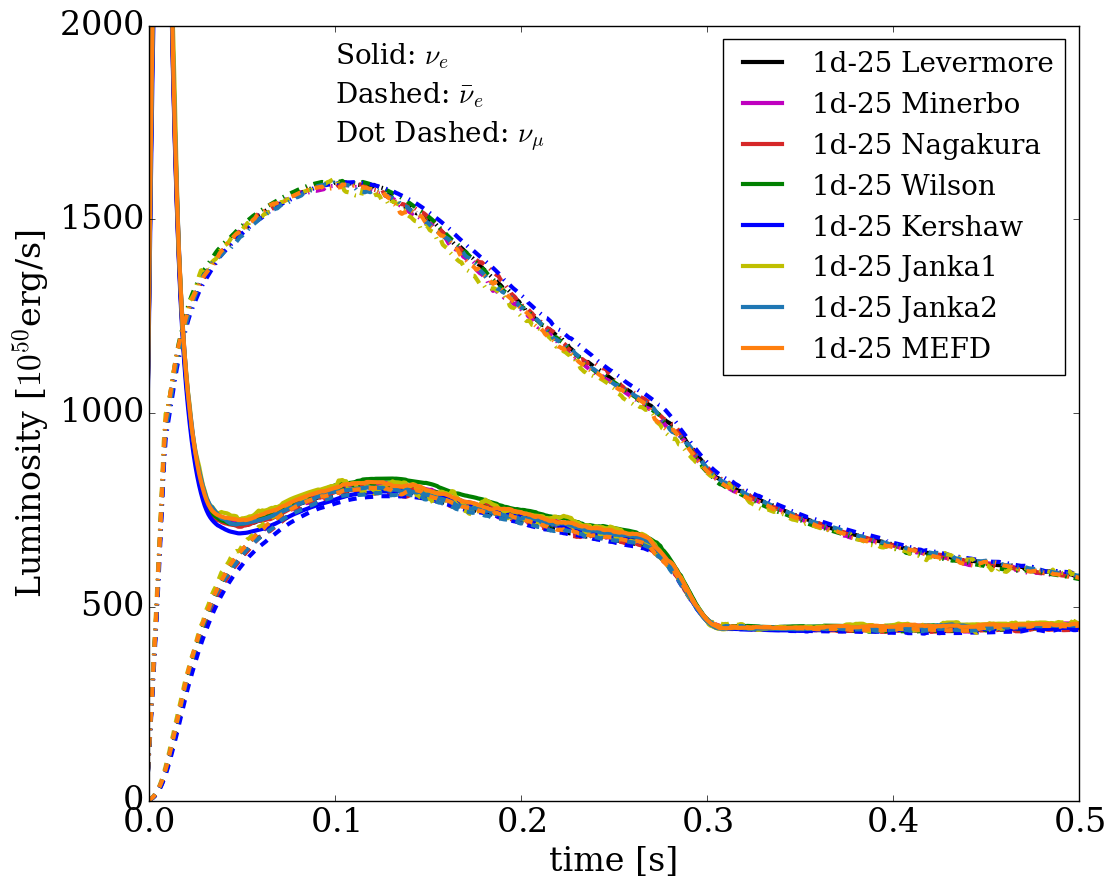}
    \caption{Luminosity evolution (measured at infinity) versus time after bounce (in seconds) of simulations with progenitor ZAMS masses 13 $M_\odot$ (top left), 16 $M_\odot$ (top right), 20 $M_\odot$ (bottom left), and 25 $M_\odot$ (bottom right).  Solid lines are electron-type neutrinos, dashed lines are anti-electron-type neutrinos, and dot-dashed lines are ``$\mu$-type" neutrinos. The differences in luminosities caused by the various closures are generally $<2\%$. }
    \label{fig:L-all}
\end{figure}

\begin{figure}
    \centering
    \includegraphics[width=0.48\textwidth]{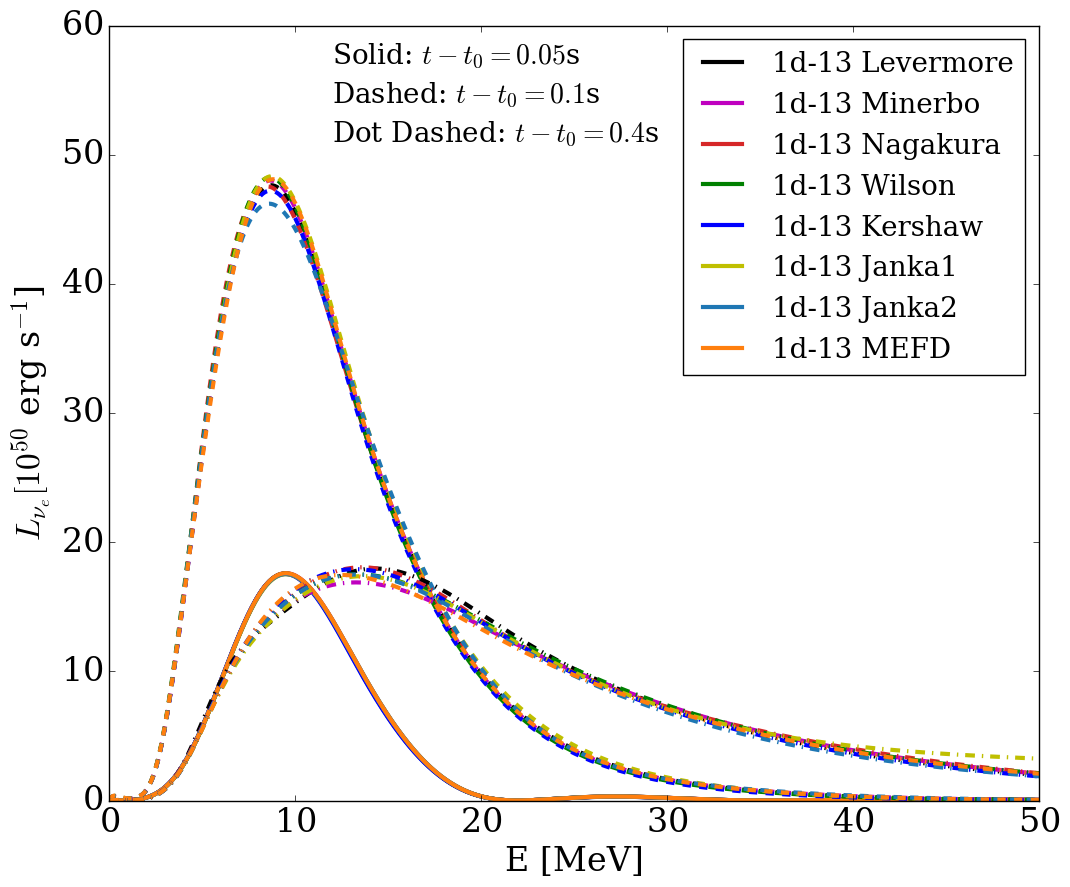}
    \includegraphics[width=0.48\textwidth]{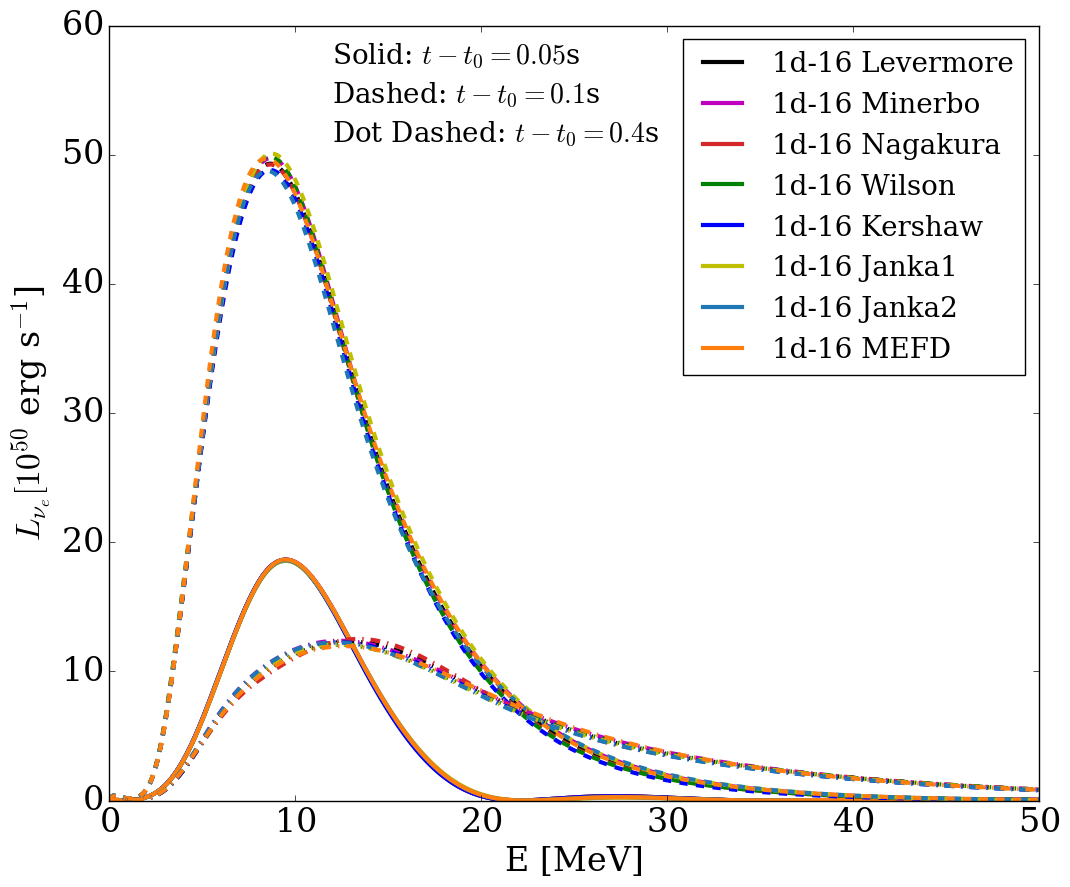}
    \includegraphics[width=0.48\textwidth]{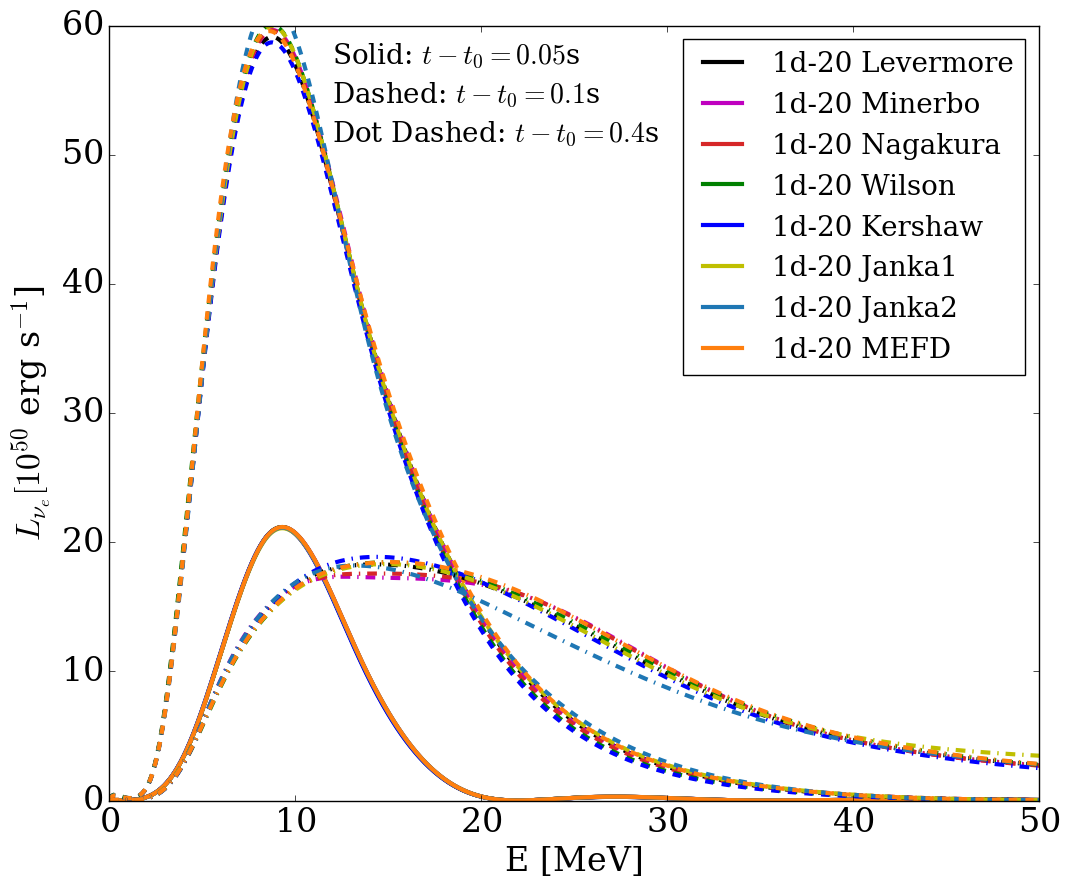}
    \includegraphics[width=0.48\textwidth]{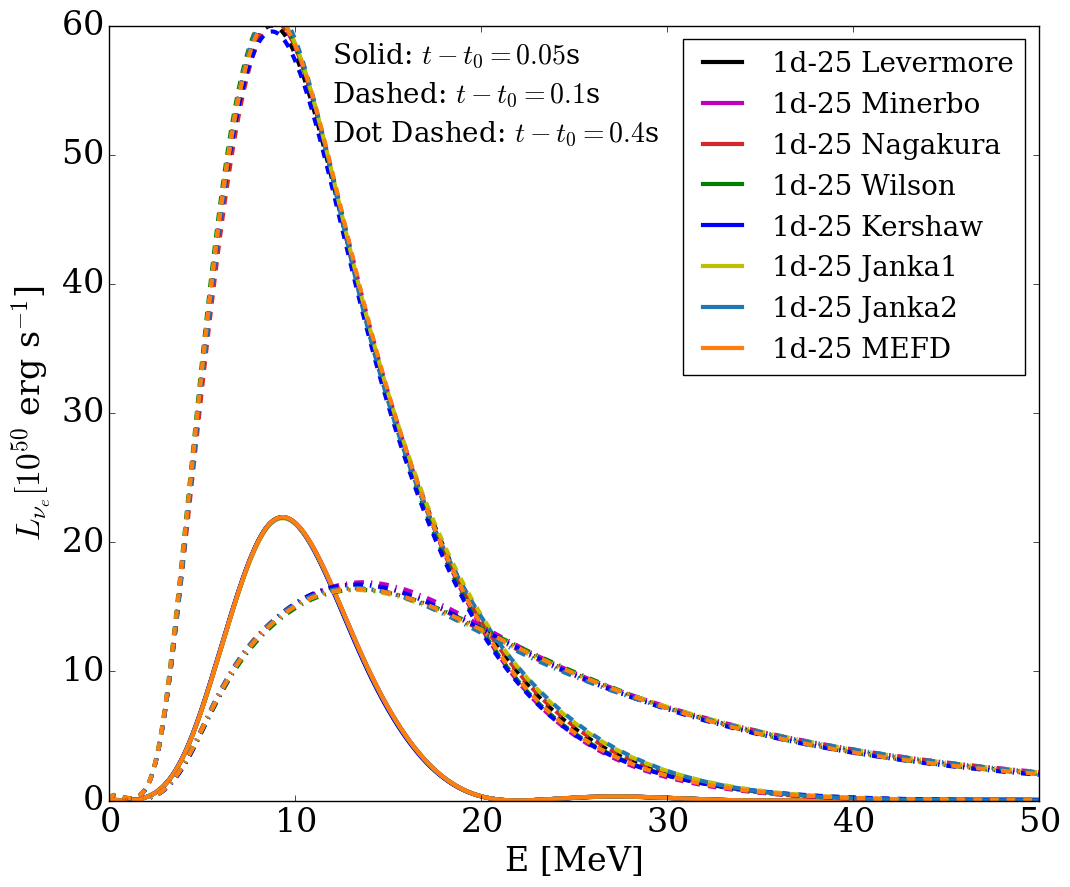}
    \caption{Electron-type neutrino spectra at 0.05 seconds (solid lines), 0.1 seconds (dashed lines), and 0.4 seconds (dot dashed lines) after bounce. {Differences caused by closure choices are generally small and can be ignored.}}
    \label{fig:spec0}
\end{figure}

\begin{figure}
    \centering
    \includegraphics[width=0.48\textwidth]{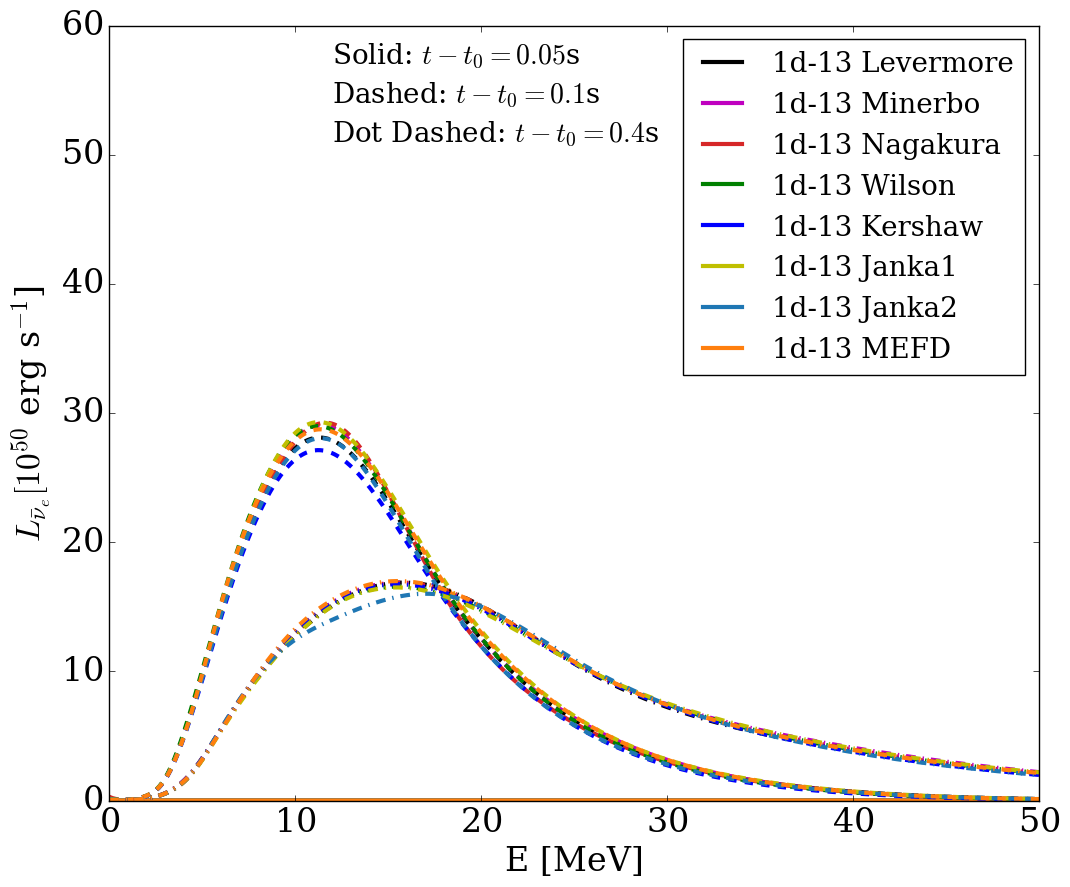}
    \includegraphics[width=0.48\textwidth]{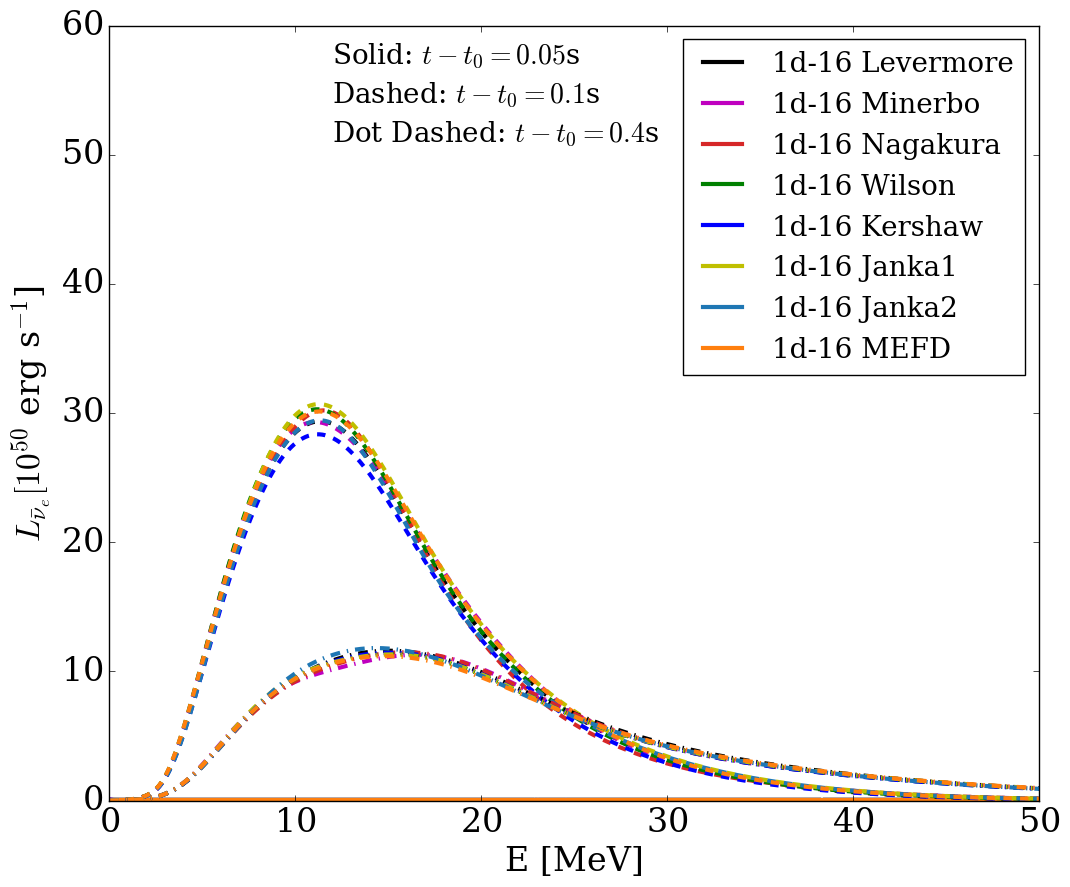}
    \includegraphics[width=0.48\textwidth]{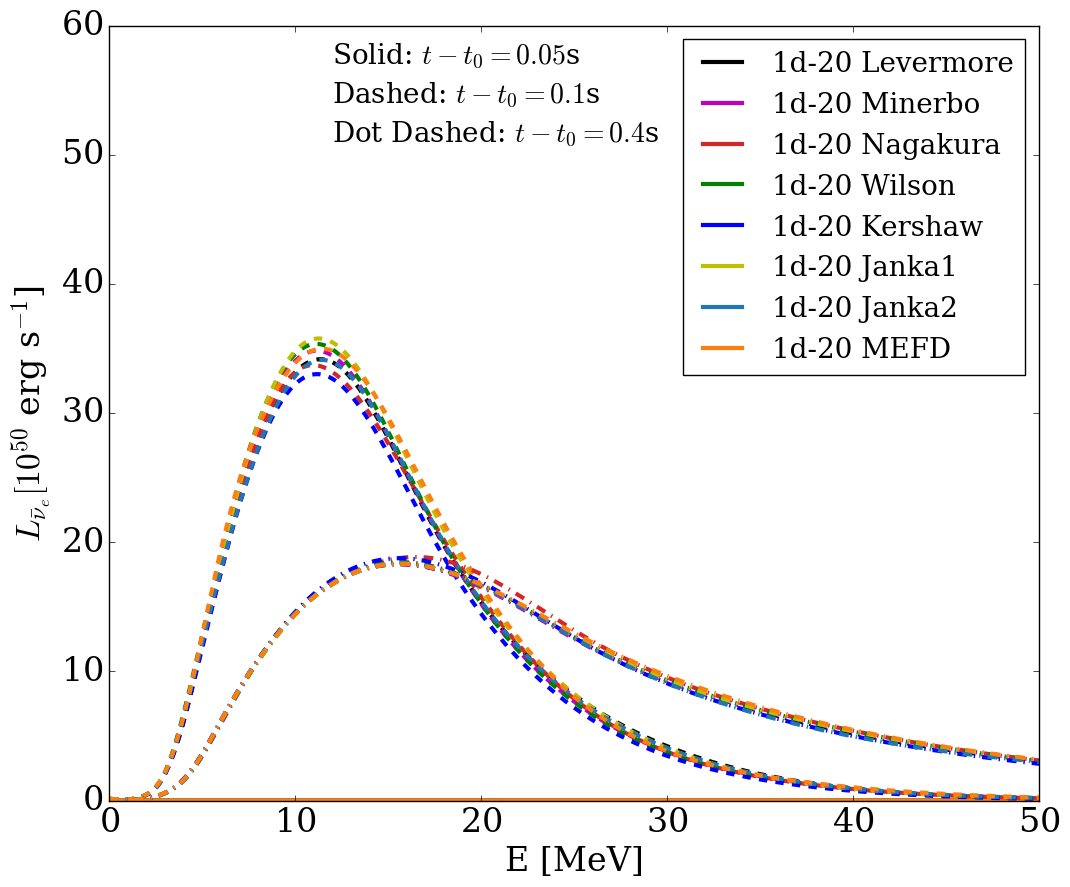}
    \includegraphics[width=0.48\textwidth]{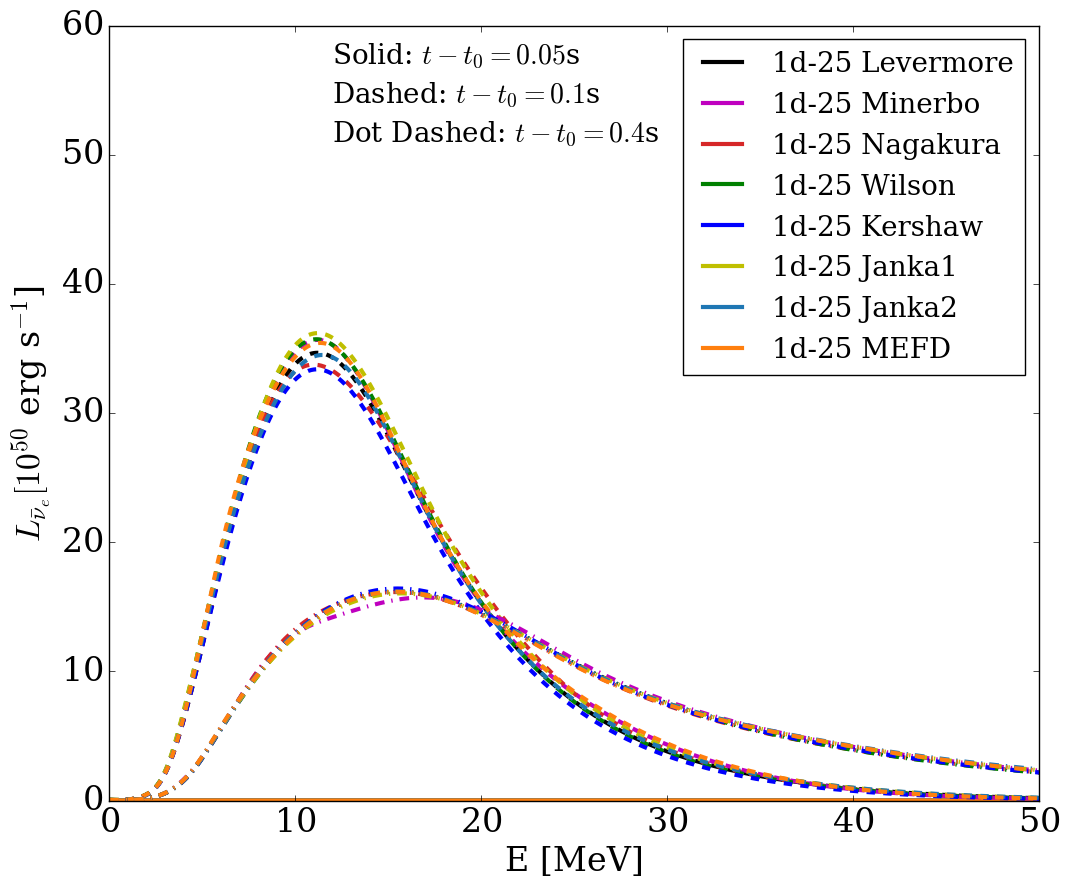}
    \caption{{Similar to Figure \ref{fig:spec0}, but for anti-electron-type neutrinos.}}
    \label{fig:spec1}
\end{figure}

\begin{figure}
    \centering
    \includegraphics[width=0.48\textwidth]{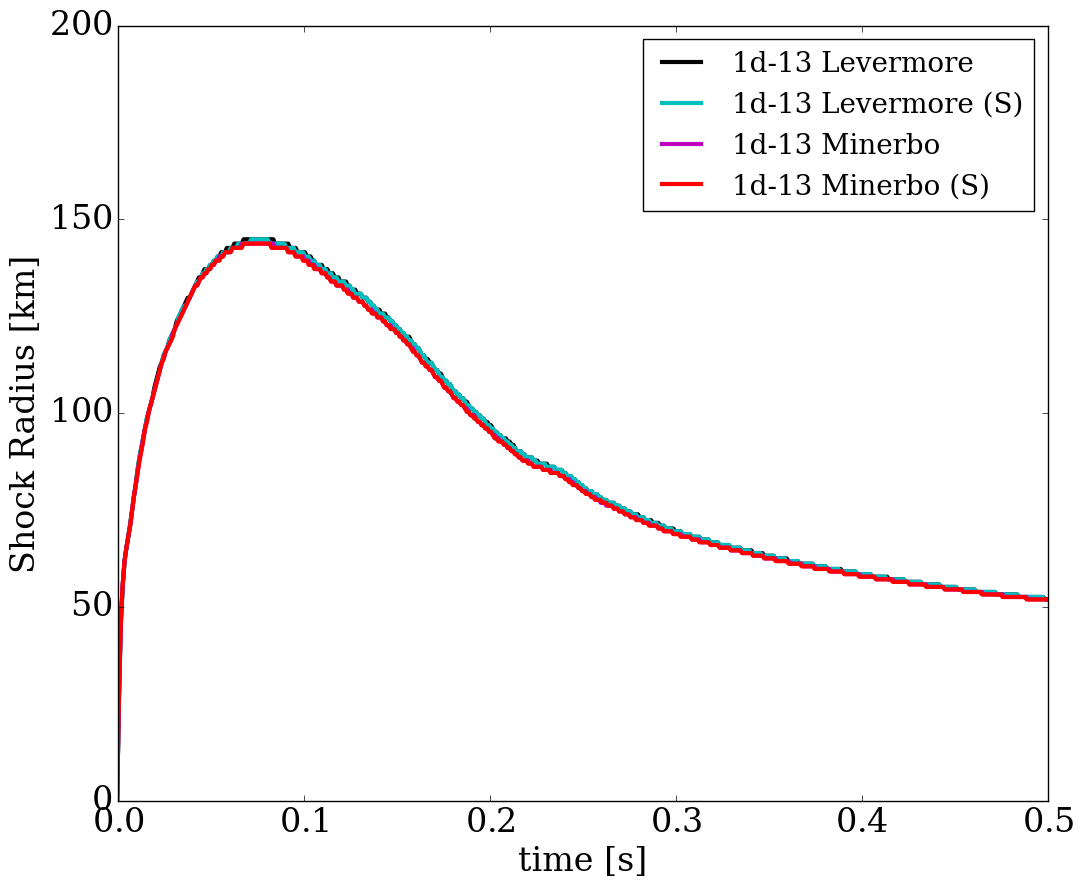}
    \includegraphics[width=0.48\textwidth]{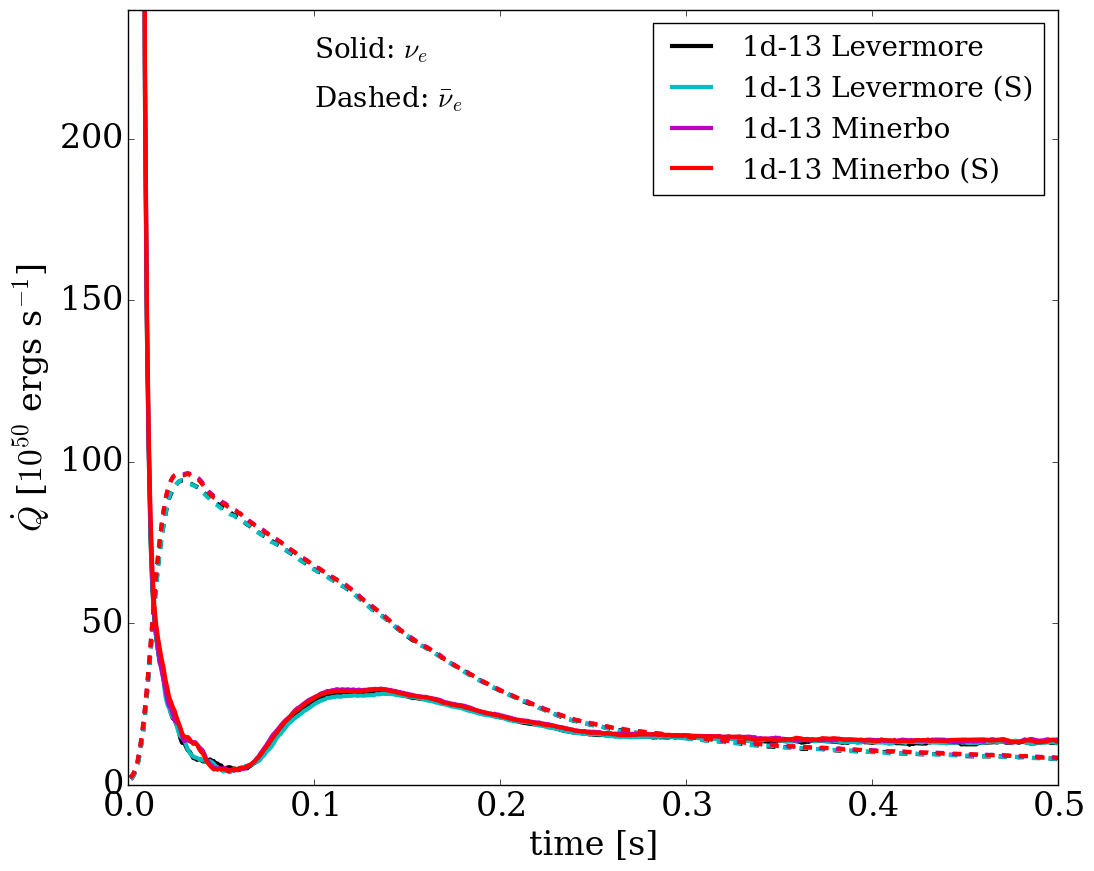}
    \caption{The shock radius and total heating rate versus time for the 13 $M_\odot$ progenitor. Here, we compare the Shibata interpolation {(denoted by (S))} and the self-consistent 3rd-order closure relation of the Levermore and the Minerbo closures. Differences between the Shibata interpolation and the self-consistent relation are smaller than the differences between the Levermore and Minerbo closures.}
    \label{fig:shibata-compare}
\end{figure}

From our 1D simulations, we find that some physical quantities are only weakly influenced by different closure choices. Figure \ref{fig:L-all} shows the neutrino luminosities measured at 10,000 km. Figure \ref{fig:spec0} and \ref{fig:spec1} show the spectra of electron-type and anti-electron-type neutrinos. For these quantities, differences caused by closures are seen to be negligible. {The similarity of these neutrino spectra means that they are only weakly influenced by differences in the $\partial_\nu$ terms in Equations \ref{equ:1d-E} and \ref{equ:1d-F}. Since the $q$ factors occur only in the $\partial_\nu$ terms, this indicates that differences caused by 3rd-moment closure choices will be small\footnote{This point is confirmed later in Figure \ref{fig:shibata-compare}.}.}

\begin{figure}
    \centering
    \includegraphics[width=0.48\textwidth]{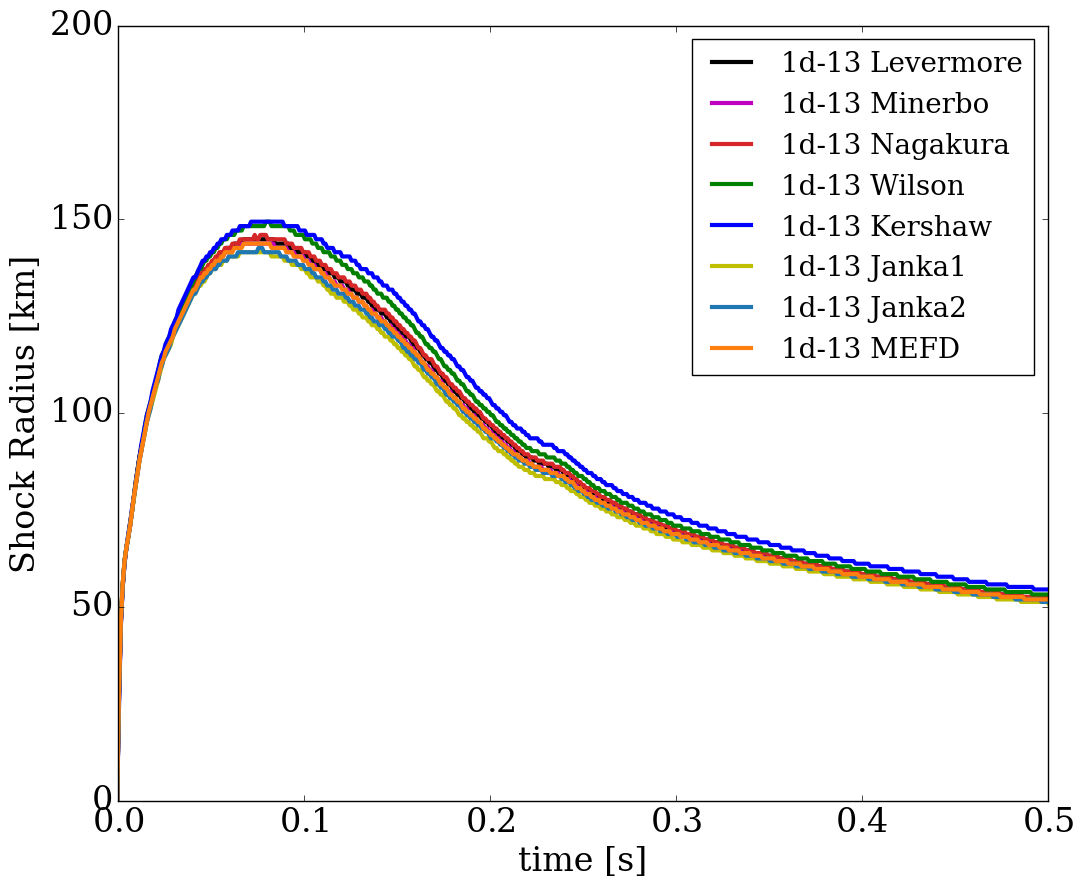}
    \includegraphics[width=0.48\textwidth]{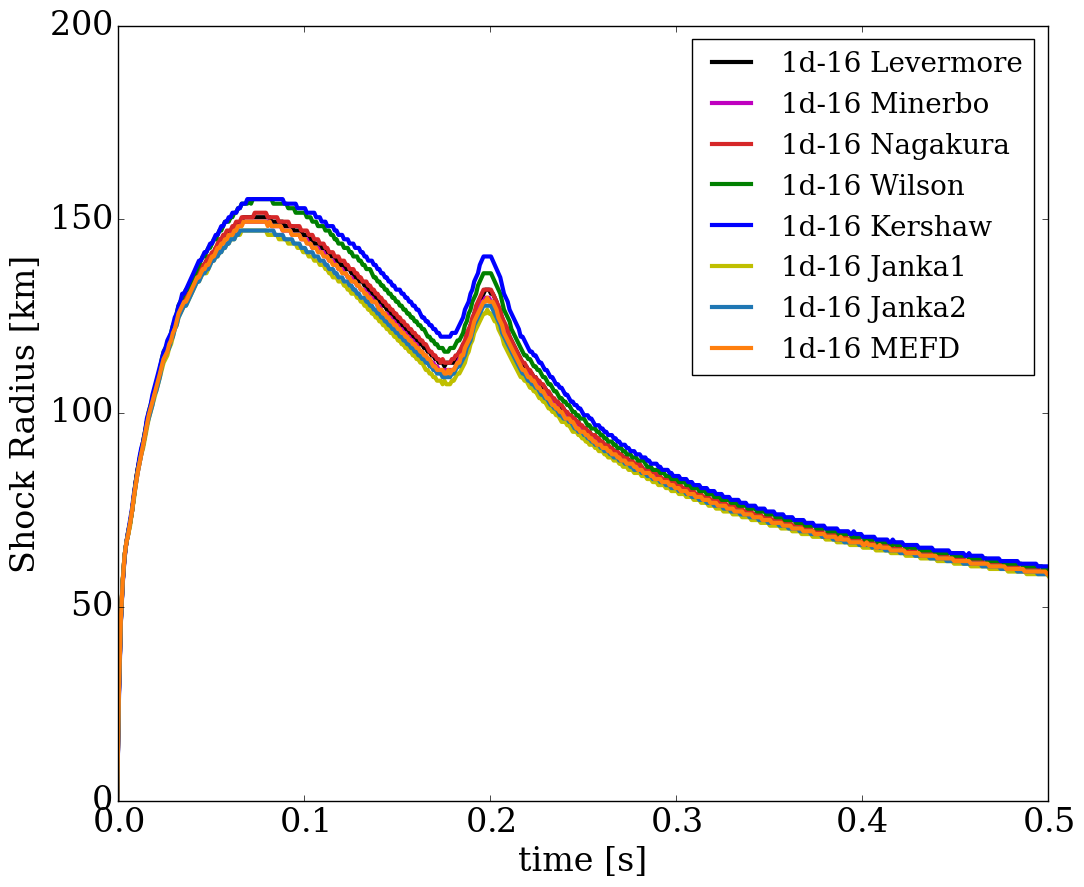}
    \includegraphics[width=0.48\textwidth]{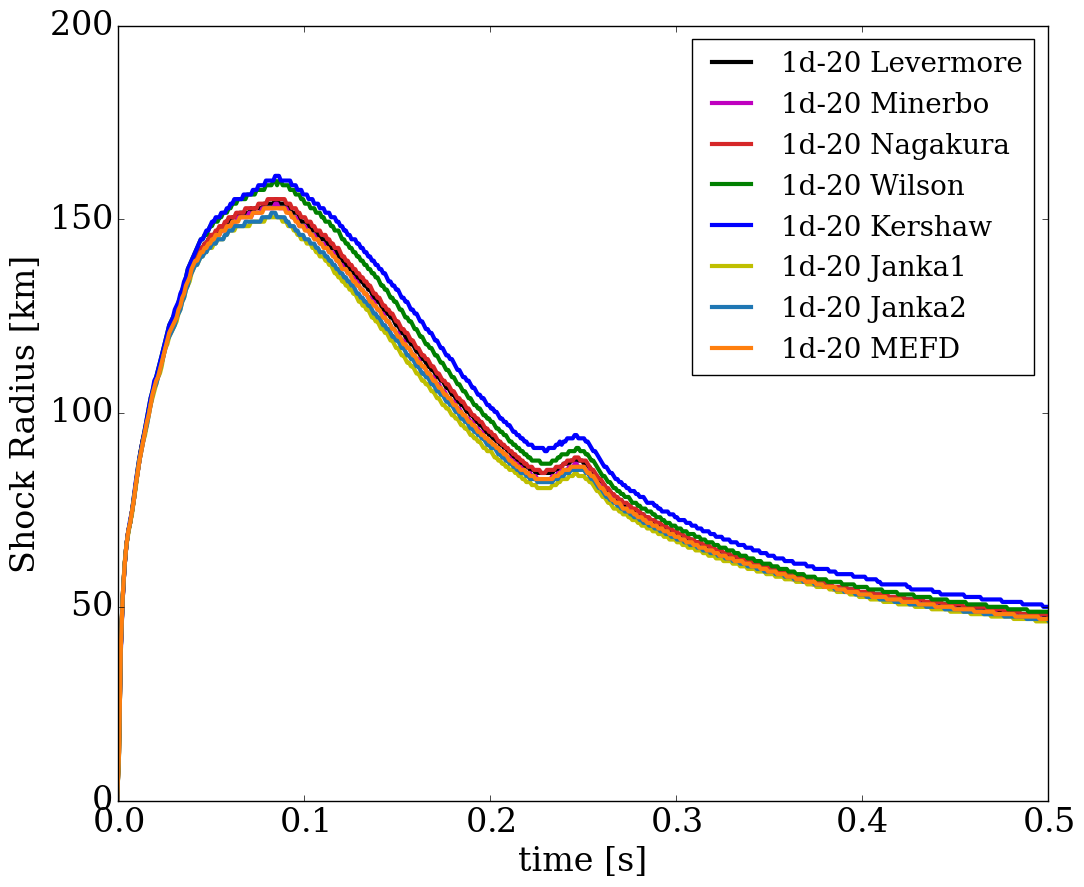}
    \includegraphics[width=0.48\textwidth]{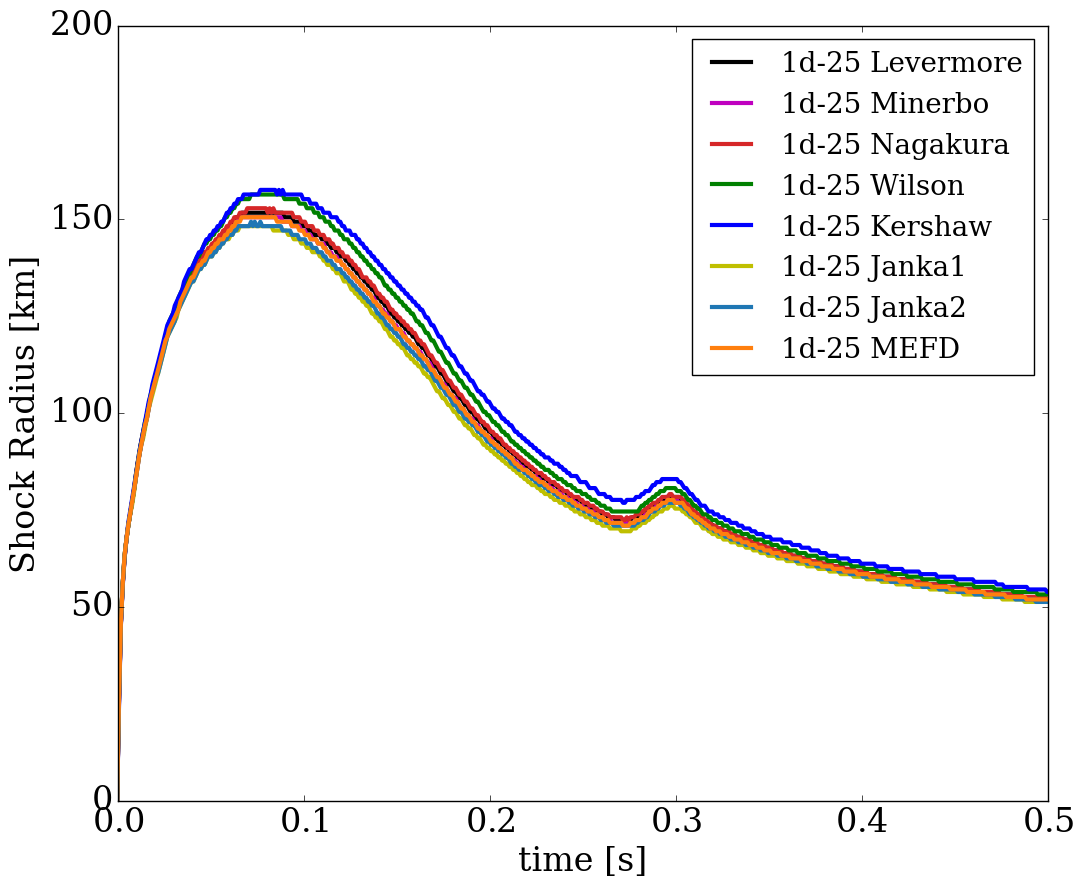}
    \caption{Shock radius versus time. For each simulation, the shock radius differences due to the different closures are less than 2\%, except for the Kershaw and Wilson closures, for which they are about 5-10\%. {The bumps at 0.2-0.3 seconds are due to the accretion of the Si-O interfaces.}}
    \label{fig:rs-all}
\end{figure}

\begin{figure}
    \centering
    \includegraphics[width=0.48\textwidth]{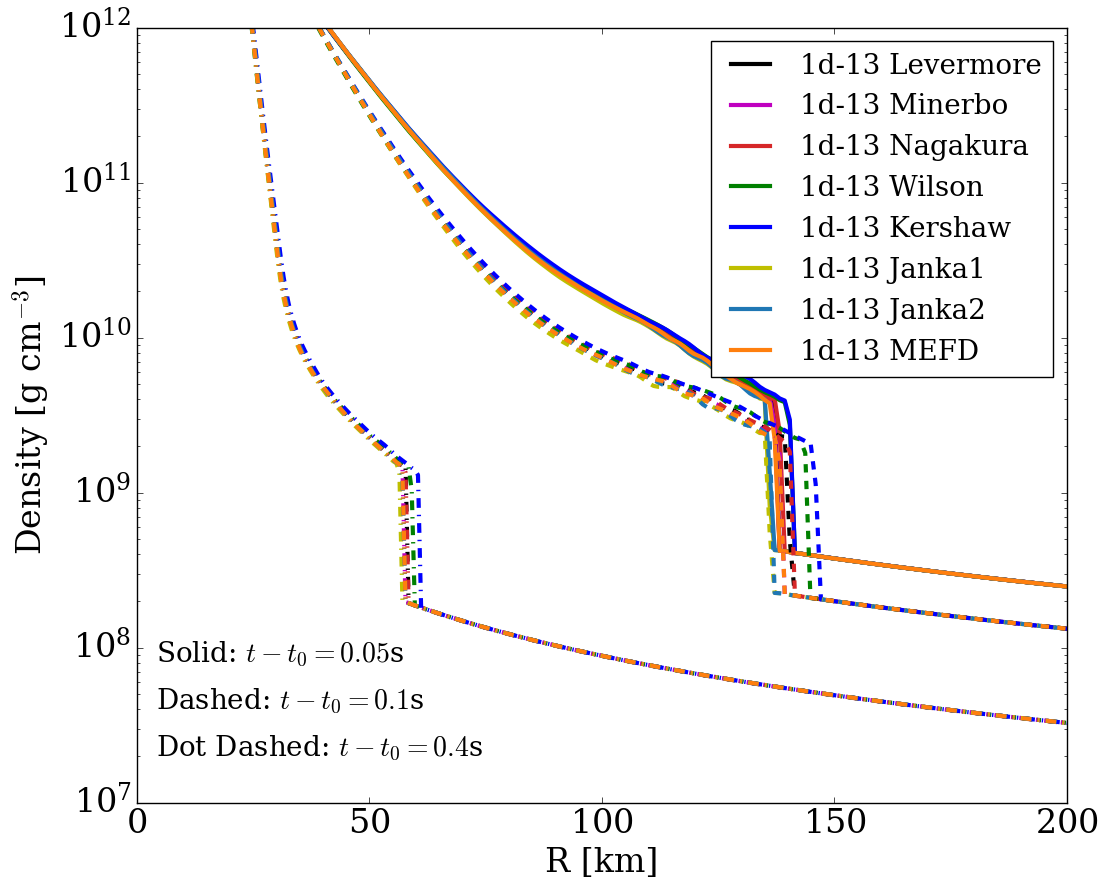}
    \includegraphics[width=0.48\textwidth]{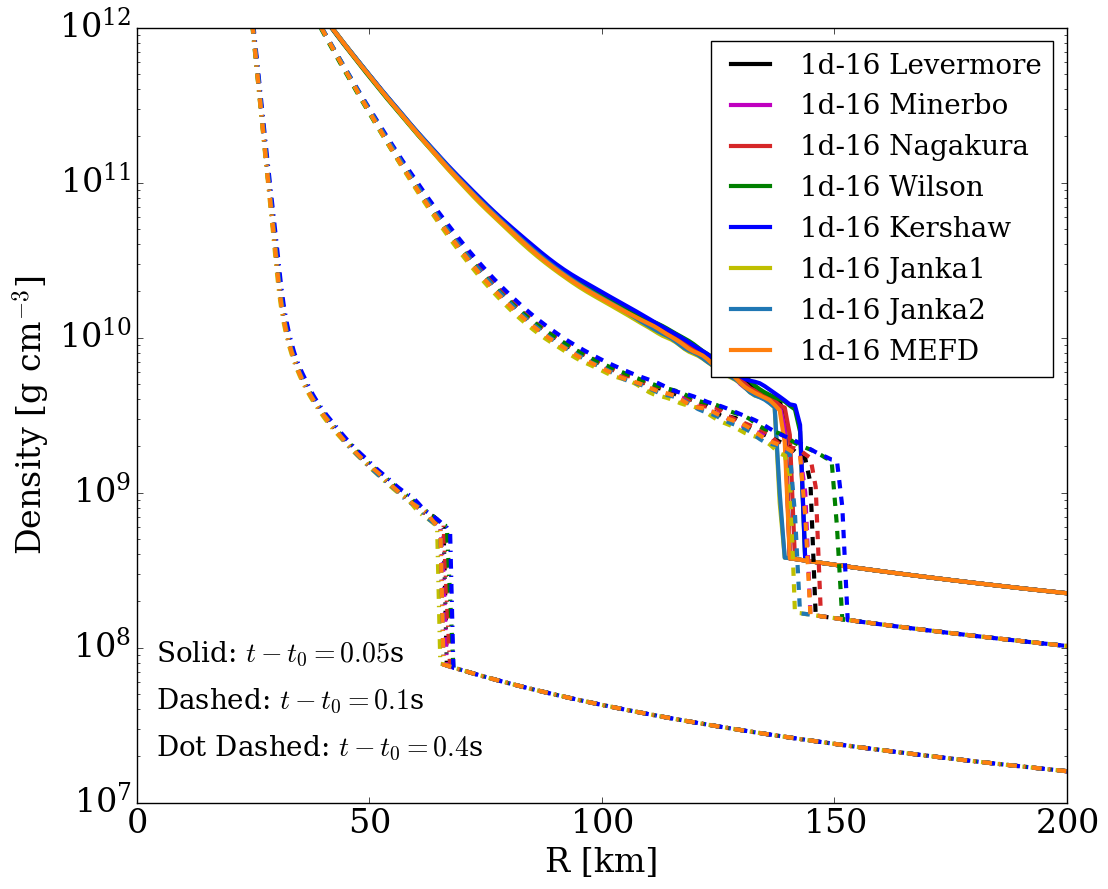}
    \includegraphics[width=0.48\textwidth]{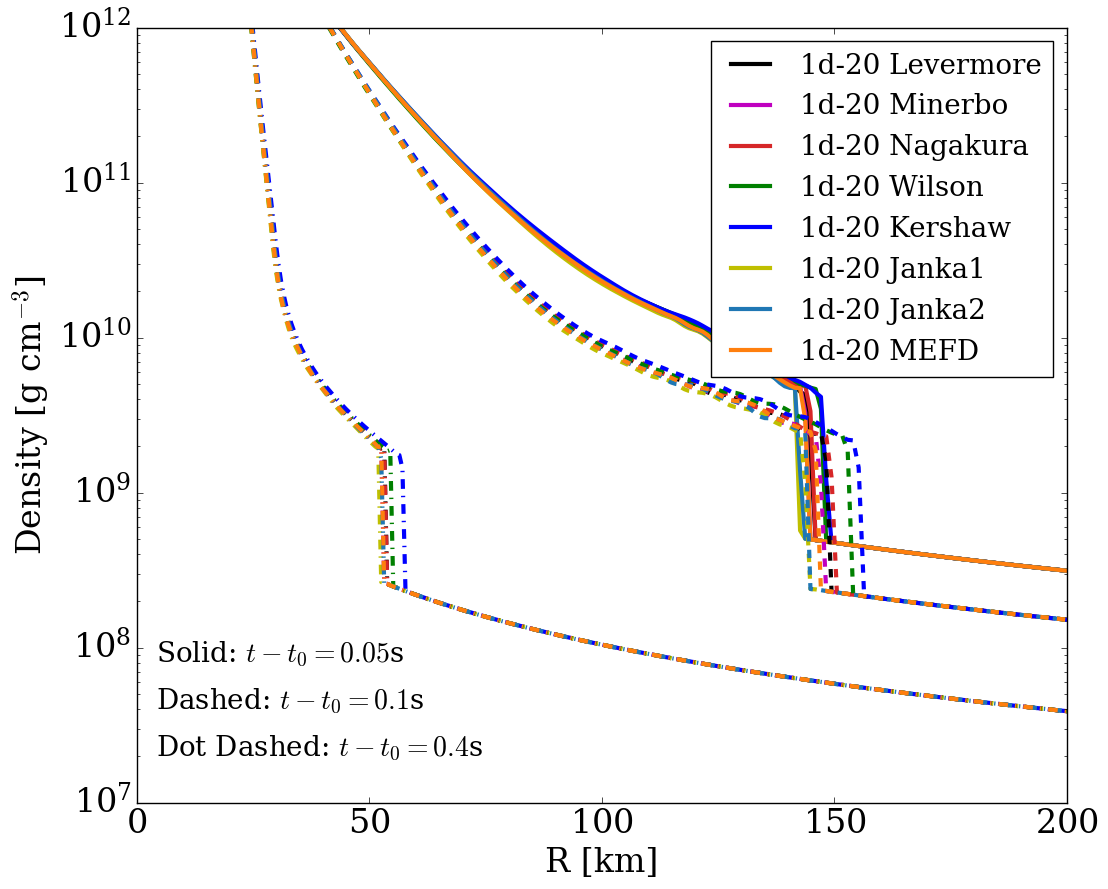}
    \includegraphics[width=0.48\textwidth]{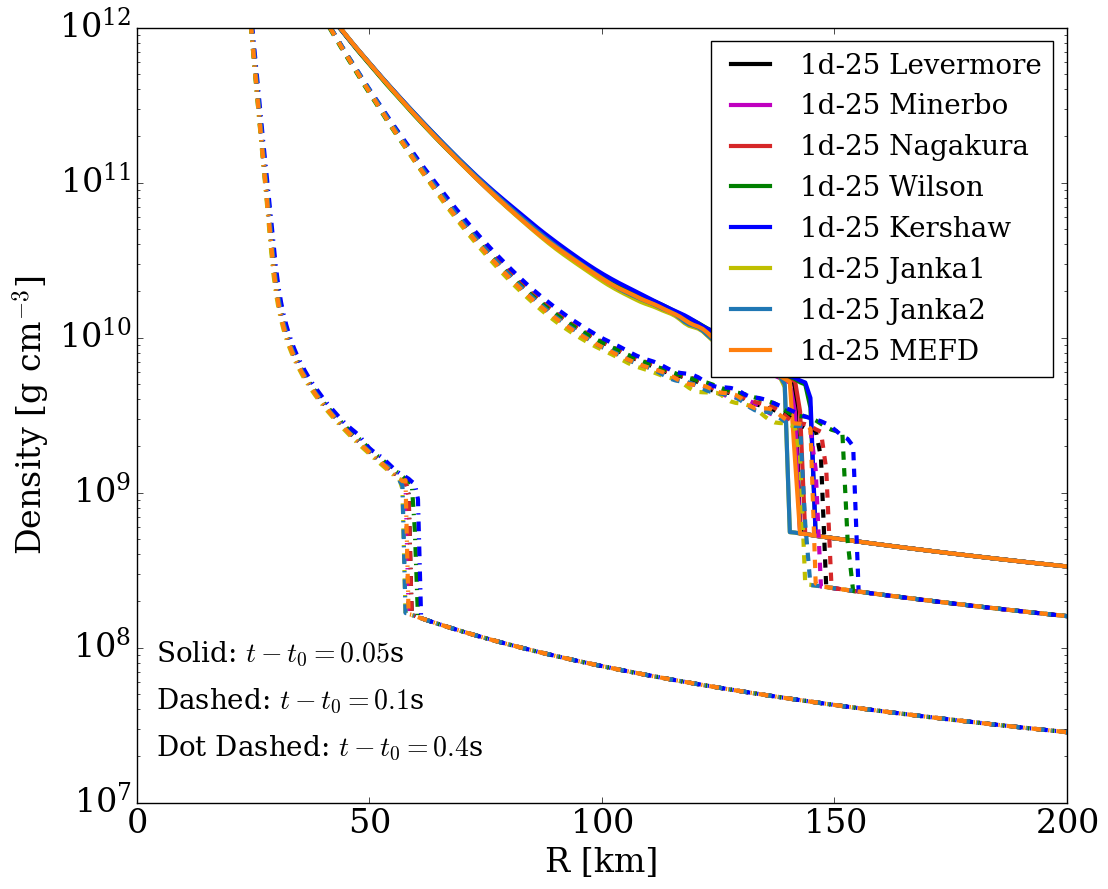}
    \caption{Density profiles at 0.05 s (solid lines), 0.1 s (dashed lines), and 0.4 s (dot dashed lines) after bounce. {The post-shock densities are simlar between simulations with different closures, but their shock positions are slightly different. Therefore, the profiles are "stretched" or "compressed" to comport the shock radius. The relatively large $\dot{Q}$ differences are partly caused by the density differences shown here, but are dominated by the gain region size differences.}}
    \label{fig:rho-profile-all}
\end{figure}

\begin{figure}
    \centering
    \includegraphics[width=0.48\textwidth]{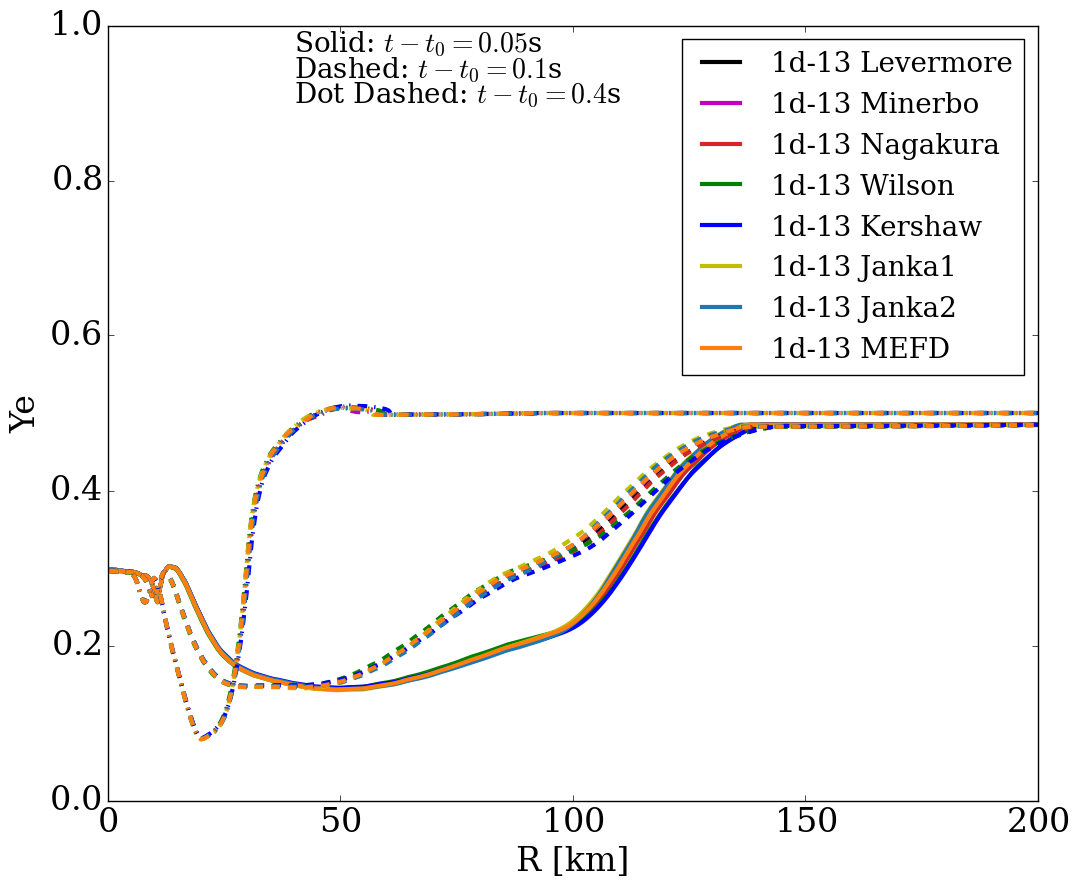}
    \includegraphics[width=0.48\textwidth]{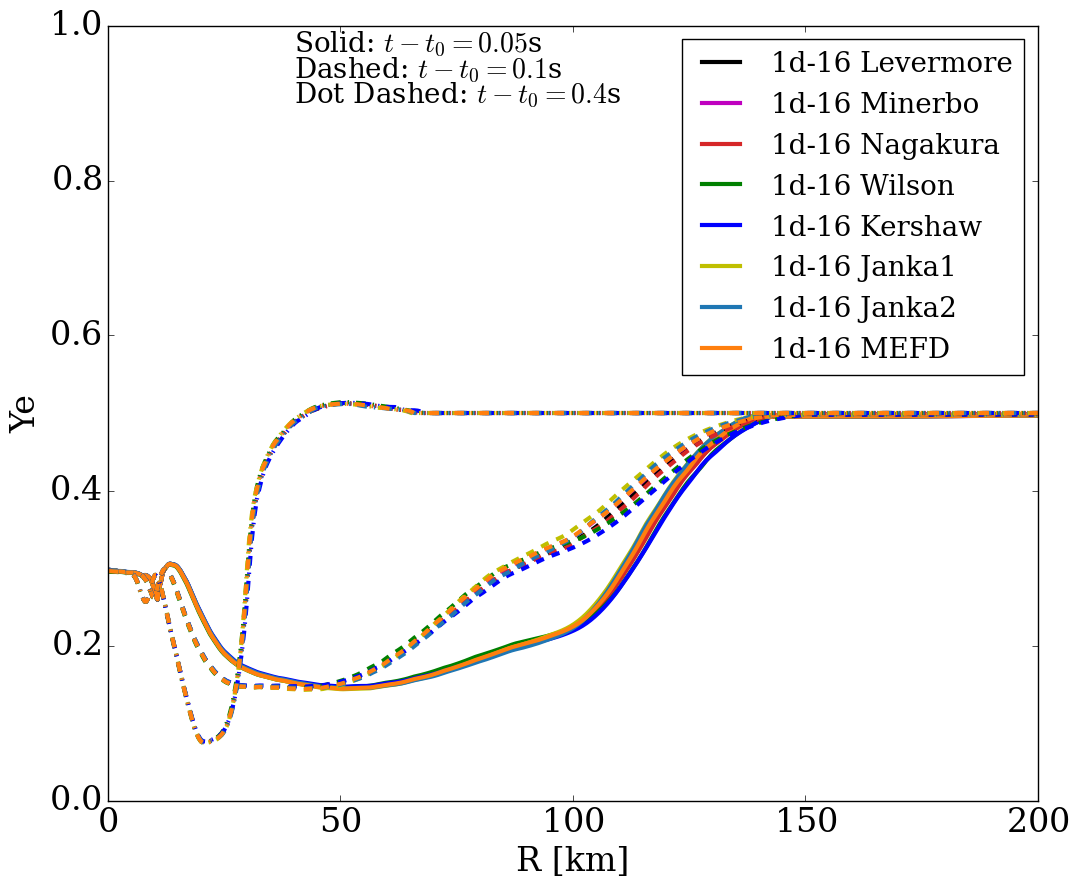}
    \includegraphics[width=0.48\textwidth]{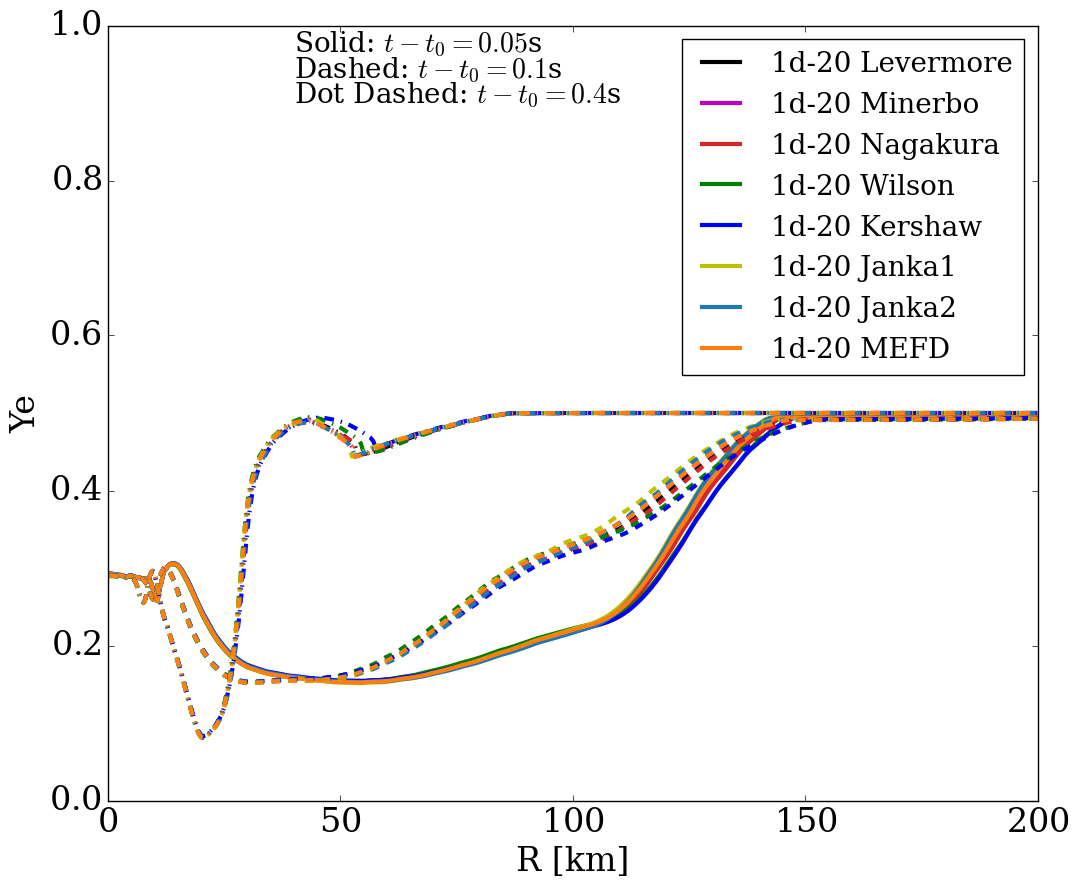}
    \includegraphics[width=0.48\textwidth]{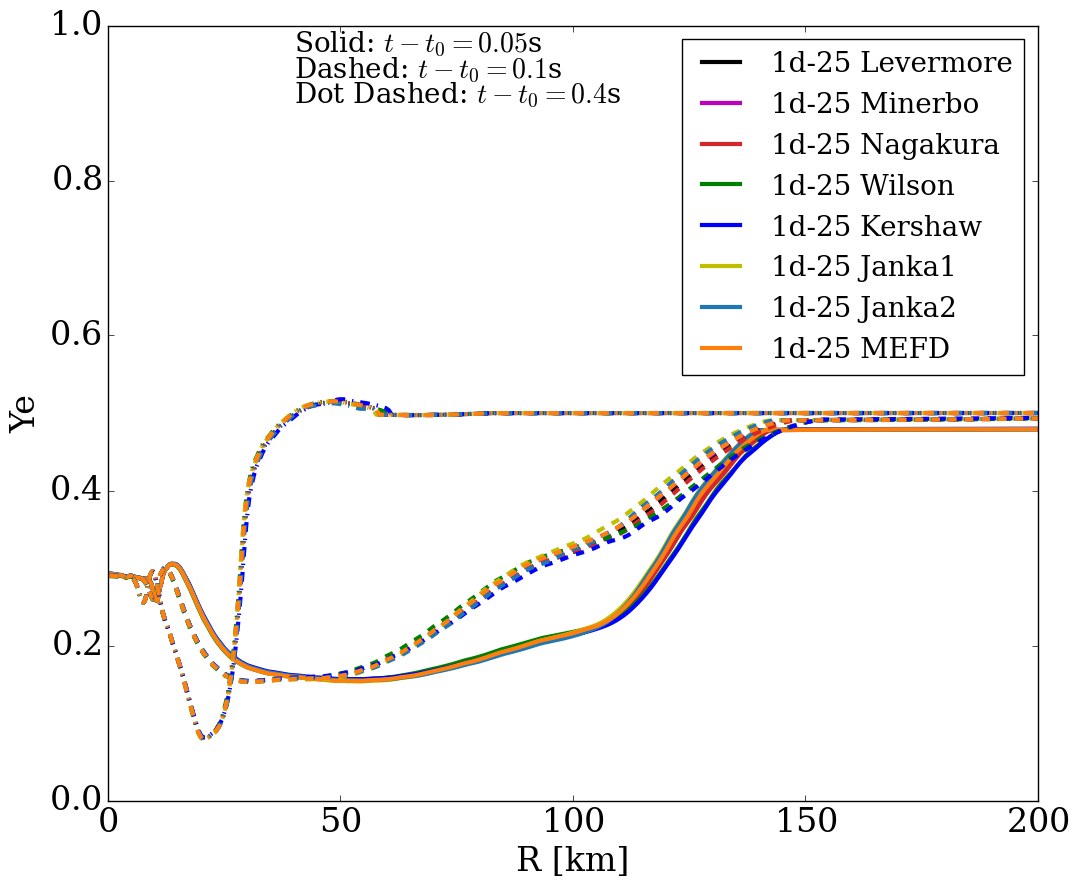}
    \caption{{Similar to Figure \ref{fig:rho-profile-all}, but for $Y_e$ profiles.} $Y_e$ differences are probably the result of differences in $\dot{Q}$s, which translate into altered shock radii and, hence, an extended region of electron capture on newly-liberated protons.}
    \label{fig:ye-profile-all}
\end{figure}

\begin{figure}
    \centering
    \includegraphics[width=0.48\textwidth]{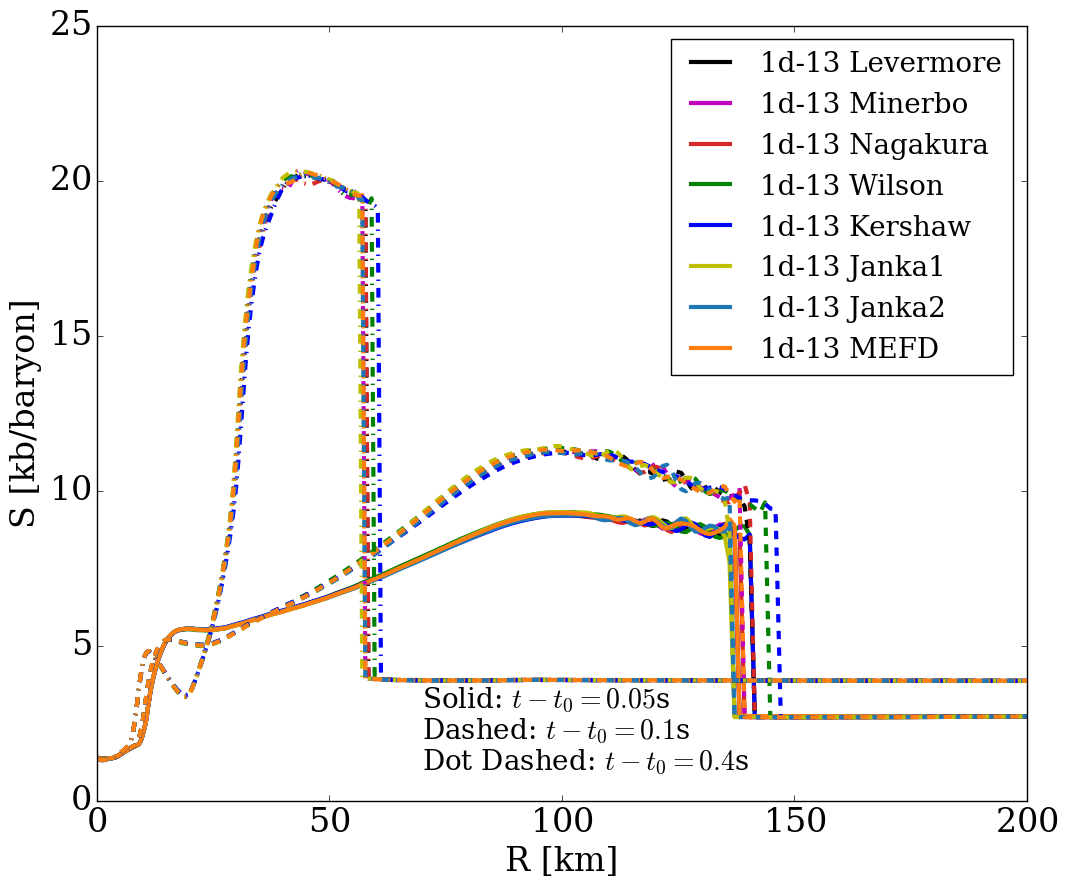}
    \includegraphics[width=0.48\textwidth]{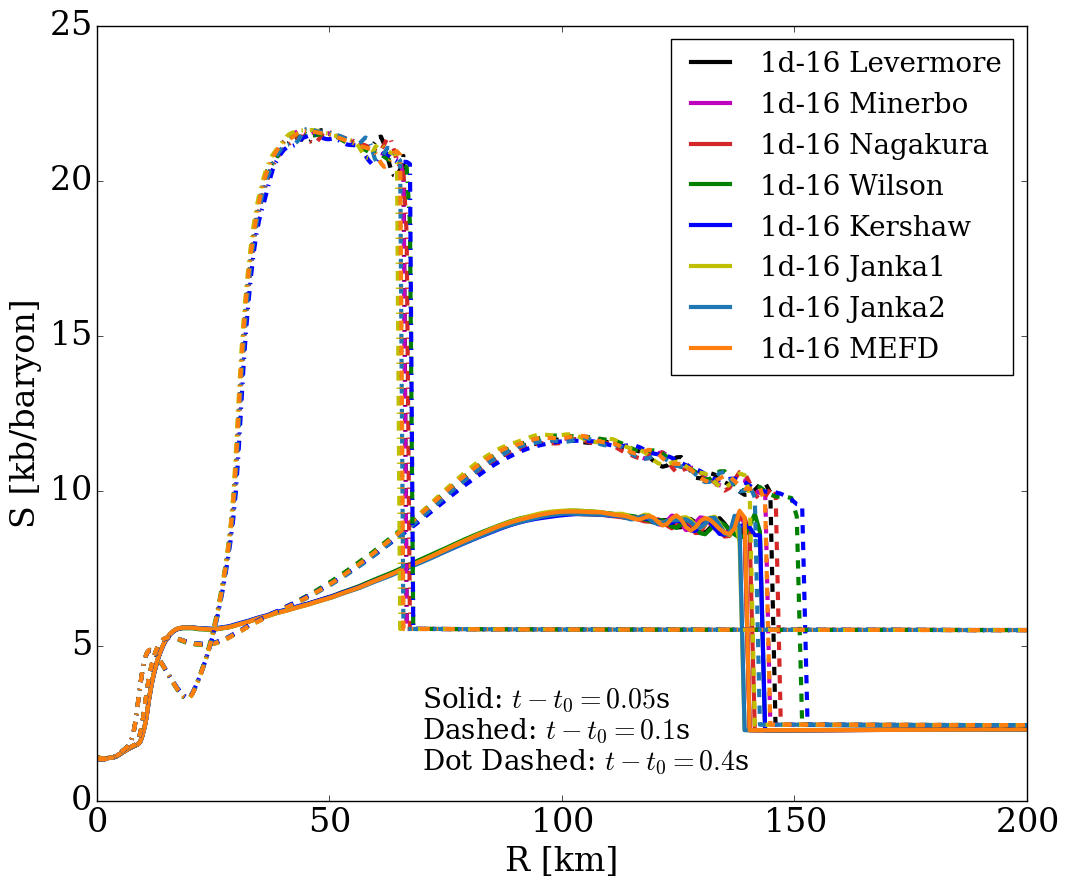}
    \includegraphics[width=0.48\textwidth]{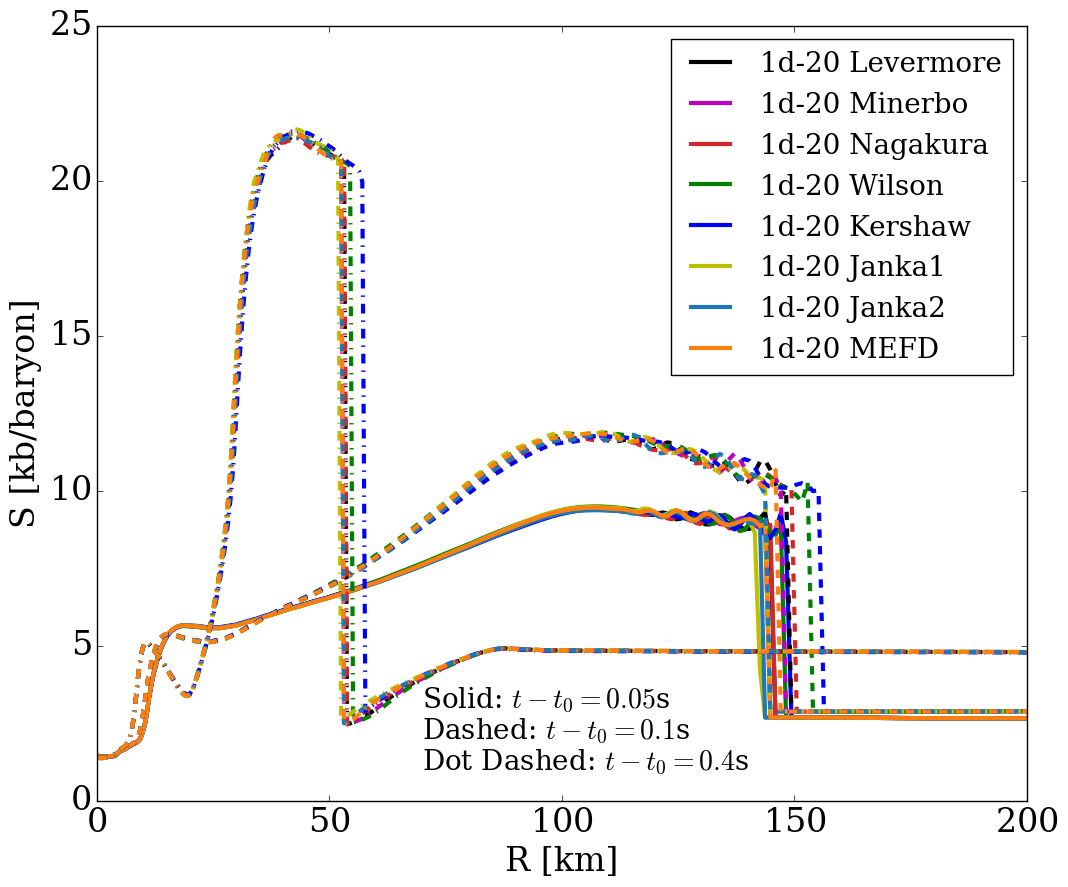}
    \includegraphics[width=0.48\textwidth]{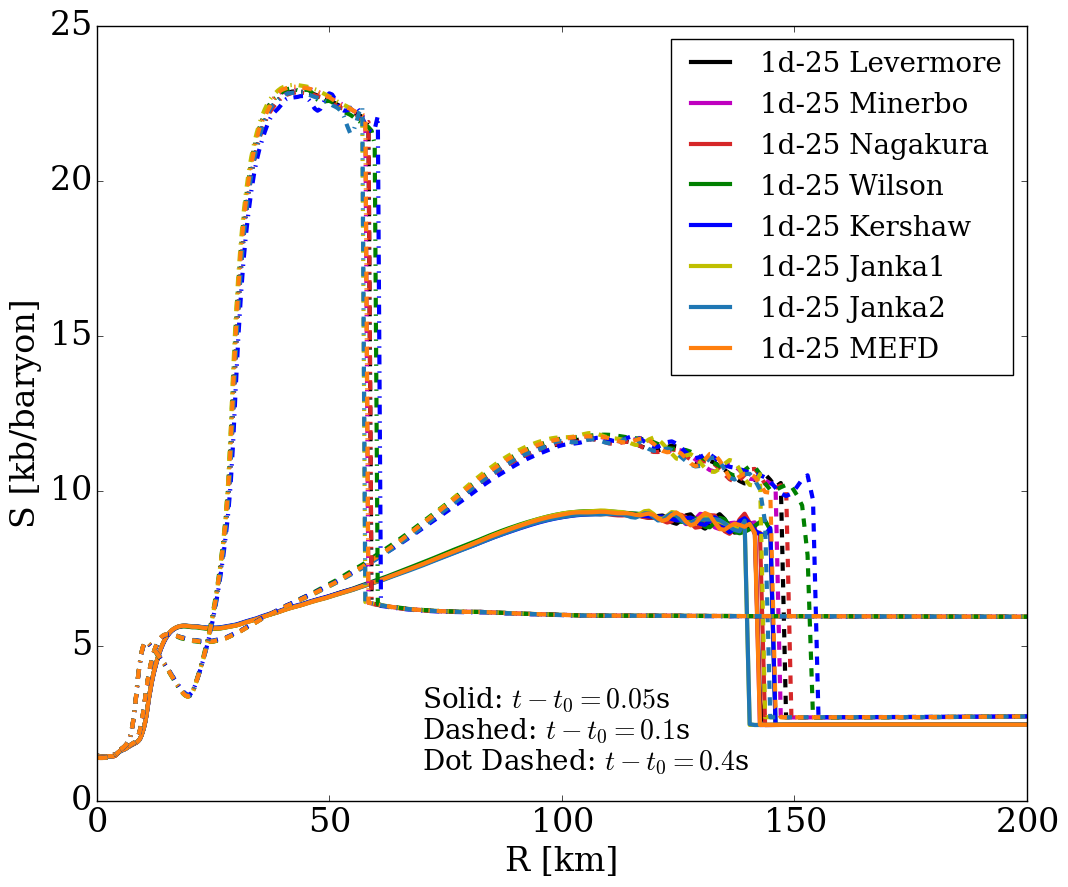}
    \caption{{Similar to Figure \ref{fig:rho-profile-all}, but for entropy profiles. Differences only occur at the stalled shock.}}
    \label{fig:S-profile-all}
\end{figure}

\begin{figure}
    \centering
    \includegraphics[width=0.48\textwidth]{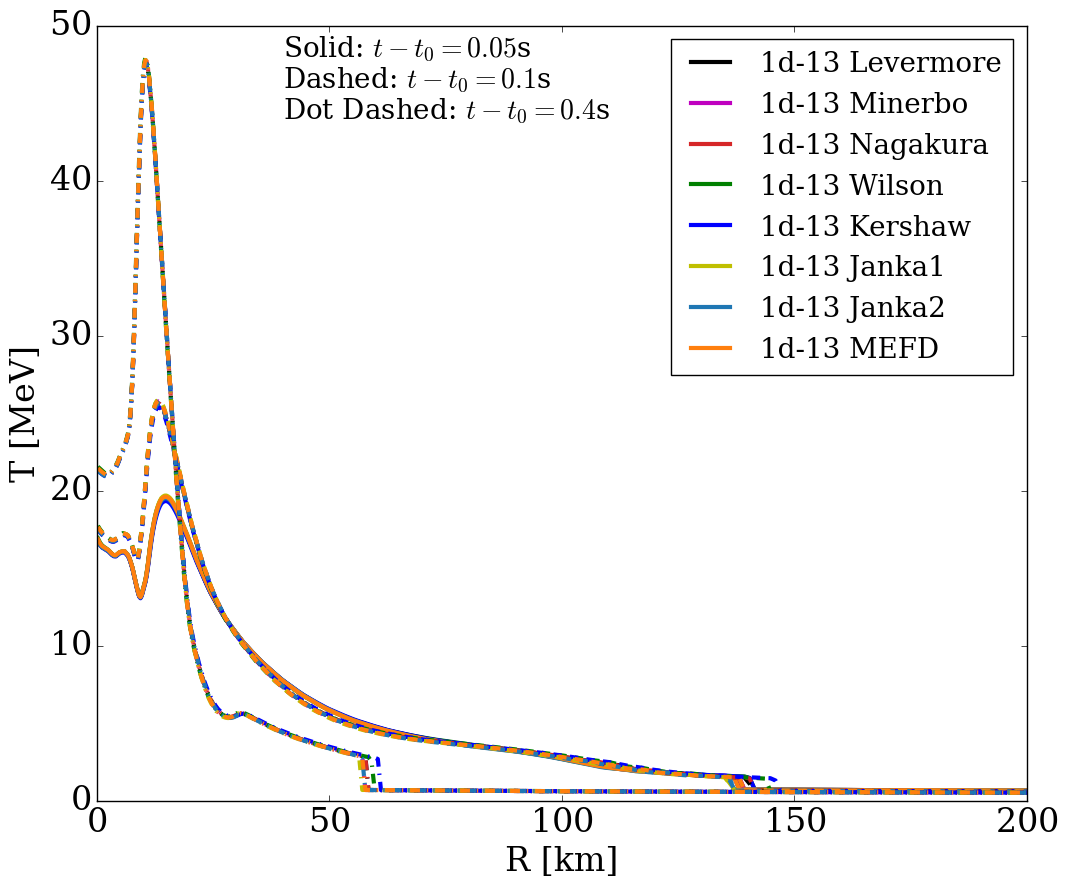}
    \includegraphics[width=0.48\textwidth]{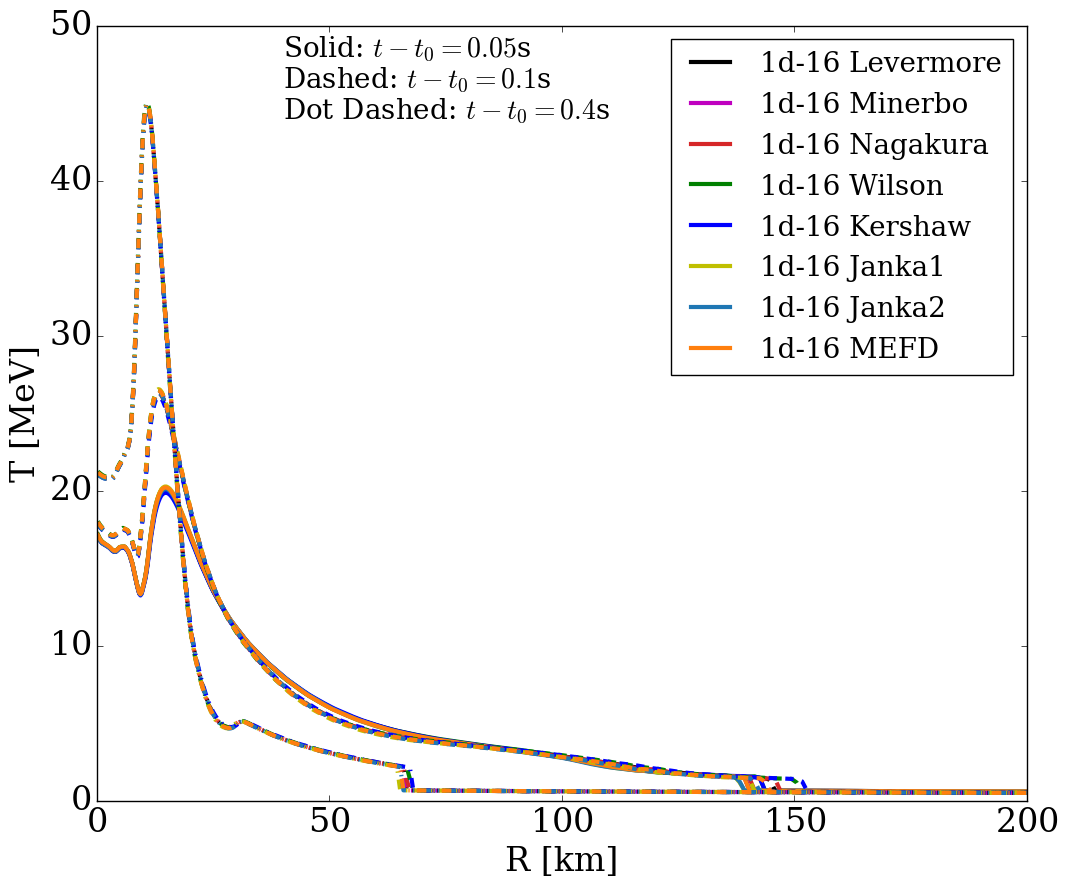}
    \includegraphics[width=0.48\textwidth]{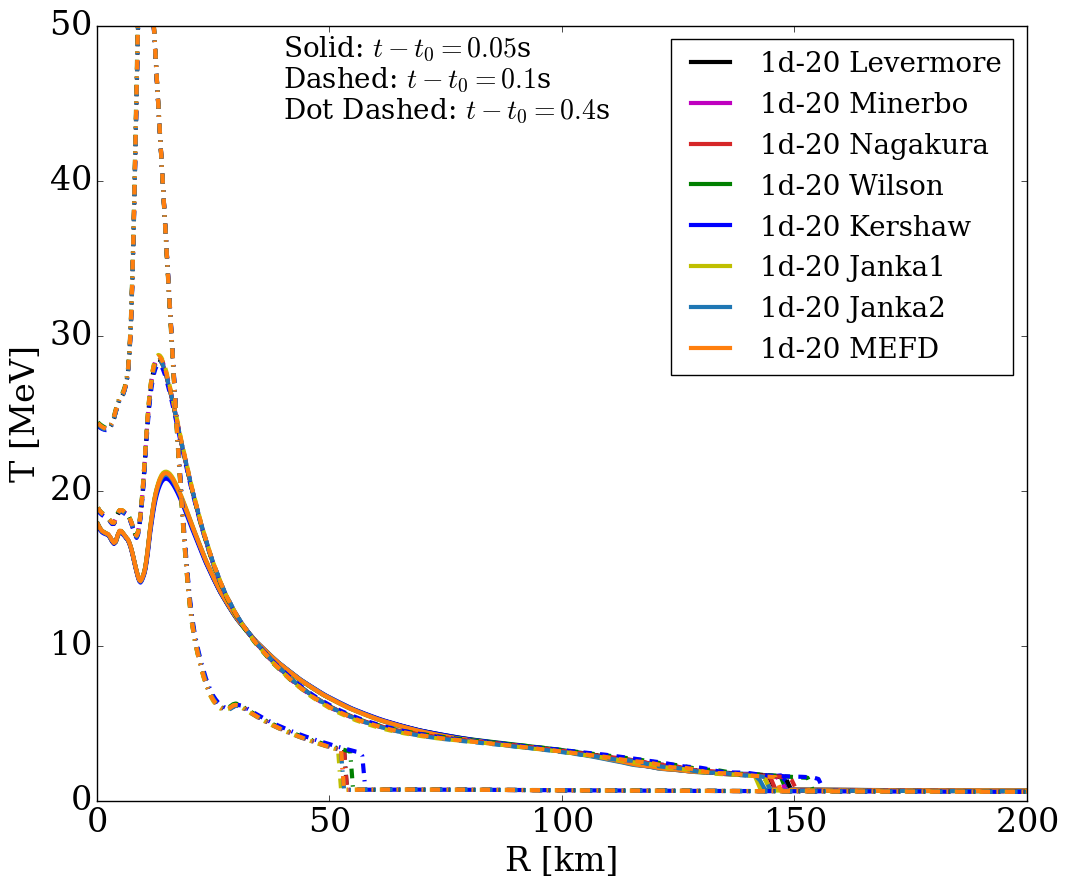}
    \includegraphics[width=0.48\textwidth]{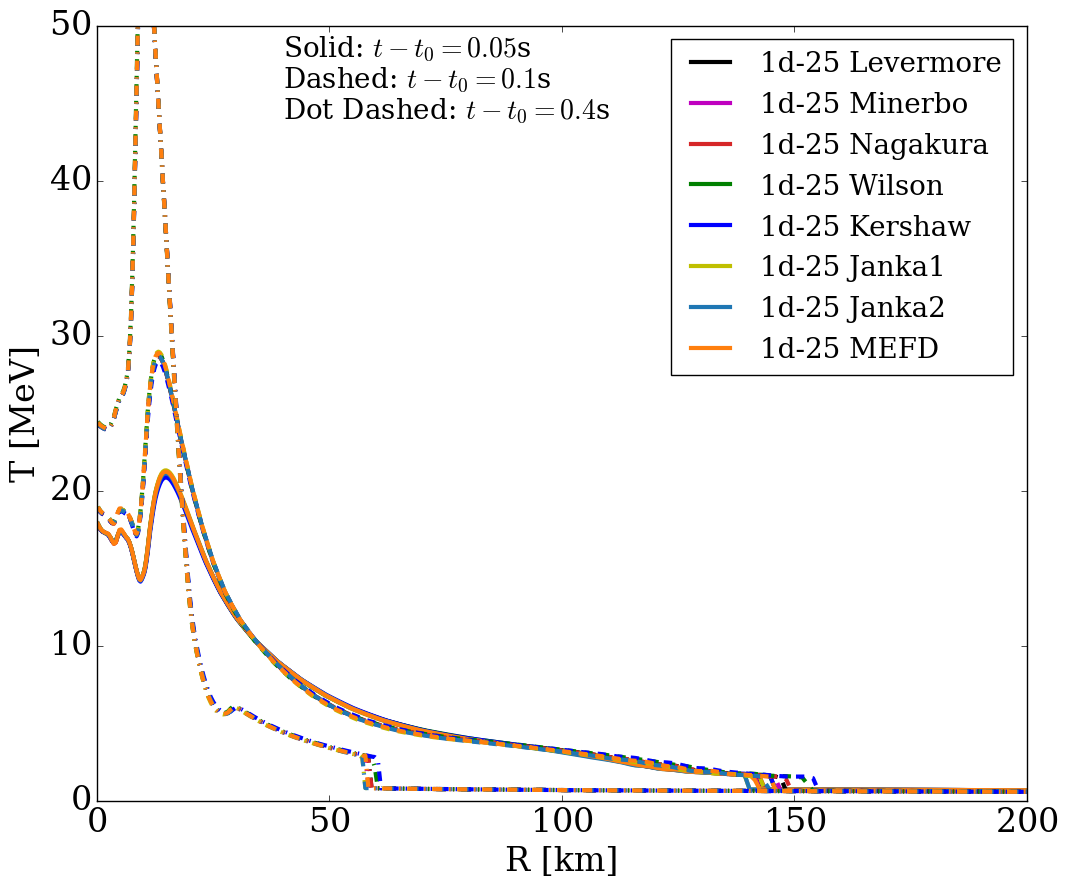}
    \caption{{Similar to Figure \ref{fig:rho-profile-all}, but for temperature profiles. Differences only occur at the stalled shock.}}
    \label{fig:T-profile-all}
\end{figure}

Figure \ref{fig:rs-all} shows the shock radius evolution. Most closures give similar shock positions, but the Kershaw and Wilson closures lead to $5\%-10\%$ larger shock radii. Figures \ref{fig:rho-profile-all}, \ref{fig:ye-profile-all}, \ref{fig:S-profile-all}, and \ref{fig:T-profile-all} show the profiles of density, $Y_e$, entropy, and temperature. Profiles at $t-t_0=0.05$, $0.1$ and $0.4$ seconds after bounce are plotted. Differences seen in these figures between closures are closely related to the positions of the shock. In the entropy and temperature profiles, differences occur only at the stalled shock. Slightly larger differences can be found in the density and $Y_e$ profiles. From these profiles, we see that the post-shock density and $Y_e$ are similar, but that the shock positions vary with the closure. In a sense, the density and $Y_e$ profiles can be regarded as ``stretched" to comport with their larger shock radii and the observed secondary differences occur only between the proto-neutron star surface (arbitrarily defined as $\rho=10^{11}$ g cm$^{-1}$) and the shock.

\begin{figure}
    \centering
    \includegraphics[width=0.48\textwidth]{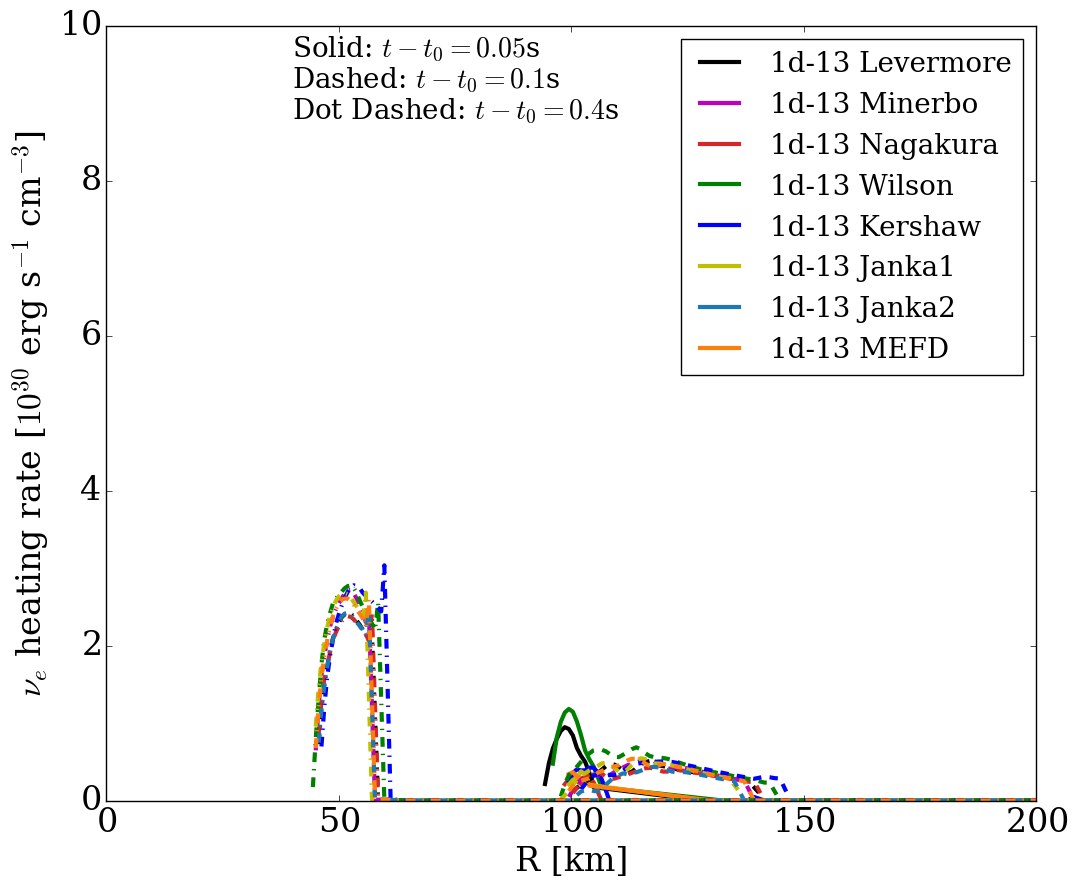}
    \includegraphics[width=0.48\textwidth]{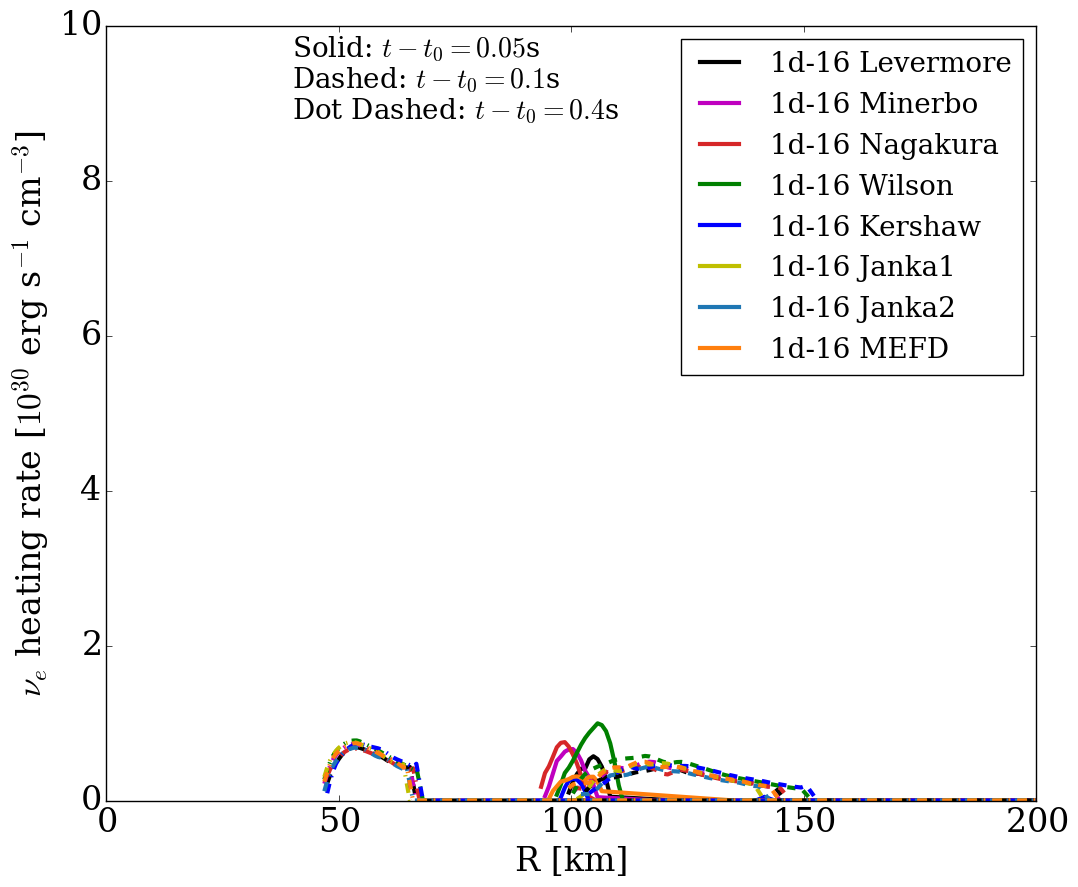}
    \includegraphics[width=0.48\textwidth]{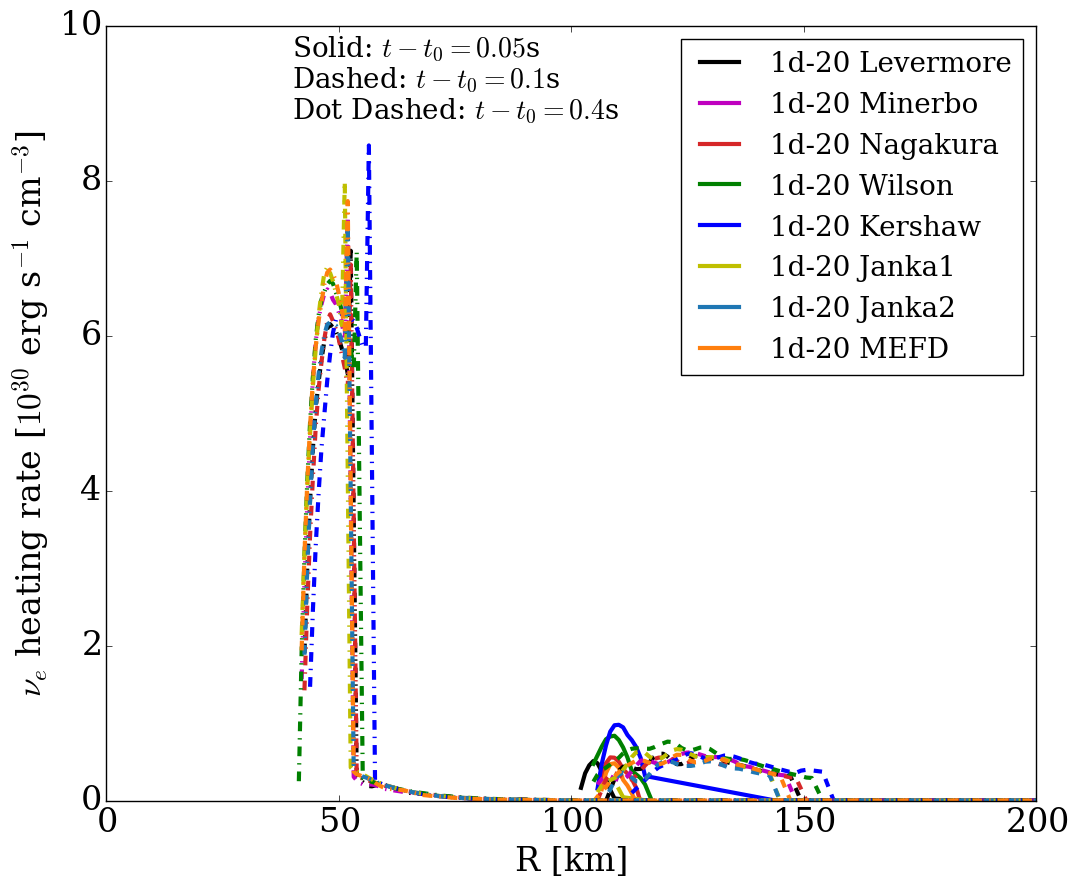}
    \includegraphics[width=0.48\textwidth]{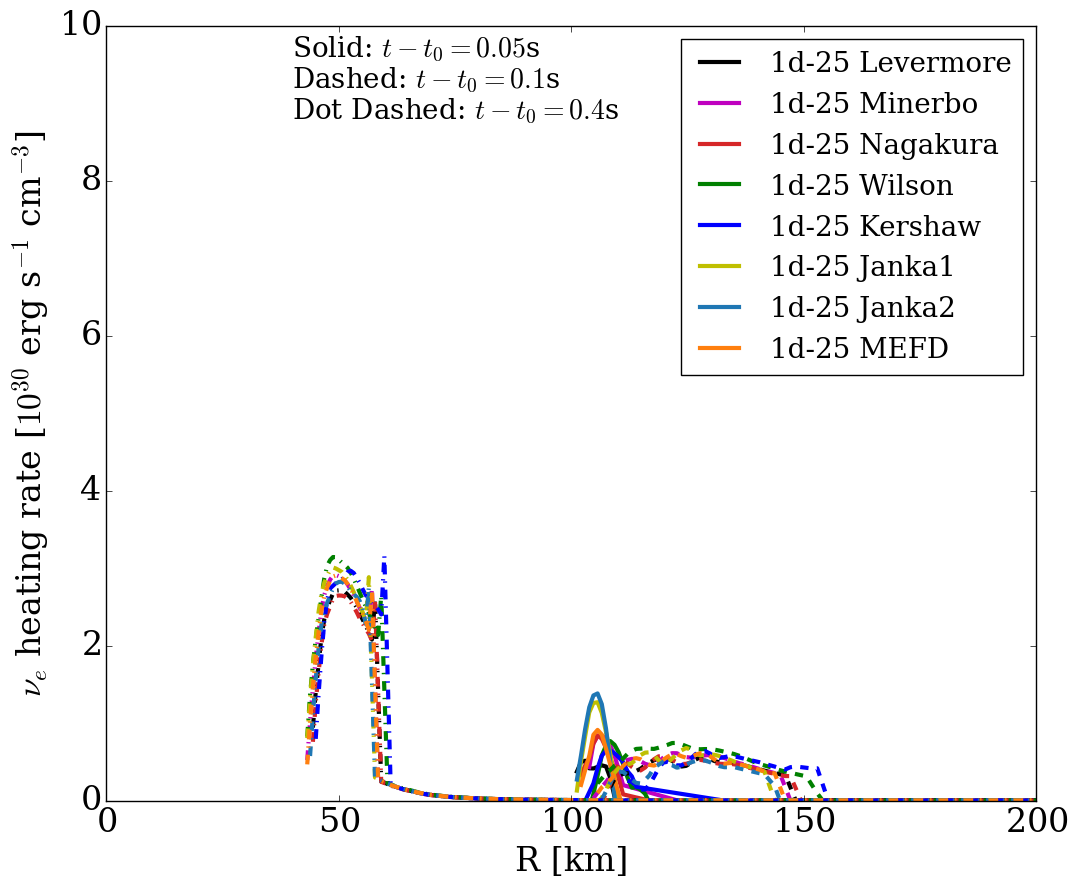}
    \caption{Electron-type neutrino heating rate profiles at 0.05 s (solid lines), 0.1 s (dashed lines) and 0.4 s (dot dashed lines) after bounce. {At early times the profiles look different because the electron-type neutrino heating is  small and heating is dominated by anti-electron-type neutrinos. However, we can still see that the gain radii (left boundary of the gain region) are roughly at the same positions for the different closure choices, while the shock radius (right boundary) varies slightly. }}
    \label{fig:qdot0-profile-all}
\end{figure}

\begin{figure}
    \centering
    \includegraphics[width=0.48\textwidth]{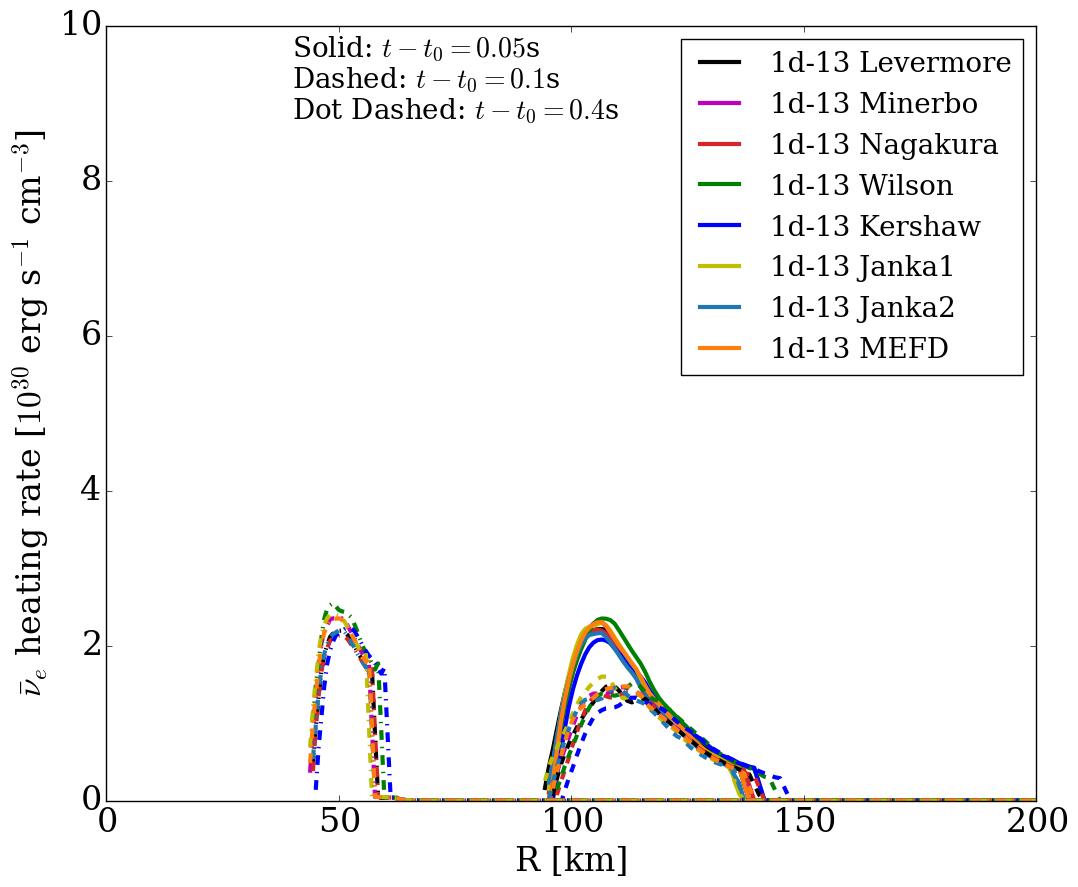}
    \includegraphics[width=0.48\textwidth]{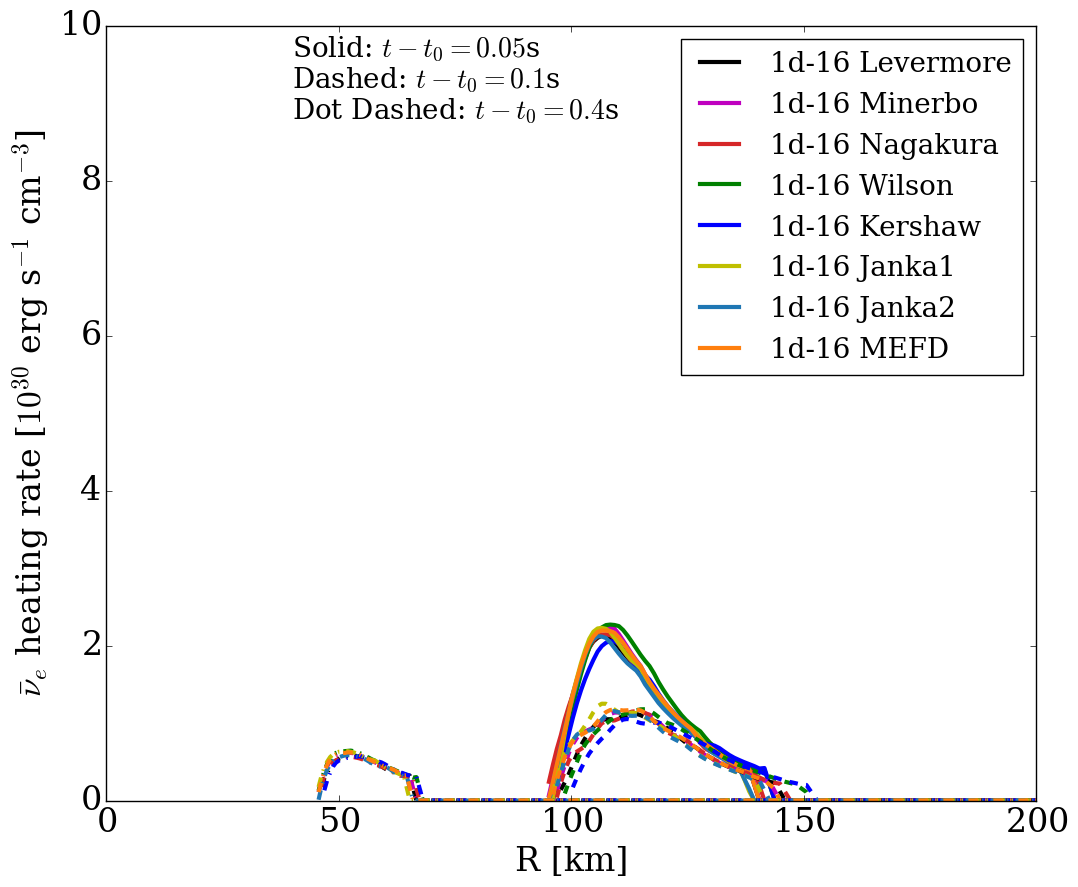}
    \includegraphics[width=0.48\textwidth]{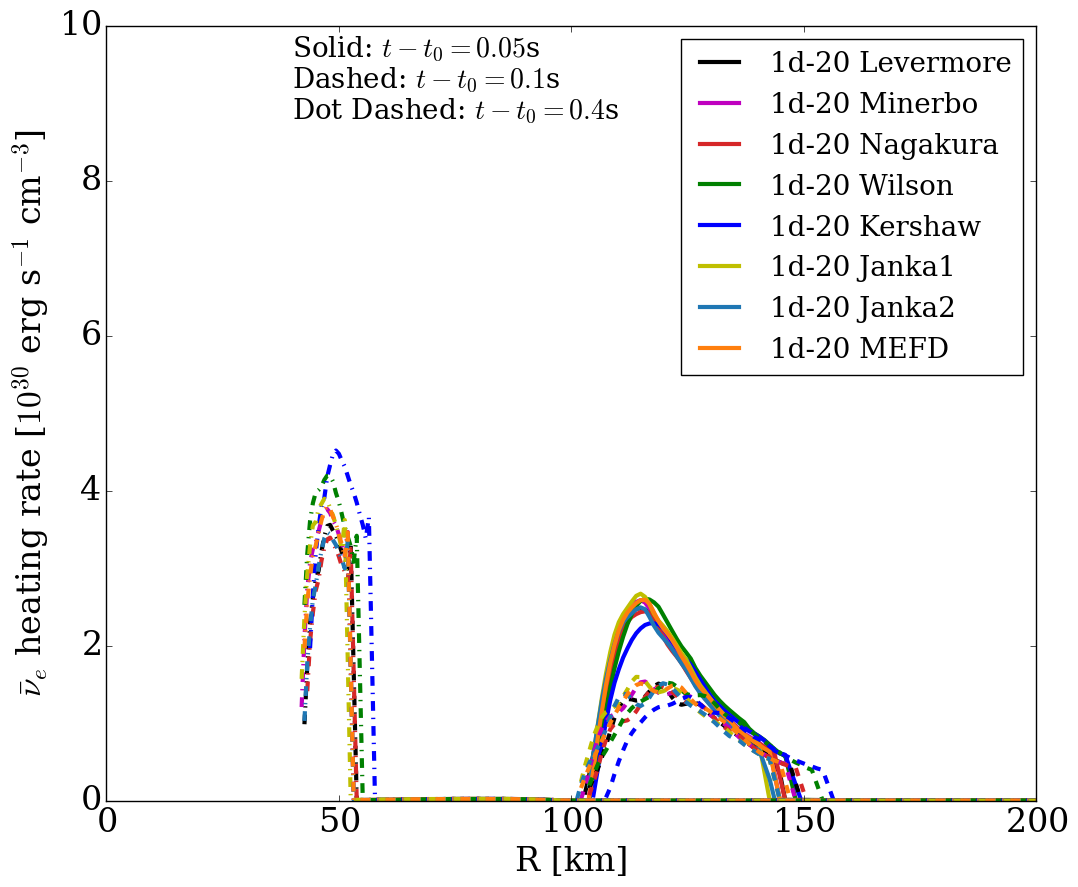}
    \includegraphics[width=0.48\textwidth]{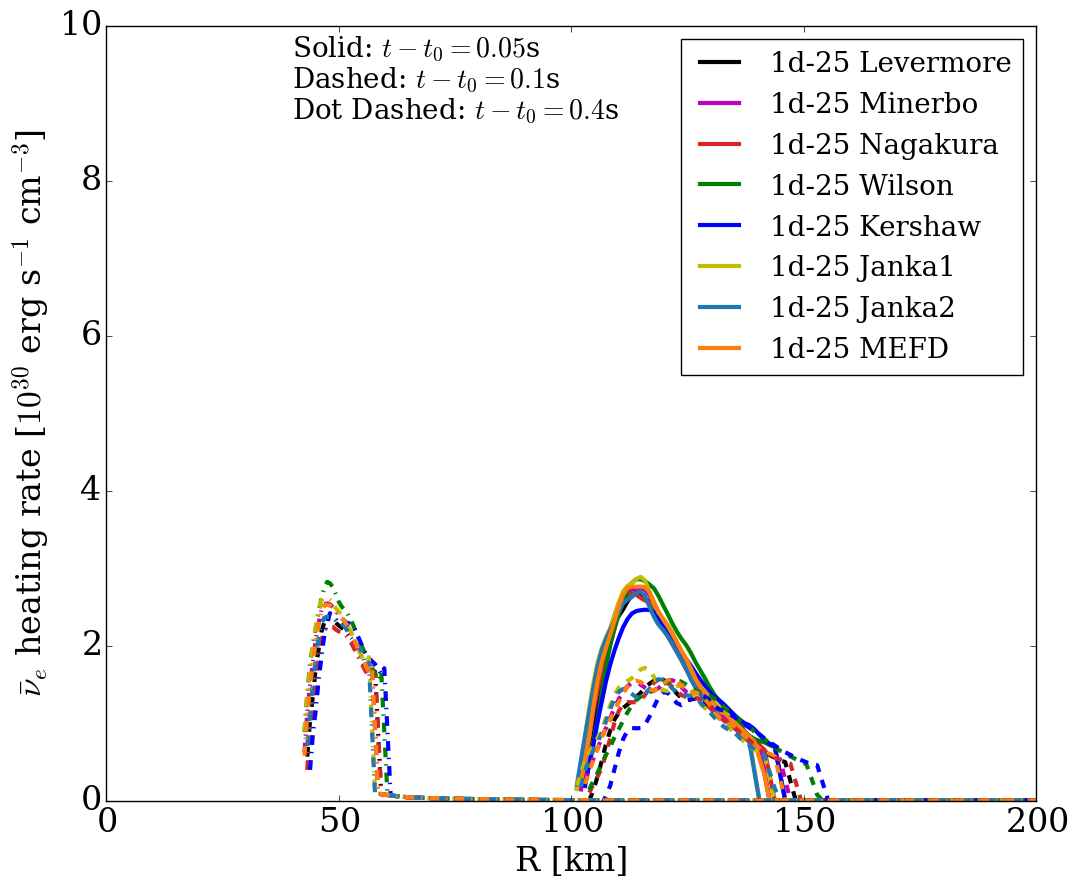}
    \caption{{Similar to Figure \ref{fig:qdot0-profile-all}, but for anti-electron-type neutrino heating rate profiles. The gain region can be clearly seen in this figure. We can see that the gain radii (left boundary of the gain region) are roughly the same for the different closure choices, while the shock radii (right boundary) vary. This leads to gain region volume differences of up to 50\%.}}
    \label{fig:qdot1-profile-all}
\end{figure}

\begin{figure}
    \centering
    \includegraphics[width=0.45\textwidth]{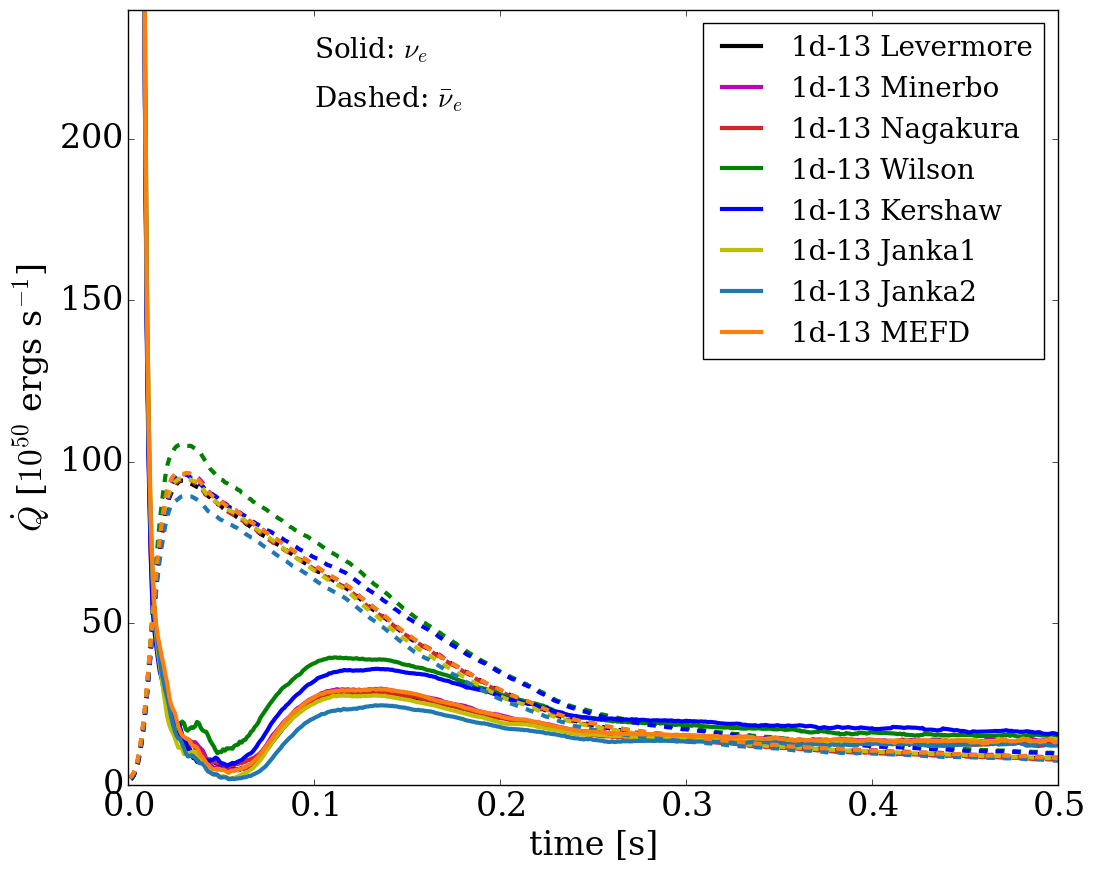}
    \includegraphics[width=0.45\textwidth]{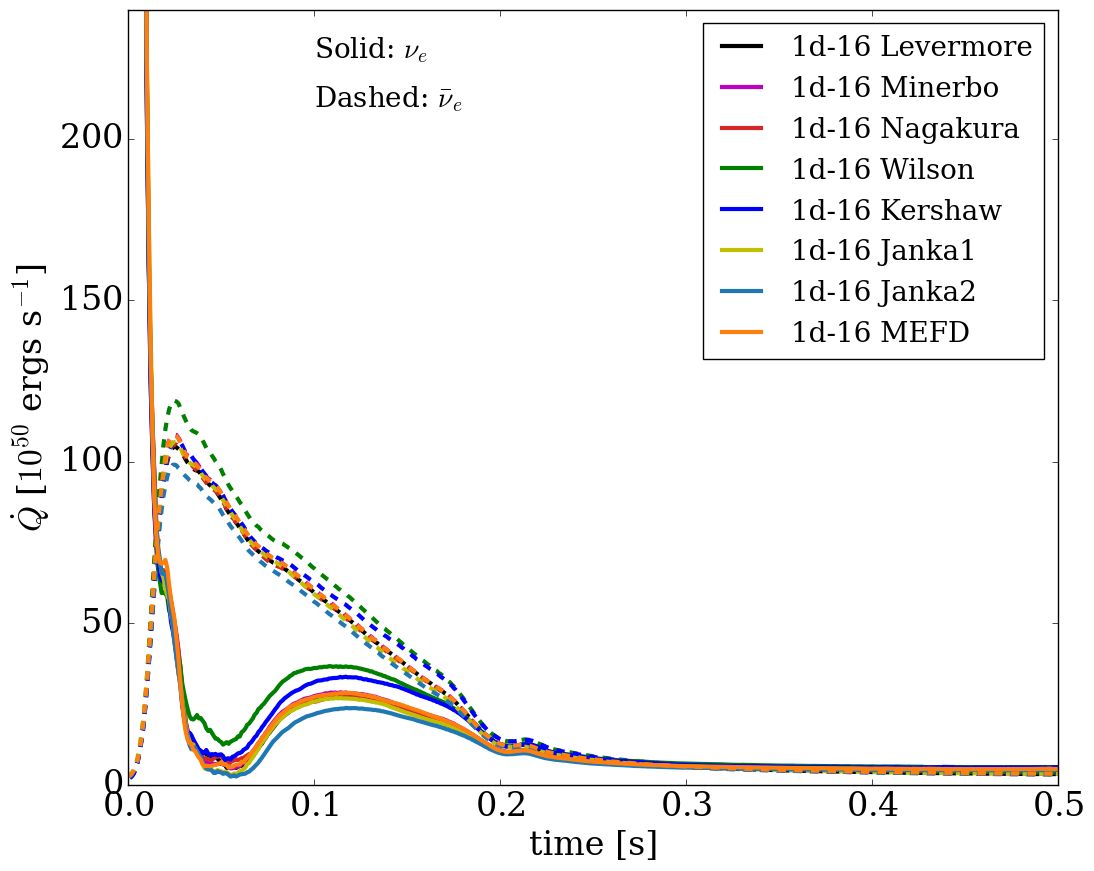}
    \includegraphics[width=0.45\textwidth]{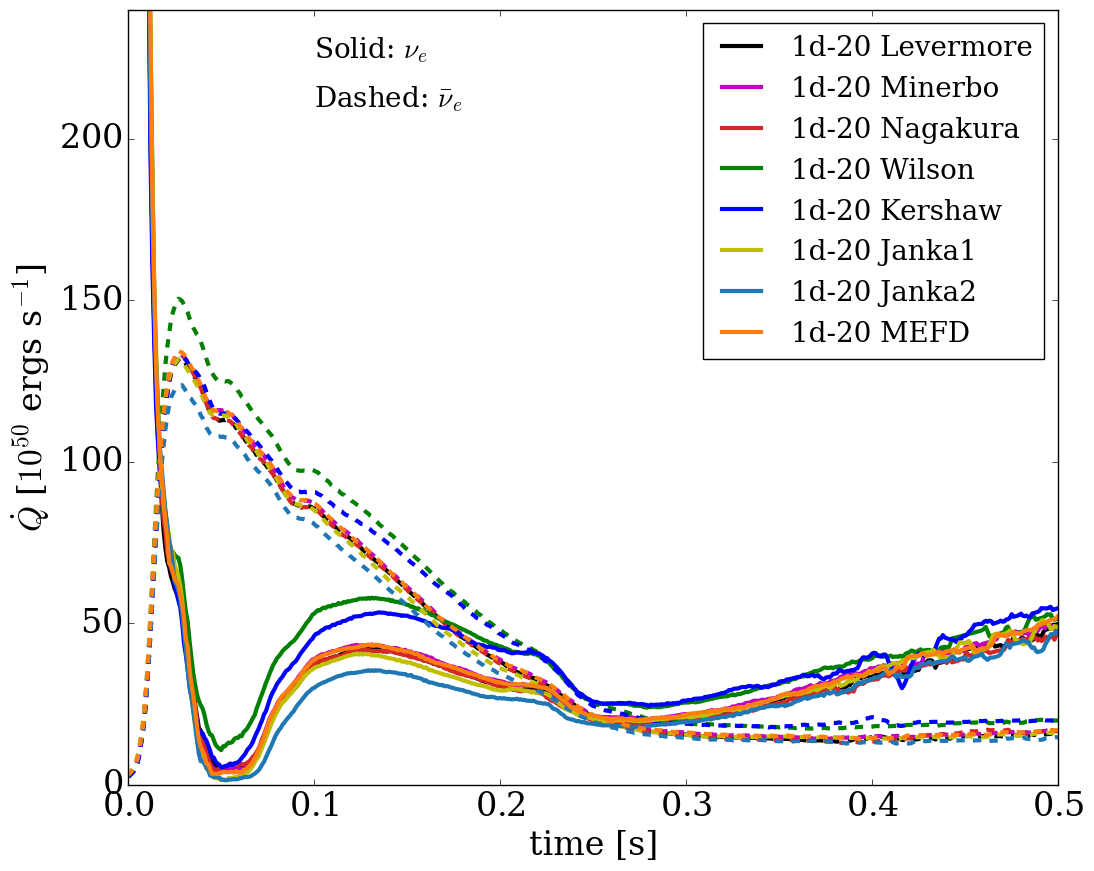}
    \includegraphics[width=0.45\textwidth]{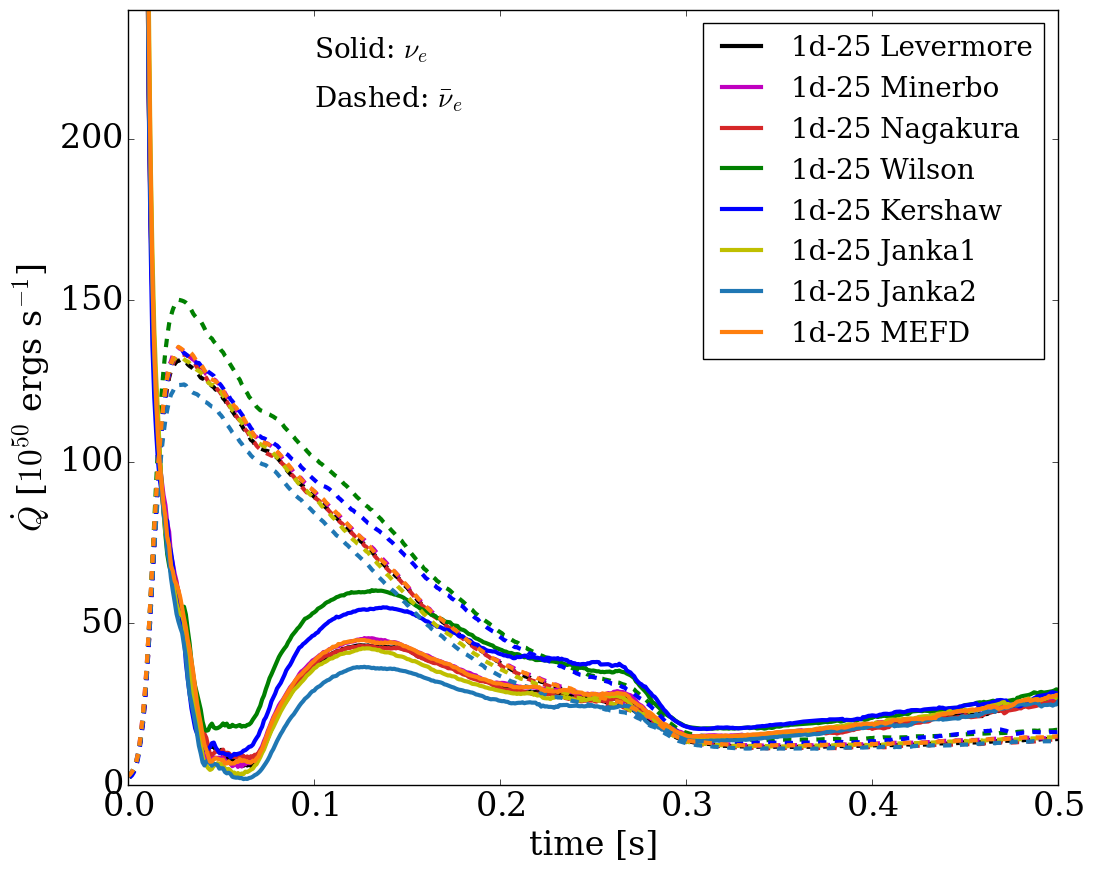}
    \caption{Total heating rate versus time. Solid lines are for electron-type neutrinos and dashed lines are for anti-electron-type neutrinos. $\mu$-type neutrinos are ignored since their heating rate is much smaller. Most closures give similar total heating rate, while the Kershaw and Wilson closures result in higher total heating rates than those of the others by up to $50\%$. The heating rate of the Janka2 closure is $\sim10\%$ lower than those of the others.}
    \label{fig:Qdot-all}
\end{figure}

Figures \ref{fig:qdot0-profile-all} and \ref{fig:qdot1-profile-all} depict the profiles of $\nu_e$ and $\bar{\nu}_e$ neutrino heating rates per unit volume ($\frac{d\dot{Q}}{dV}$). The heating rate of ``$\mu$-type" neutrinos is much smaller and is here ignored. From these plots, we note that most closures have a similar gain radius and that the size of gain region is mostly determined by the shock radius.  Closures leading to larger shock radii, such as the Kershaw and Wilson choices, can lead to gain regions $\sim50\%$ larger in volume. The result of larger gain regions is shown in Figure \ref{fig:Qdot-all}, which shows the integrated total heating rate, $\dot{Q}$. Differences in the total heating rates are significantly larger than those seen in Figures \ref{fig:qdot0-profile-all} and \ref{fig:qdot1-profile-all}; these differences are mainly caused by the shock position differences.  

\begin{figure}
    \centering
    \includegraphics[width=0.48\textwidth]{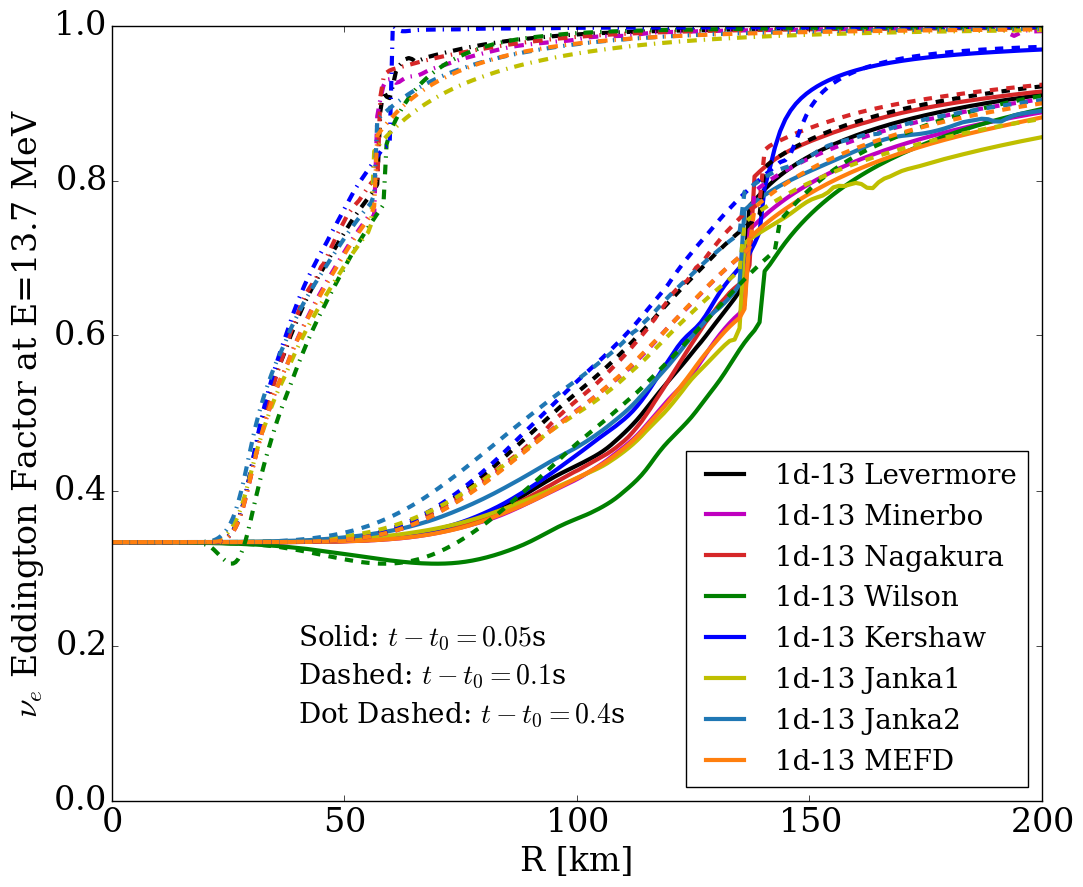}
    \includegraphics[width=0.48\textwidth]{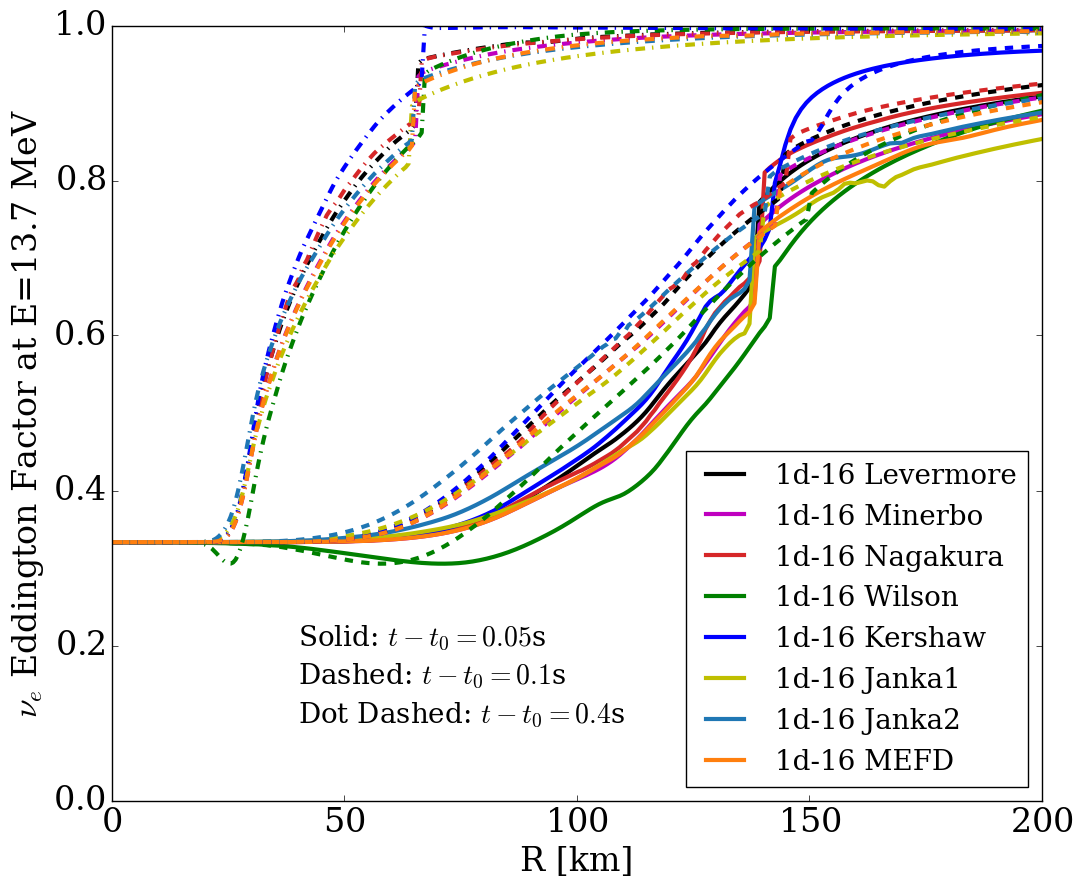}
    \includegraphics[width=0.48\textwidth]{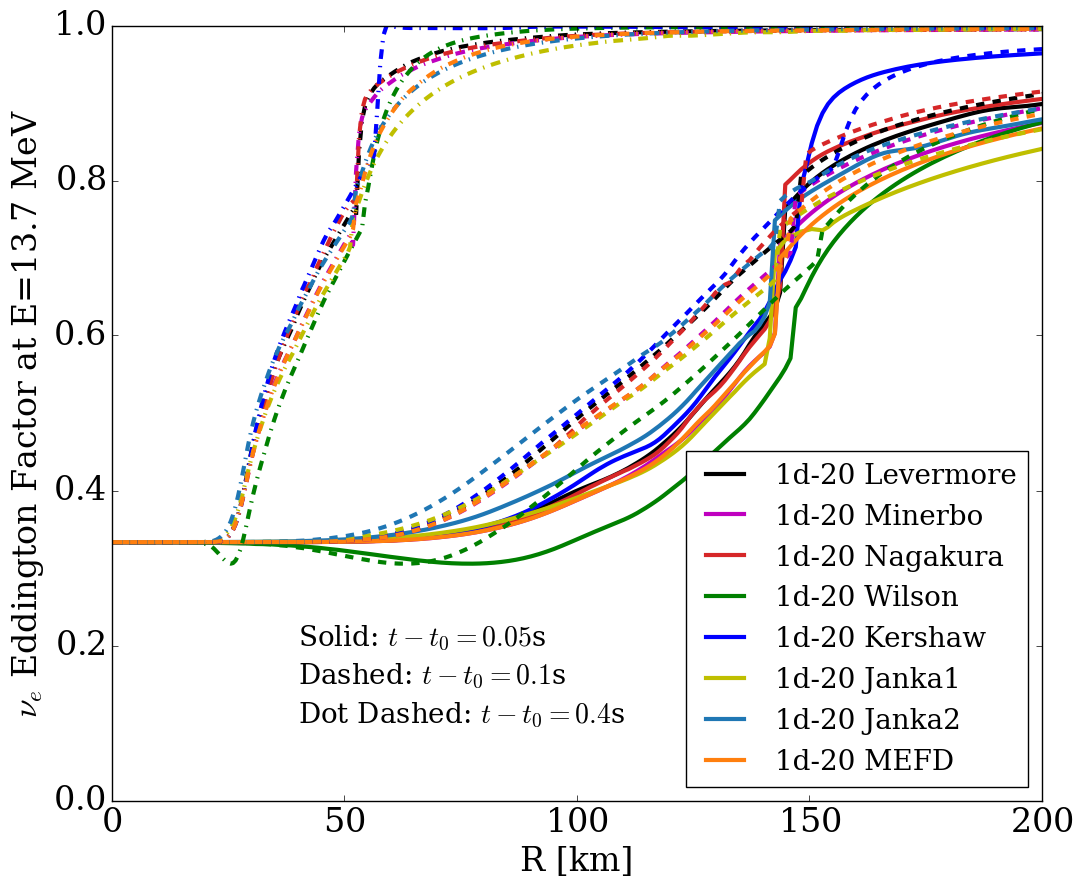}
    \includegraphics[width=0.48\textwidth]{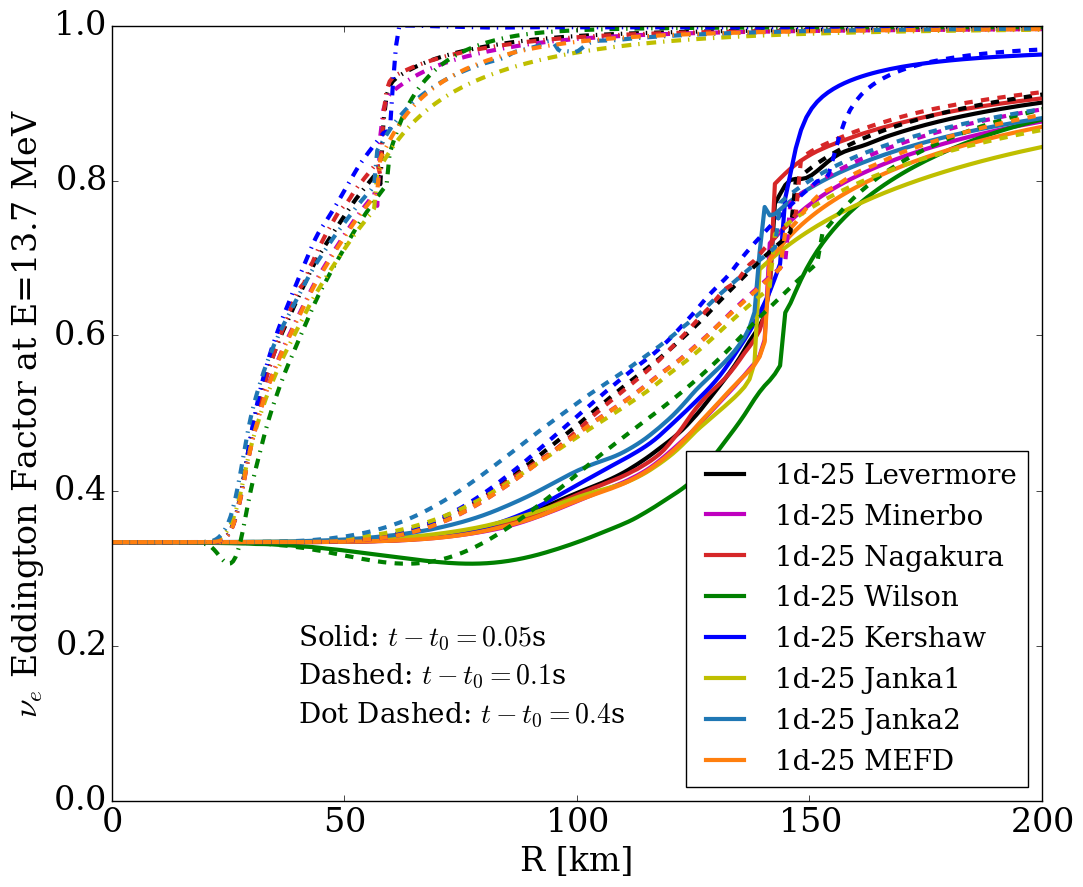}
    \caption{Electron-type neutrino Eddington factor at 13.7 MeV at 0.05 s (solid lines), 0.1 s (dashed lines), and 0.4 s (dot dashed lines) after bounce. {We chose to highlight the 13.7-MeV group instead of that nearer the peak of the electron-type neutrino spectrum because the heating rate is proportional to $E^2\mathcal{F}$ and 13.7 MeV is closer to the peak of the energy deposition. Eddington factor profiles of the energy group closest to the spectra peak show similar results.}}
    \label{fig:p0-profile-all}
\end{figure}

\begin{figure}
    \centering
    \includegraphics[width=0.48\textwidth]{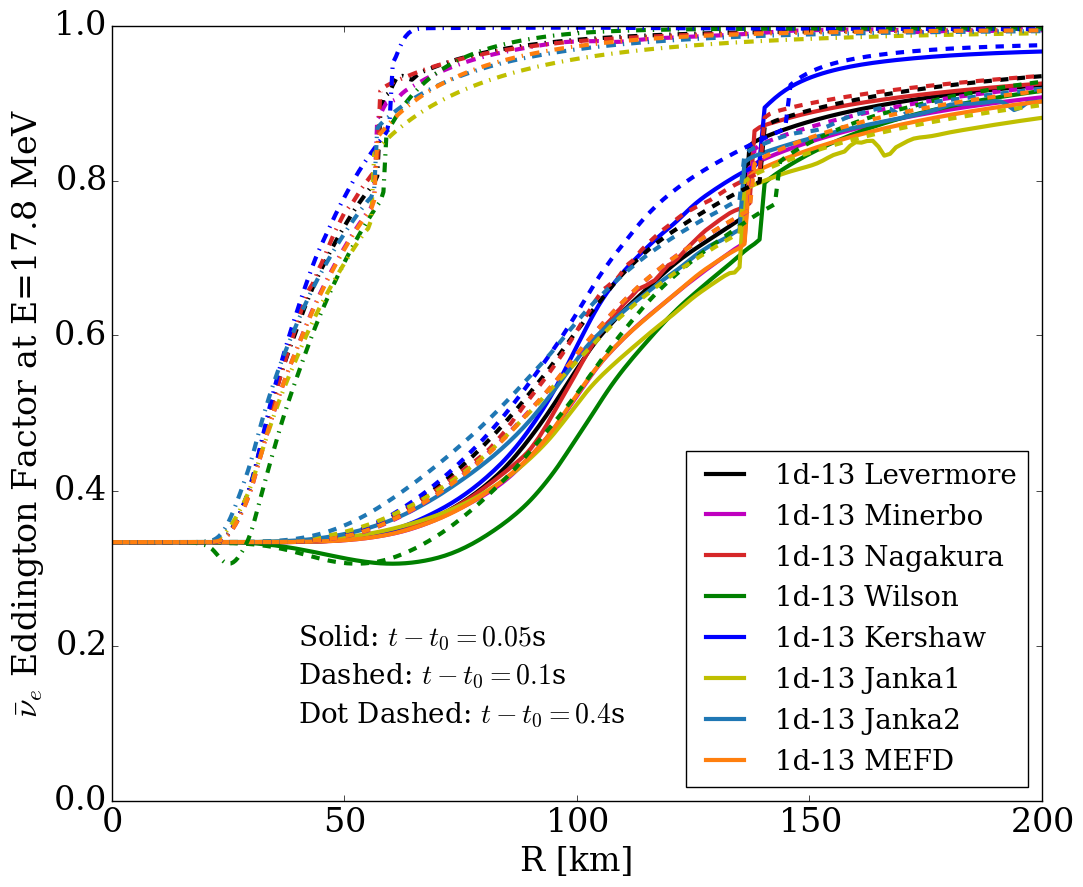}
    \includegraphics[width=0.48\textwidth]{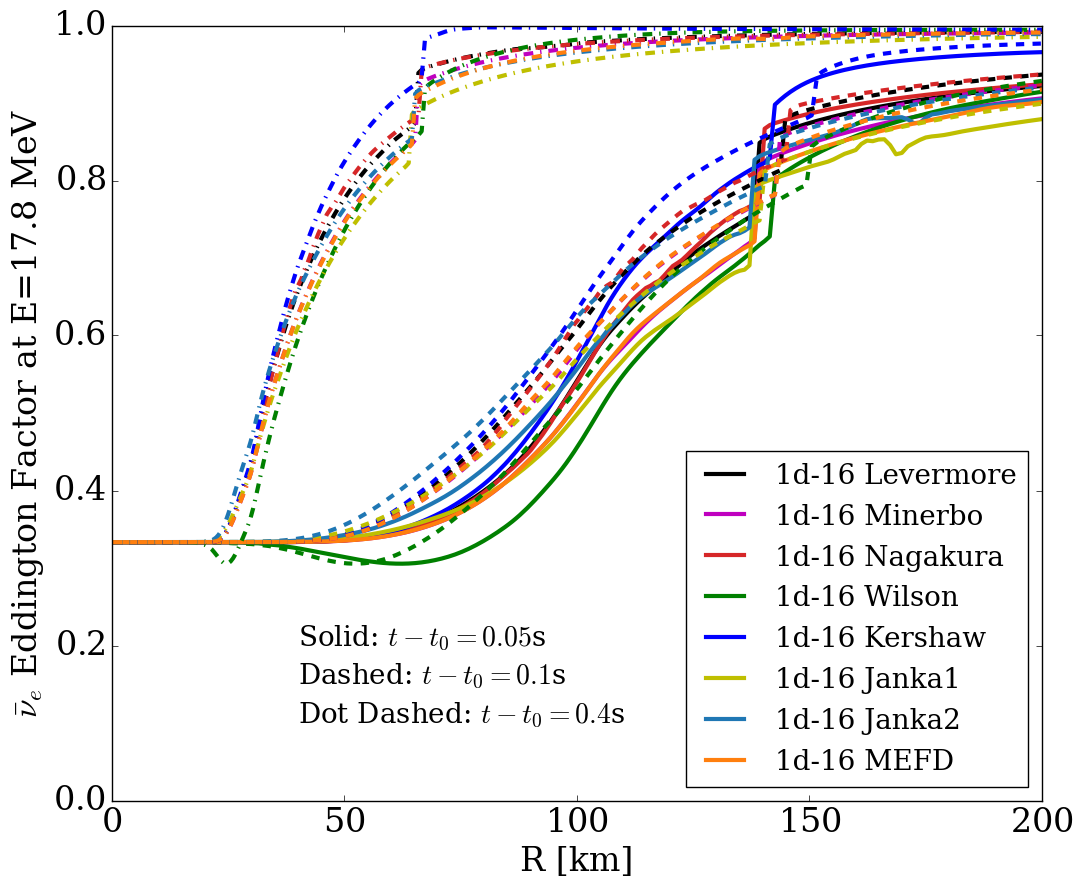}
    \includegraphics[width=0.48\textwidth]{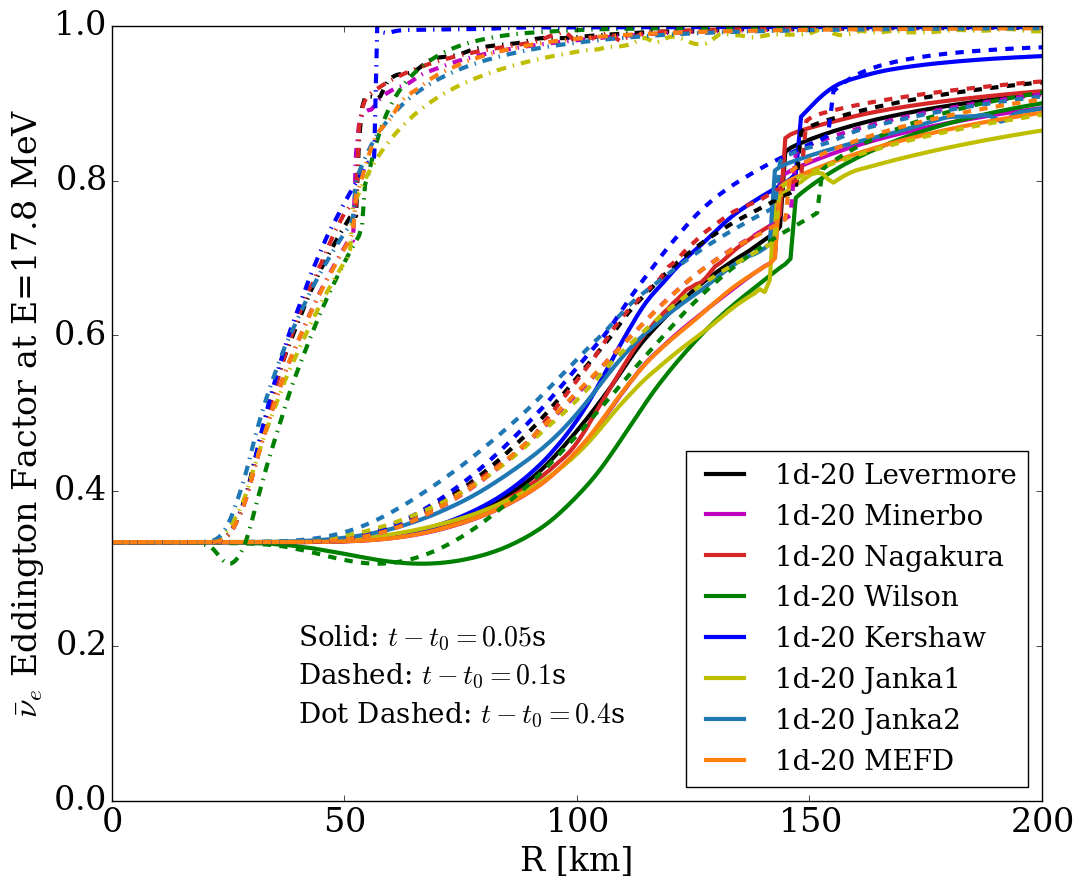}
    \includegraphics[width=0.48\textwidth]{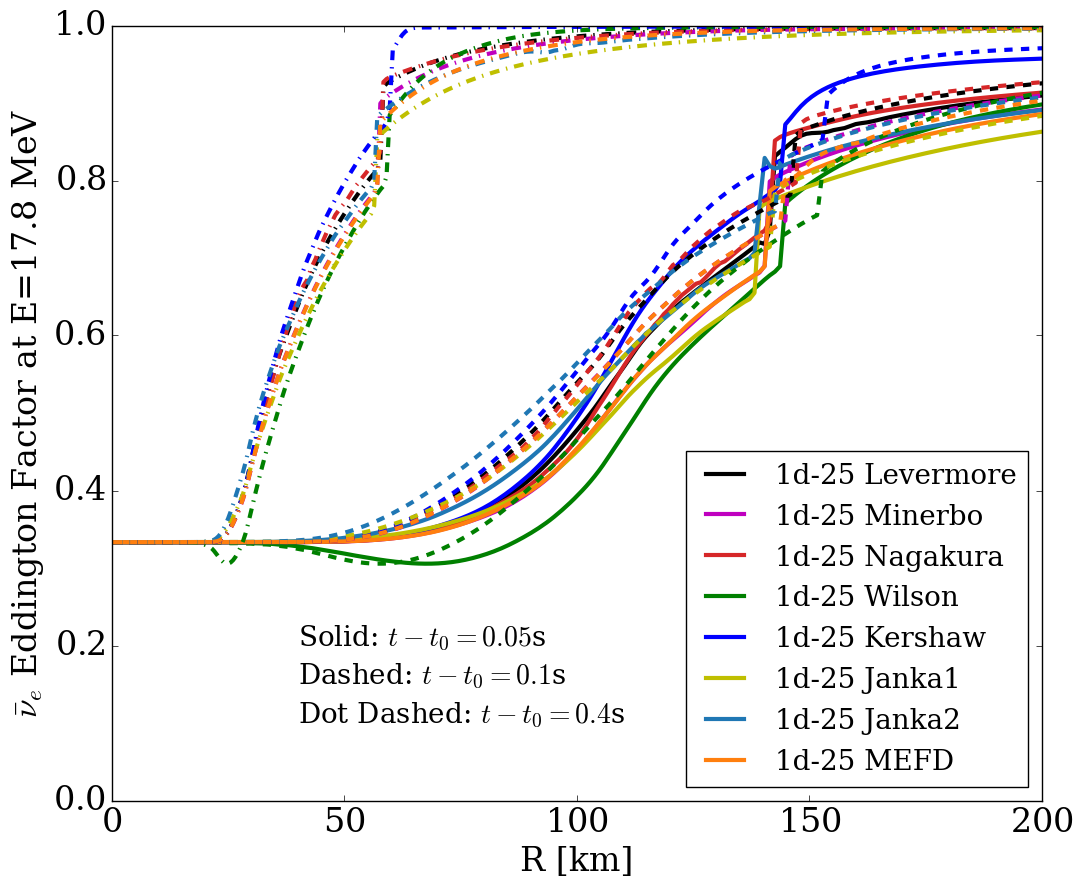}
    \caption{{Anti-e}lectron-type neutrino Eddington factor at 17.8 MeV at 0.05 s (solid lines), 0.1 s (dashed lines), and 0.4 s (dot dashed lines) after bounce. {We chose to show 17.8-MeV group instead of that near the peak of the electron-type neutrino spectrum because the heating rate is proportional to $E^2\mathcal{F}$ and 17.8 MeV is closer to the peak of energy deposition. Eddington factor profiles of the energy group closest to the spectra peak show similar results.}}
    \label{fig:p1-profile-all}
\end{figure}

The $\nu_e$ and $\bar{\nu}_e$ Eddington factor profiles of the energy group closest to the energy deposition peaks are shown in Figure \ref{fig:p0-profile-all} and \ref{fig:p1-profile-all}. Although these profiles deviate from each other significantly, we have already noted that the profiles of other variables show much weaker differences all closely related to the relative shock positions. One thing to notice is that larger deviations in the Eddington factor profiles don't necessarily translate into larger differences in other quantities. The Kershaw and Janka1 closures deviate from the Levermore closure by a similar fraction. However, unlike the Kershaw closure, the Janka1 closure manifests little difference from the Levermore closure in all other figures shown.

\begin{figure}
    \centering
    \includegraphics[width=0.48\textwidth]{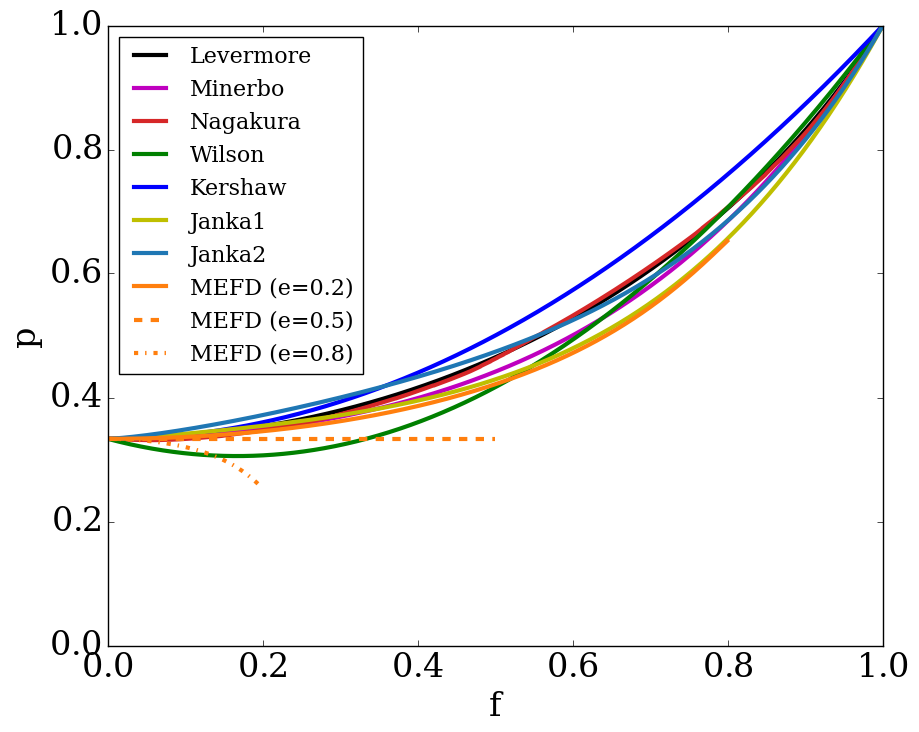}
    \includegraphics[width=0.48\textwidth]{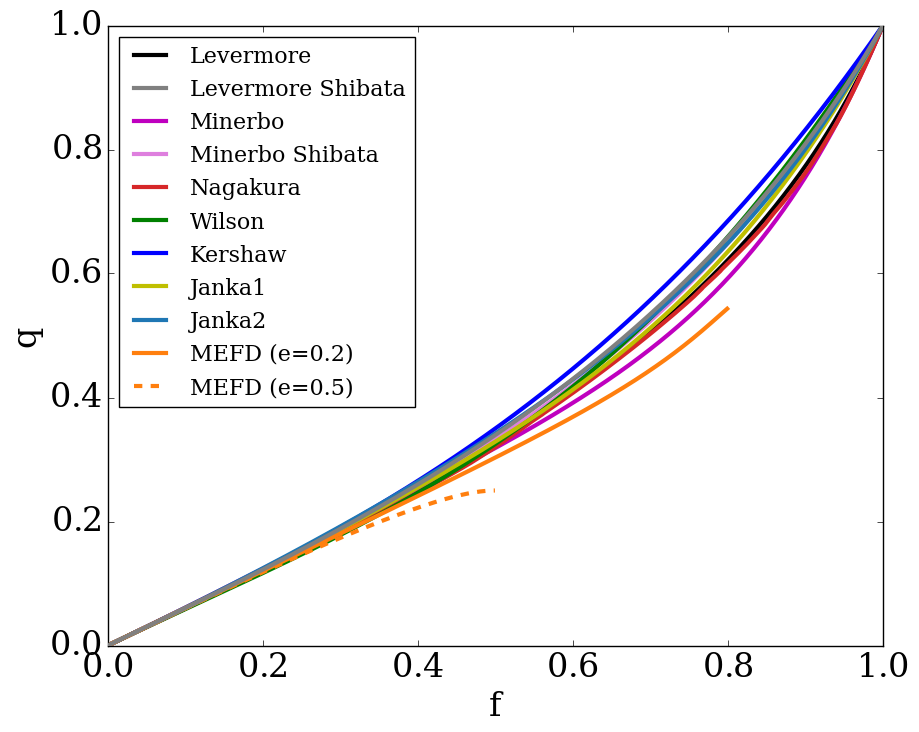}
    \caption{The $p$ and $q$ parameters given by different closures. Left panels are $p$ parameters and right panels are $q$ parameters. Most closures don't depend on $e$ except for the maximum entropy closure with a Fermi-Dirac distribution (MEFD), so we plot MEFD with $e=0.2$, $e=0.5$ and $e=0.8$. The $e=0$ case is the same as the Minerbo closure. Notice that the flux factor $f$ in the MEFD closure can not exceed $1-e$. }
    \label{fig:pq}
\end{figure}

Figure \ref{fig:shibata-compare} compares different treatments of the 3rd-order closure relation. For closures like the Levermore and the Minerbo, there are two ways to calculate $q$. The first way is to use the self-consistent form of the 3rd closure their formalism provides, and the other way is to use the Shibata interpolation in which $q=\frac{(3p+2)f}{5}$. In the right panel of Figure \ref{fig:pq}, we can observe that the Shibata interpolation overestimates $q$ by $10\%-15\%$ for the Levermore and Minerbo closures. However, these differences in the 3rd-order closure result in little difference in the shock positions and in the total heating rates. This can be interpreted to mean that the Shibata 3rd-order interpolation is good enough for time-dependent CCSN calculations. This is understandable, since the $q$ factor occurs only in the $\partial_\nu$ term in Equation \ref{equ:1d-F}, and will have a smaller effect than the choice of Eddington factor. The 3rd-order moment relations in the Levermore and MEFD closures are much more difficult to calculate compared to their 2nd order relations. Hence, we suggest that the Shibata interpolation may be used without significant error if so desired. 

Although different choices of 3rd-order closures seem to have little effect on simulation results, one cannot simply ignore the need for non-trivial 3rd-order closures. In one 1D test case, we set $H_{ijk}=0$ and the stalled shock radius was 20-30\% larger. In 2D, this made a non-exploding model explode. We have also tested using $q=f$, which is a straight diagonal line in the right panel of Figure \ref{fig:pq}. This choice led to comparable deviations in various physical quantities that we witnessed when we compared the various 2nd-order closures studied here.

\subsection{2D Simulations}
\begin{figure}
    \centering
    \includegraphics[width=0.48\textwidth]{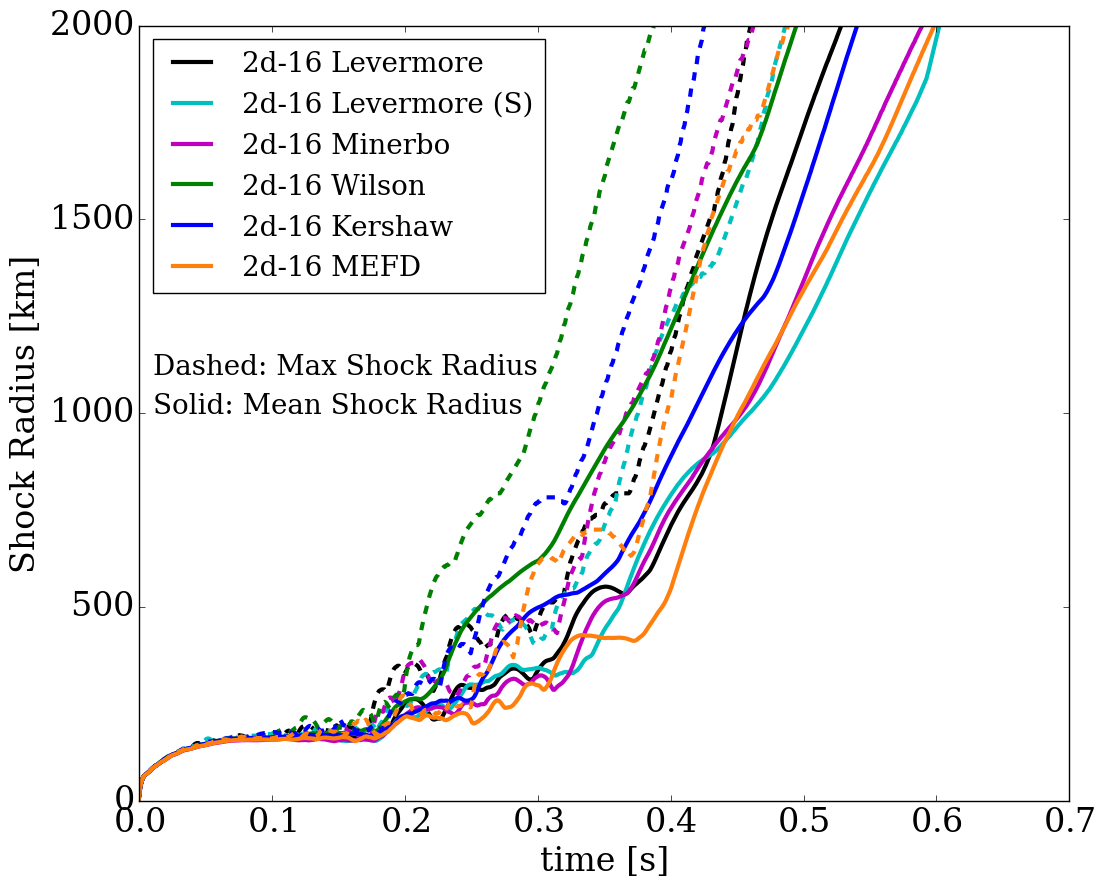}
    \includegraphics[width=0.48\textwidth]{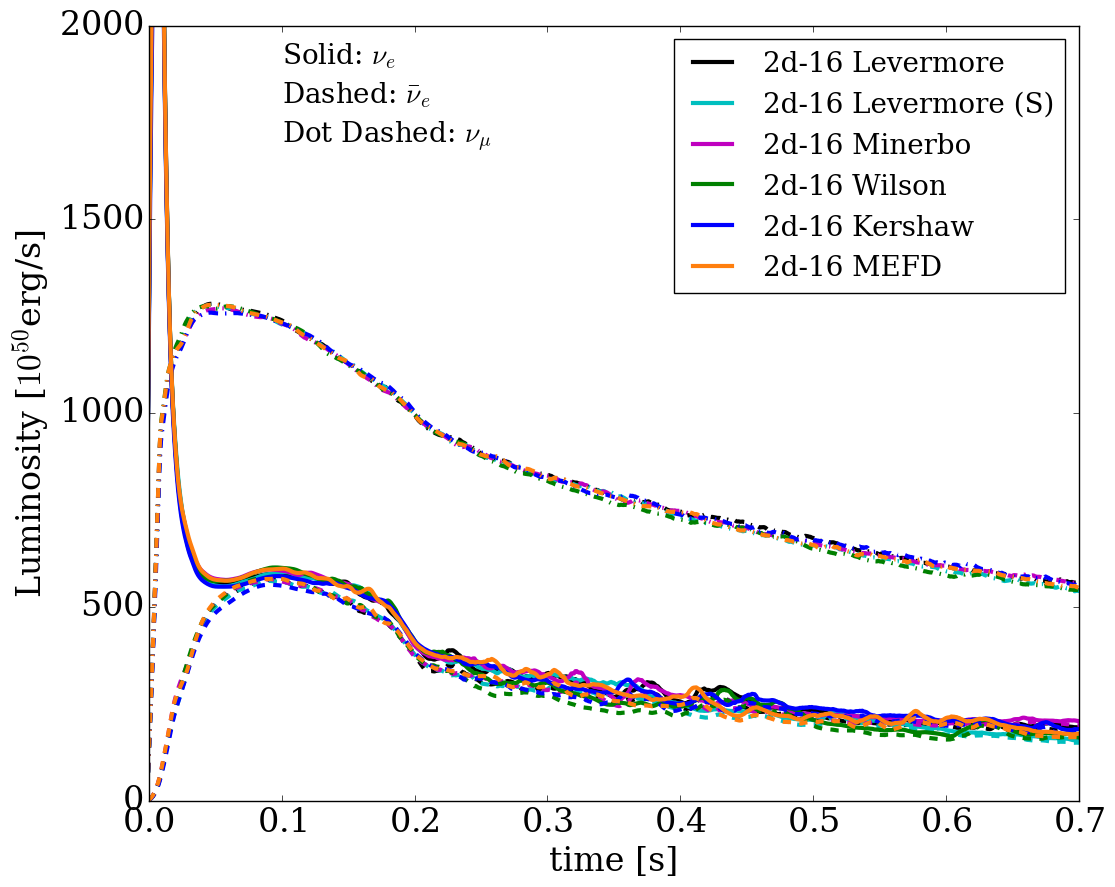}
    \includegraphics[width=0.48\textwidth]{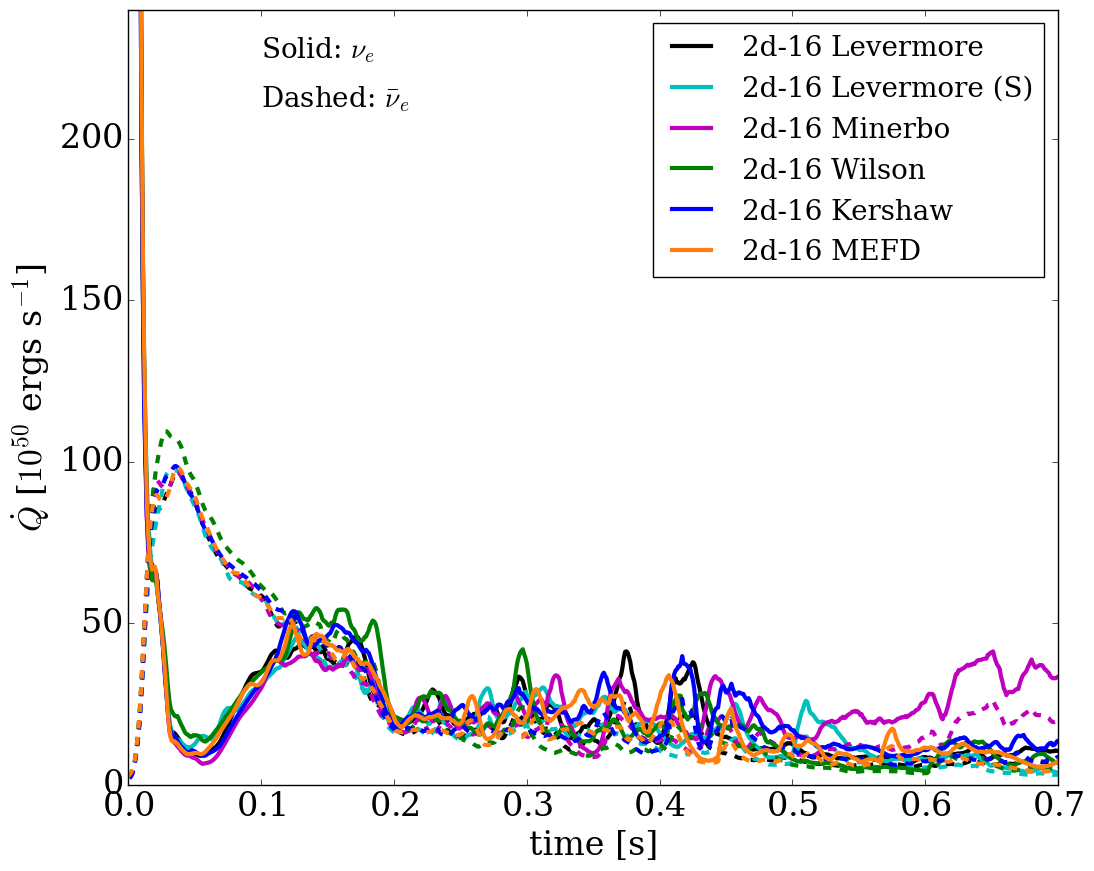}
    \includegraphics[width=0.48\textwidth]{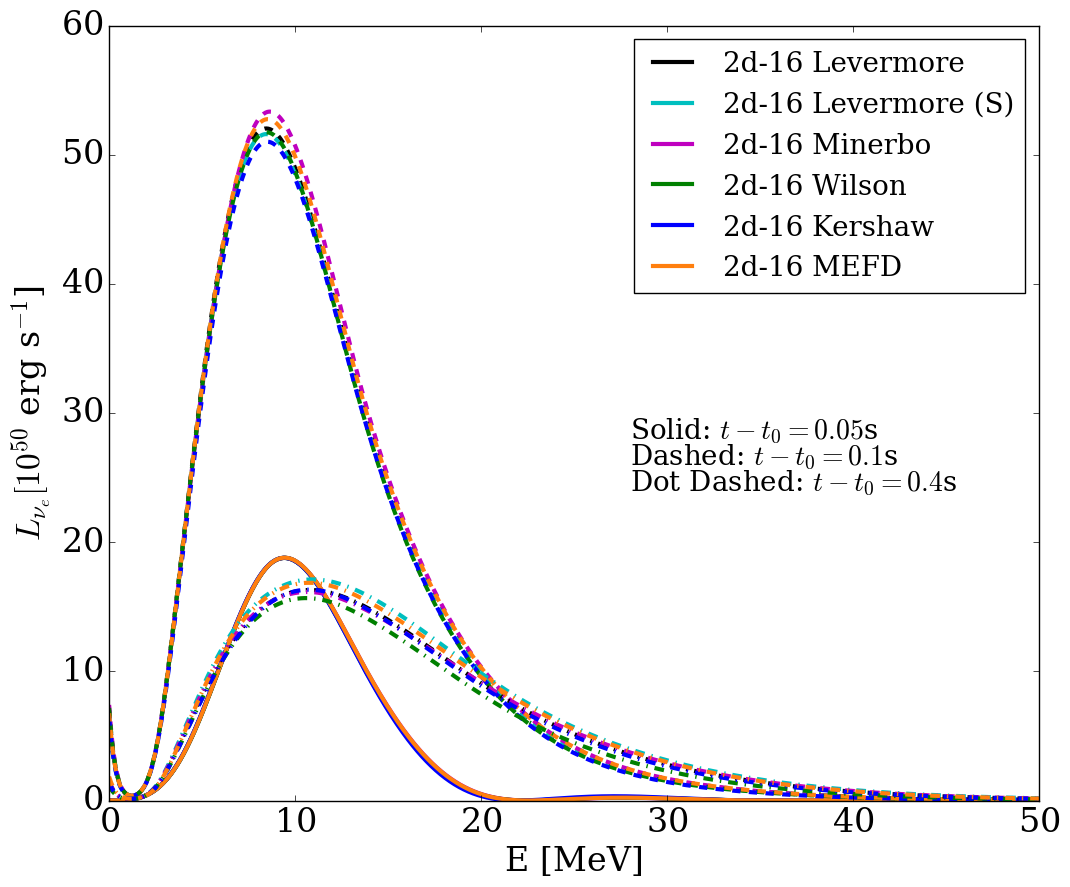}
    \caption{{The shock radius (top left), luminosity (top right), total heating rate (bottom left), and electron-type neutrino spectrum (bottom \tianshu{right}) evolution in the 2D simulations of the 16-$M_\odot$ progenitor. {All simulations show a similar shock propagation speed, while the Wilson and Kershaw closures explode a bit earlier. The Wilson closure has stronger heating compared to others before 200 ms after bounce.} %The Levermore, Levermore with Shibata interpolation (Levermore (S)), and Minerbo closures give similar results. They explode at roughly the same time with the same shock velocity. The Wilson closure behaves differently, leading to stronger heating, and an earlier explosion.
    } }
    \label{fig:2D-evolution}
\end{figure}

\begin{figure}
    \centering
    \includegraphics[width=0.48\textwidth]{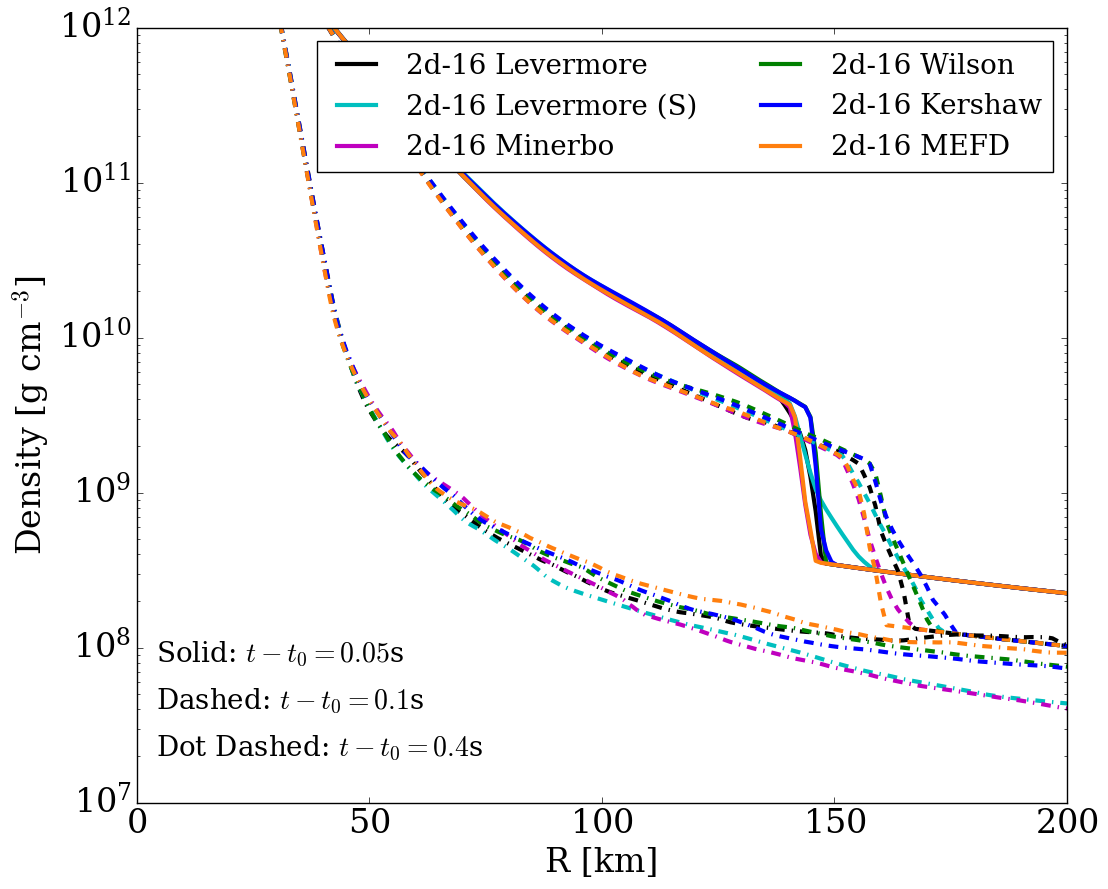}
    \includegraphics[width=0.48\textwidth]{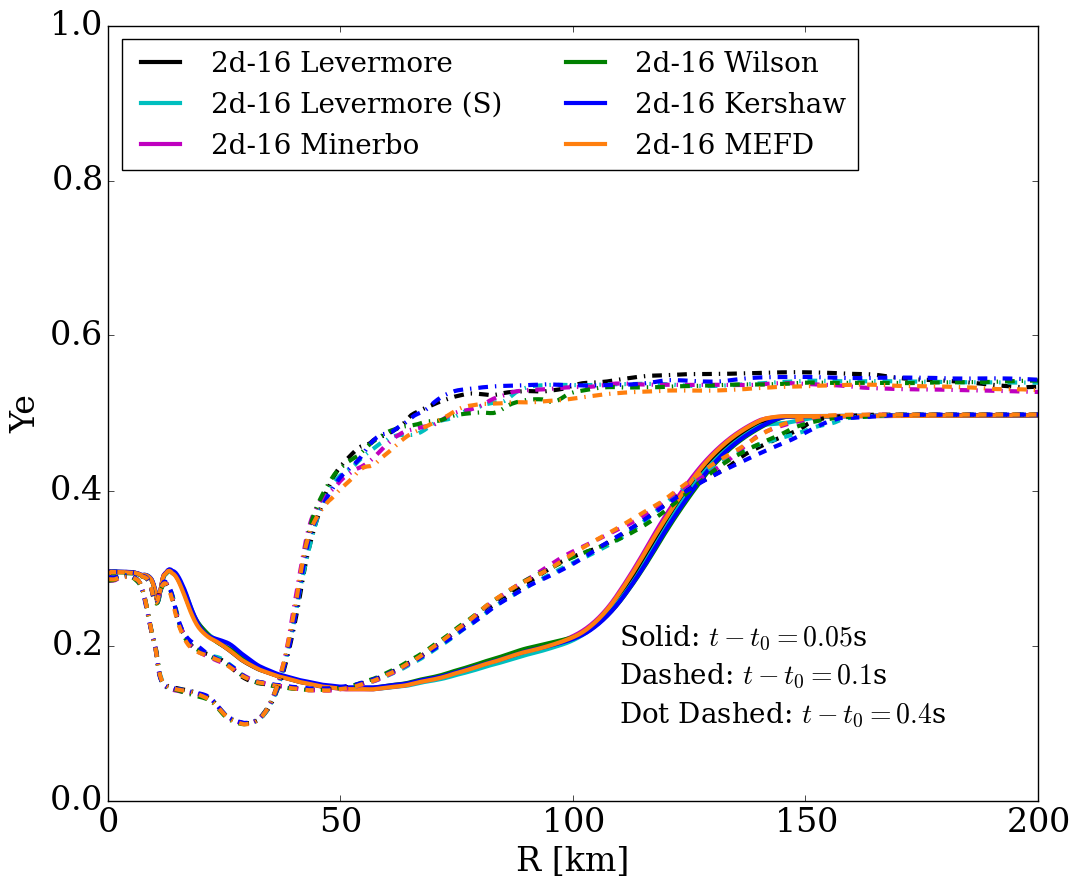}
    \includegraphics[width=0.48\textwidth]{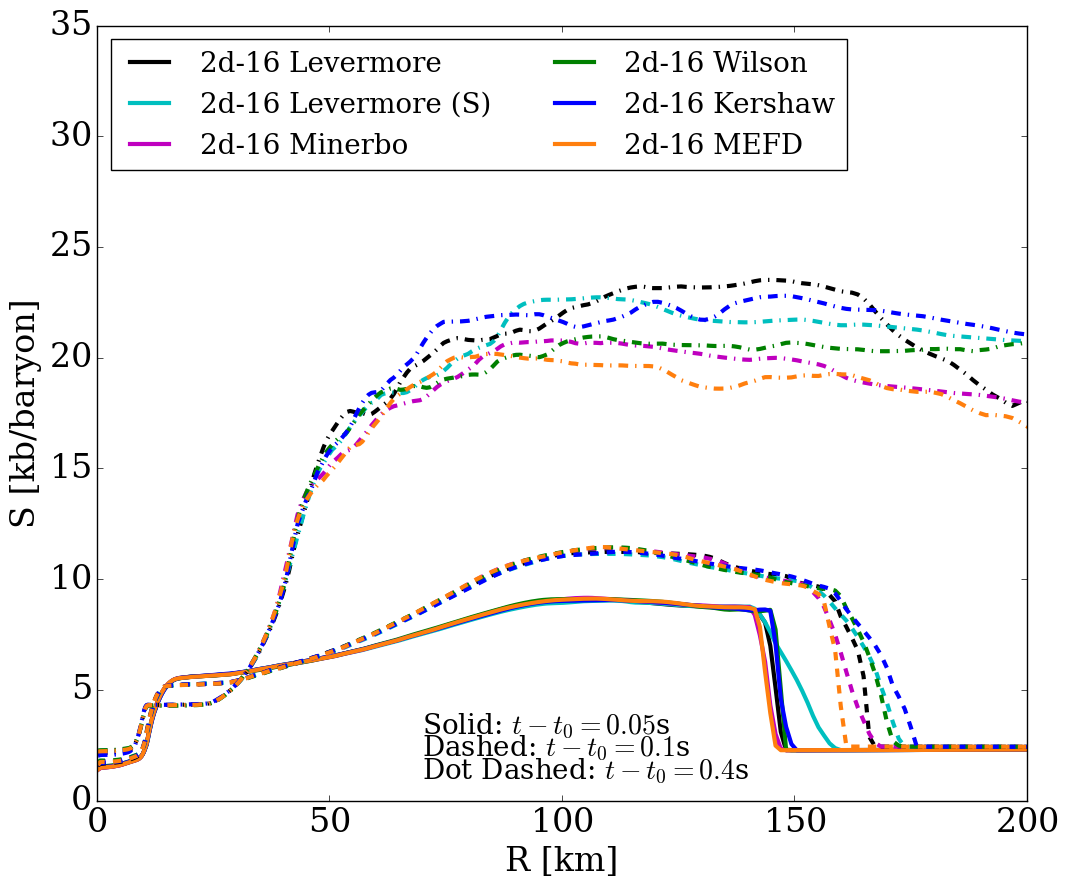}
    \includegraphics[width=0.48\textwidth]{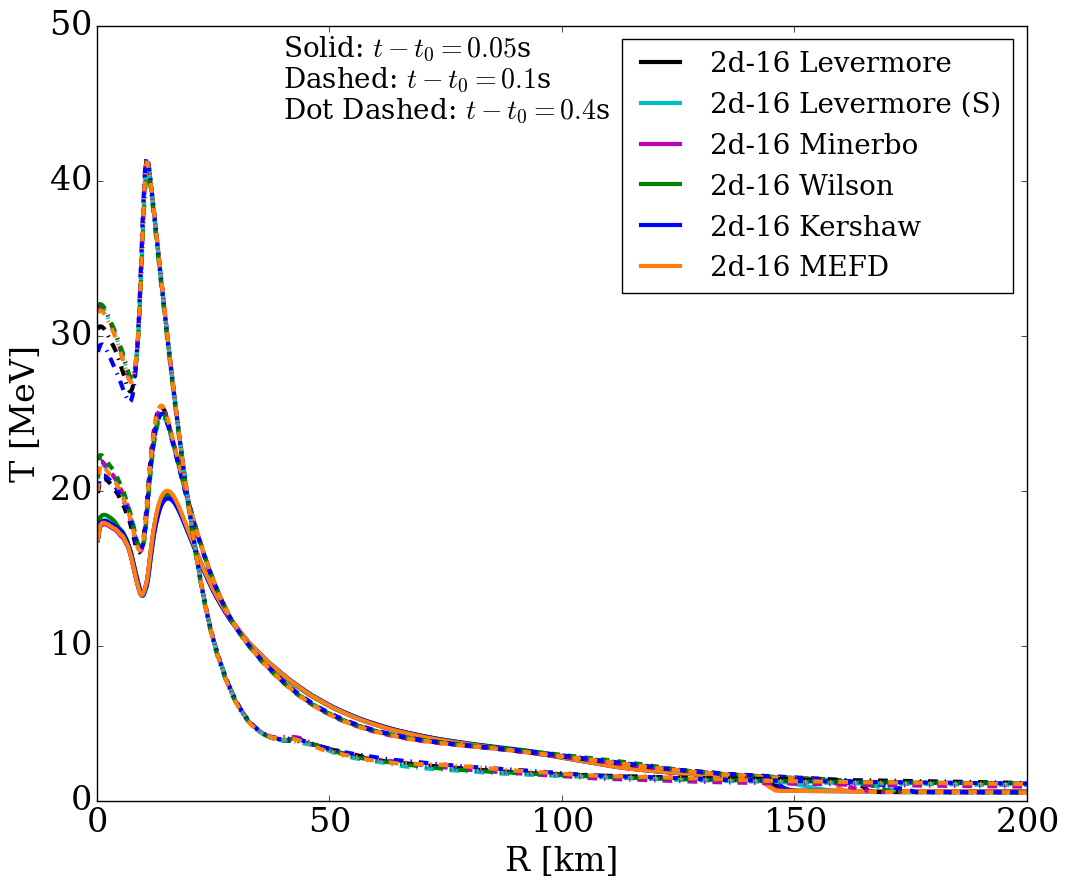}
    \caption{{The density (top left), $Y_e$ (top right), entropy (bottom left), and temperature (bottom \tianshu{right}) angle-averaged profiles at 0.05, 0.1, and 0.4 seconds after bounce in the 2D simulations of the 16-$M_\odot$ progenitor. Profiles before the explosion show similar behavior as seen in 1D, while the differences in profiles at 0.4 seconds are probably amplified by the different explosion morphology caused by hydrodynamic chaos.}}
    \label{fig:2D-profiles}
\end{figure}

{
We test our findings in 1D with a few 2D simulations. They were done with the 16 $M_\odot$ progenitor of \citet{sukhbold2018} using the 1) Levermore, 2) Levermore with Shibata interpolation (denoted by Levermore (S) hereafter), 3) Minerbo, 4) Wilson, {5) Kersahw, and 6) MEFD} closures as representative closures. Figure \ref{fig:2D-evolution} shows the temporal evolution of the shock radii, neutrino luminosities, heating rates, and spectra of electron-type neutrinos. Same as in 1D, differences in luminosities and spectra are quite minor. In the shock radius panel, {we can see that all simulations have comparable shock velocities. The Levermore, Levermore (S), Minerbo, and MEFD closures explode at roughly the same time, while the Wilson and Kershaw closures seem to explode a bit earlier based on the mean shock radii}. From the 1D shock evolution plot (top right panel in Figure \ref{fig:rs-all}), we know that the Si-O interface accretes at about 0.2 seconds after bounce. 
The differences between the Wilson and other closures can also be seen in the heating rate panel, in which the Wilson closure has a higher heating rate after 0.03 seconds and until it explodes at $\sim$0.2 seconds. The Minerbo closure has a higher heating rate after $\sim$0.5 seconds; this is probably due to the different explosion morphology, itself in part a consequence of the chaotic nature of the turbulent flow.

Figure \ref{fig:2D-profiles} shows the angle-averaged profiles of density, $Y_e$, entropy, and temperature. Profiles at $t-t_0=0.05$, $0.1$ and $0.4$ seconds after bounce are plotted. For profiles at 0.05 and 0.1 seconds after bounce, the conclusion is the same as in 1D. Differences in $S$ and $T$ occur only near the stalled shock. The $\rho$ and $Y_e$ profiles are stretched to comport with their shock radii. The profiles at $\sim$0.4 seconds have larger differences, especially the entropy profile. However, this is likely due to the different explosion morphologies caused by chaos, rather than due to the differences in the closures, since that model too explodes at $\sim$0.2-0.3 seconds after bounce. 

In general, these 2D simulations manifest only small differences between closure choices. As shown in Figure \ref{fig:shibata-compare}, the Levermore (S) model deviates much less from the Levermore closure than from the Minerbo closure in 1D. Moreover, the 2D simulation results deviate from those employing the Levermore closure at roughly the same level. This suggests that the differences caused by closure choices are smaller than the differences due to the chaotic nature of the hydrodynamics.
}

\section{Conclusion}
\label{sec:conclusion}

In this paper, multiple 1D {and 2D} time-dependent CCSN simulations are compared to study the effects of different closure choices in moment methods such as M1. We find that most closures behave similarly, while the Kershaw and the Wilson closures give $5\%-10\%$ larger shock radii {before the explosion} compared to others. Differences that arise in other physical quantities are all closely related to this shock position difference. Differences in time-dependent simulations between two types of the most widely used closures, the Levermore closure and the maximum entropy closures (MEFD or Minerbo), are quite slight. {Our 2D simulations also indicate that the closure-induced differences are not amplified in calculations at higher dimensions.}
In general, we suggest that differences caused by closure choices are small compared to effects due to choices of, for example, the nuclear equation of state {\citep{couch2013,yasin2020,Boccioli2022}}. %\tianshu{For example, \citet{yasin2020} compared different equations of state with 1D CCSN simulations. Equations of state such as LS220 lead more easily to an explosion, while equations of state such Shen or SKShen often fail to explode models, and the shock radii of the non-exploding simulations are different by up to 100\%.} 

%\tianshu{Deleted: Moreover, differences caused by closure choices are stable. In some {1D} test simulations, we evolved the first 100 milliseconds after bounce with one closure and then change to another one. The physical quantities quickly converge to the corresponding values in the simulations done purely with the second closure. This indicates that changes made in very early post-bounce phases are not the main cause of differences between closures, even though the differences in the Eddington factors at that time due to the different closures are larger (e.g. \citet{iwakami2022}). As a result, closure comparisons made in the very early post-bounce phases may not be able to capture the main effects.}

Deviations in Eddington factor profiles are not easily translated into deviations in other quantities. This means that simply comparing the Eddington factors calculated by closure relations to the results given by the MC and SN methods may not inform one which closure behaves better, since small deviations in the Eddington factor can still cause larger differences in a time-dependent CCSN simulation. To better understand how well the two-moment approximation and closure choices work, longer term CCSN simulations done by the MC and the SN methods are called for.

We have also compared the Shibata interpolation with the self-consistent 3rd-order closure relations for the Levermore and the Minerbo closures. We find that the differences caused by these two treatments are quite small. Therefore, for closures with complex 3rd-order relations like Levermore and MEFD, the Shibata interpolation $q=\frac{(3p+2)f}{5}$ may be substituted if so desired. 

Finally, we need to mention caveats to the results of this paper. {Our comparisons are made here only between 1D/2D simulations and we presume that the 3D simulations will behave similarly. We haven't looked at how closure choices may influence the turbulence field, which is different in 2D and 3D. Although we anticipate that closure choices may change the turbulence field only very slightly, testing this awaits comparisons between 3D simulations.} In addition, we have only compare closures to each other. Even though most closures give consistent results, there is still the possibility that they all deviate from a Monte Carlo or an SN time-dependent CCSN simulation in some yet-to-be-determined way. Future work should address this possibility. 

\section*{Acknowledgements}

%begin{acknowledgements}
We thank Christopher White, Matt Coleman, and David Vartanyan for many useful discussions and insights. We gratefully acknowledge support from the U.S. Department of Energy Office of Science and the Office of Advanced Scientific Computing Research via the Scientific Discovery through Advanced Computing (SciDAC4) program and Grant DE-SC0018297 (subaward 00009650) and support from the U.S. National Science Foundation (NSF) under Grants AST-1714267 and PHY-1804048 (the latter via the Max-Planck/Princeton Center (MPPC) for Plasma Physics).  We also acknowledge access to the Frontera cluster (under awards AST20020 and AST21003), and this research is part of the Frontera computing project at the Texas Advanced Computing Center \citep{Stanzione2020}. Frontera is made possible by NSF award OAC-1818253. Additionally, a generous award of computer time was provided by the INCITE program, enabling this research to use resources of the Argonne Leadership Computing Facility, a DOE Office of Science User Facility supported under Contract DE-AC02-06CH11357. Finally, the authors acknowledge computational resources provided by the high-performance computer center at Princeton University, which is jointly supported by the Princeton Institute for Computational Science and Engineering (PICSciE) and the Princeton University Office of Information Technology, and our continuing allocation at the National Energy Research Scientific Computing Center (NERSC), which is supported by the Office of Science of the U.S. Department of Energy under contract DE-AC03-76SF00098.

%\end{acknowledgements}

\bibliography{sample}

\section*{Appendix}
\label{sec:appendix}
Here, we summarize the derivations of the 3rd-moment closures of the Levermore, Minerbo, and MEFD closures.
\subsection*{Levermore}
Assume that there exists an inertial frame in which the radiation field is isotropic, i.e., $\mathcal{F}(E,\mu)=\mathcal{F}(E)$. Under a Lorentz transformation, a photon with energy $E$ and direction $\mu$ becomes $\tilde{E}=\gamma(1+\beta\mu)E$ and $\tilde{\mu}=\frac{\mu+\beta}{1+\beta\mu}$. Since $Ed^3x$, $\frac{d^3p}{E}$ and $dN=\mathcal{F}(E,\mu)d^3xd^3p$ are both Lorentz invariant, we get $\tilde{\mathcal{F}}(\tilde{E},\tilde{\mu})=\mathcal{F}(E,\mu)$. Therefore, the flux factor is
\equ{
f&&=\frac{\int \tilde{E}\tilde{\mu}\tilde{\mathcal{F}}(\tilde{E},\tilde{\mu})d^3\tilde{p}}{\int \tilde{E}\tilde{\mathcal{F}}(\tilde{E},\tilde{\mu})d^3\tilde{p}} 
=\frac{\int \tilde{E}^2\tilde{\mu}\mathcal{F}(E)d^3p}{\int \tilde{E}^2\mathcal{F}(E)d^3p} \nonumber\\
&&=\frac{\int (1+\beta\mu)(\mu+\beta)d\mu}{\int (1+\beta\mu)^2d\mu}
=\frac{4\beta}{3+\beta^2}\, .
}
Similarly, the $p$ and $q$ parameters are given by
\equ{
p&&=\frac{1+3\beta^2}{3+\beta^2} \nonumber\\
q&&=\frac{\beta(3-8\beta^2+9\beta^4)-3(1-\beta^2)^3\text{arctanh}(\beta)}{\beta^4(3+\beta^2)}\, .
}
Next, solve for $\beta$ from $f$ we get $\beta = \frac{2-\sqrt{4-3f^2}}{f}$ and 
\equ{
p=&&\frac{5-2\sqrt{4-3f^2}}{3} \nonumber\\
q=&&\frac{1}{\left(-2+a\right)^5}\Bigg(4 f^3 \left(286-89 a\right)+576 f \left(-2+a\right)+3 f^5 \left(-80+9
a\right) \nonumber\\&&
\left.-48 \left(f^6+f^2 \left(42-15 a\right)+3 f^4 \left(-5+a\right)+16 \left(-2+a\right)\right)
\text{\rm arctanh}\left[\frac{-2+a}{f}\right]\right)\nonumber\\
a&=&\sqrt{4-3 f^2}\, .
}
These formulae are equivalent to those provided in \citet{vaytet2011}.

\subsection*{Minerbo and MEFD}
Let $\mathcal{F}(\mu)$ be the angular distribution and $\mu=\cos(\theta)$. The entropy of the radiation field is given by
\equ{
S[\mathcal{F}(\mu)] \propto -k(1-k\mathcal{F}(\mu))\ln(1-k\mathcal{F}(\mu))-\mathcal{F}(\mu)\ln \mathcal{F}(\mu)
}
where k=-1 and 1 corresponds to Bose-Einstein and Fermi-Dirac statistics. For Maxwell-Boltzmann statics, the entropy is
\equ{
S[\mathcal{F}(\mu)] \propto -\mathcal{F}(\mu)\ln \mathcal{F}(\mu)\, .
}
By maximizing this functional under the constraints that the zeroth and first angular moment of the distribution function are equal to the phase space occupancy and the flux factor is given by:
\equ{
e&&=\frac{1}{4\pi}\int_{0}^{2\pi}d\phi\int_{-1}^1\mathcal{F}(\mu)d\mu \nonumber\\
f&&=\frac{1}{4\pi e}\int_{0}^{2\pi}d\phi\int_{-1}^1\mu \mathcal{F}(\mu)d\mu
}
the general form of the maximum entropy distribution is found to be \citep{minerbo1978,cernohorsky1994}
\equ{
\mathcal{F}(\mu) = \frac{1}{e^{\eta-\alpha\mu}+k}\, ,
}
where the values of $\eta$ and $\alpha$ are solved from the constraints. Generally these values can only be solved numerically, but in certain limits they can be calculated analytically.

In the Maxwell-Boltzmann limit, $k=0$ and the constraints become
\equ{
f=\coth\alpha-\frac{1}{\alpha}\, .
}
Thus, the 2nd and 3rd moment closures are given by
\equ{
p&&=\frac{\alpha^2+2-2\alpha\coth\alpha}{\alpha^2}=1-\frac{2f}{\alpha} \nonumber\\
q&&=\frac{\alpha(6+\alpha^2)\coth\alpha-3(2+\alpha^2)}{\alpha^3}=f+\frac{6f}{\alpha^2}-\frac{2}{\alpha}\, .
}

The function $L(x)=\coth x-\frac{1}{x}$ is the Langevin function, and its inverse can be approximated by a polynomial fit \citep{cernohorsky1994,just2015}:
\equ{
\alpha=L^{-1}(f)=\frac{15f}{5-3f^2+f^3-3f^4}\, .
}
Substituting this approximation into the closures, we get the Minerbo closure:
\equ{
p&&= \frac{1}{3}+\frac{2}{15}(3f^2-f^3+3f^4) \nonumber\\
q&&= \frac{f}{75}(45+10f-12f^2-12f^3+38f^4-12f^5+18f^6)\, .
}

The Fermi-Dirac case is much more complex. First, we derive some useful relations. Let $\mathcal{F}(\eta,\alpha,n)=\int_{-1}^1\frac{\mu^nd\mu}{e^{\eta-\alpha\mu}+1}$. Then, we know that: 
\equ{
\mathcal{F}(\eta,-\alpha,n)&&=(-1)^n\mathcal{F}(\eta,\alpha,n) \nonumber\\
\mathcal{F}(-\eta,-\alpha,n)&&=\int_{-1}^{1}\mu^n(1-\frac{1}{e^{\eta-\alpha\mu}+1})d\mu \nonumber\\
&&=\left.\frac{1}{n+1} \right\vert_{-1}^{1}-\mathcal{F}(\eta,\alpha,n)\, .
}
Combining these two relations, we get
\equ{
\mathcal{F}(-\eta,\alpha,n)-\frac{1}{n+1}&&=-\left(\mathcal{F}(\eta,\alpha,n)-\frac{1}{n+1}\right),\,\,\,\,\text{if $n$ is even,} \nonumber\\
\mathcal{F}(-\eta,\alpha,n)&&=\mathcal{F}(\eta,\alpha,n),\,\,\,\,\text{if $n$ is odd\, .}
}
The multipliers $\eta$ and $\alpha$ are numerically solved from the constraints. Actually, one can easily show that:
\equ{
\exp(\eta)=\frac{\sinh(1-e)\alpha}{\sinh(e\alpha)}\label{equ:eta}\, .
}
Thus, $\eta'=\eta$ with fixed $\alpha$ simply means $e'=1-e$.

Now, we consider the extreme case when $\alpha\rightarrow\infty$. The angular distribution function is a step function and the integrals can be done easily \citep{cernohorsky1994}:
\equ{
f_{\rm max}&&=1-e \nonumber\\
p_{\rm max}&&=1-2e+\frac{4}{3}e^2 \nonumber\\
q_{\rm max}&&=1-3e+4e^2-2e^3\, .
}
Then, we define some auxiliary functions and replace $\eta$ with Equation \ref{equ:eta}:
\equ{
x(e,\alpha)&&=\frac{f}{f_{\rm max}} \nonumber\\
\chi(e,\alpha)&&=\frac{p-\frac{1}{3}}{p_{\rm max}-\frac{1}{3}} \nonumber\\
\psi(e,\alpha)&&=\frac{q}{q_{\rm max}}\, .
}
With the relations derived above, we see that $x(e,\alpha)=x(1-e,\alpha)$, $\chi(e,\alpha)=\chi(1-e,\alpha)$, and $\psi(e,\alpha)=\psi(1-e,\alpha)$. After numerically eliminating $\alpha$ with $x$, we have
\equ{
\chi(e,x)&&=\chi(1-e,x) \nonumber\\
\psi(e,x)&&=\psi(1-e,x)\, .
}
The analytic formula derived in \citet{cernohorsky1994} assumes that $\chi(e,x)=\chi(x)$ is independent on $e$. This approximation is accurate to $0.1\%$. Then, the functional form of $\chi(x)$ is given by the $e\rightarrow0$ case, which is $\chi(x)=L^{-1}(x)$ where $L(x)=\coth x-1/x$ is the Langevin function. However, $\psi(e,x)$ strongly depends upon $e$ and this approximation can't be generalized to the 3rd-order moment closure relation. %See Figure \ref{fig:mefd}.

\clearpage

%\begin{figure}
%    \centering
%    \includegraphics[width=0.8\textwidth]{figures/chi-psi.png}
%    \caption{$\chi$ and $\psi$ as a function of $e$ at given $x$ for the MEFD closure. Blue curves are $\chi$s and red curves are $\psi$s. Note that $\psi$ strongly depends on $e$, while $\chi$ is almost constant. There is a $0.1\%$ level difference between $e=0$ and $e=0.5$ in the $\chi$ curves.}
%    \label{fig:mefd}
%\end{figure}

\end{document}